\documentclass[journal,12pt,onecolumn,draftclsnofoot]{IEEEtran}

\IEEEoverridecommandlockouts
\usepackage{cite}
\usepackage{url}
\usepackage{enumerate}

\usepackage[T1]{fontenc}

\usepackage{booktabs}
\usepackage{multirow}
\usepackage{subcaption}

\usepackage[font=footnotesize]{caption}
\usepackage[font=footnotesize]{subcaption}

\usepackage[cmex10]{amsmath}
\usepackage{amssymb}
\usepackage{amsthm}

\usepackage[ruled,vlined]{algorithm2e}
\SetKw{Break}{break}

\usepackage{tikz}
\usetikzlibrary{matrix,arrows,shapes,positioning,chains,scopes,decorations.markings,calc,patterns,backgrounds,fit}
\tikzset{->-/.style={decoration={markings,mark=at position .5 with {\arrow{>}}},postaction={decorate}}}

\tikzset{endblk/.style={
    rounded rectangle,minimum size=6mm,
    thick, draw,
    align=center,midway,
    font=\small}
    }
\tikzset{process/.style={
    rectangle,minimum size=6mm,
    thick, draw,
    align=center,midway,
    font=\small}
}
\tikzset{conditional/.style={
    shape aspect=3,rounded corners=2mm,
    diamond,minimum size=6mm,
    thick, draw,
    align=center,midway,
    font=\small}
}

\tikzset{skip loop/.style={to path={-- ++(0,#1) |- (\tikztotarget)}}}

\tikzset{point/.style={coordinate},>=stealth',draw=black!70,
  arrow/.style={->},every join/.style={rounded corners},
  hv path/.style={to path={-| (\tikztotarget)}},
  vh path/.style={to path={|- (\tikztotarget)}},
  lyes/.style={label=177:yes},
  lno/.style={label=177:no},
  ryes/.style={label=3:yes},
  rno/.style={label=3:no},
  bno/.style={label=-93:no},
  byes/.style={label=-93:yes},
}

\usepackage{pgfplots}
\usepackage{pgfplotstable}
\usepackage{booktabs}
\usepackage{colortbl}

\makeatletter
\newcommand{\removelatexerror}{\let\@latex@error\@gobble}
\makeatother

\newtheorem{lemma}{Lemma}
\newtheorem{theorem}{Theorem}
\newtheorem{corollary}{Corollary}
\theoremstyle{definition}

\DeclareMathOperator*{\argmin}{arg\,min}
\DeclareMathOperator*{\argmax}{arg\,max}

\newcommand{\bin}{\mathsf{Binom}}
\newcommand{\bet}{\mathsf{Beta}}
\newcommand{\mle}{\hat{p}_\mathsf{MLE}}
\newcommand{\minimax}{\hat{p}_\mathsf{MM}}
\newcommand{\wbu}{\hat{p}_\mathsf{Bayes}}

\newcommand{\pgb}{p_{\mathsf{G}\mathsf{B}}}
\newcommand{\pbg}{p_{\mathsf{B}\mathsf{G}}}
\newcommand{\pg}{p_{\mathsf{G}}}
\newcommand{\pb}{p_{\mathsf{B}}}

\newcommand{\tmax}{t_\text{max}^\mathcal{L}}
\newcommand{\Eindep}{E_\mathsf{indep}}
\newcommand{\EGE}{E_\mathbf{GE}}

\begin{document}

\title{BAR: Blockwise Adaptive Recoding \\for Batched Network Coding}

\author{\IEEEauthorblockN{
		Hoover~H.~F.~Yin, Shenghao~Yang, Qiaoqiao~Zhou, Lily~M.~L.~Yung, and Ka~Hei~Ng
	}
	\thanks{This paper was presented in part at the 2016 IEEE International Symposium on Information Theory \cite{adaptive} and will be presented in part at 2021 IEEE International Symposium on Information Theory \cite{yin21impact}.}
	\thanks{H.~Yin is with the Institute of Network Coding, The Chinese University of Hong Kong, Hong Kong, China.
		S.~Yang is with the School of Science and Engineering, The Chinese University of Hong Kong, Shenzhen, Shenzhen, China.
		He is also with Shenzhen Key Laboratory of IoT Intelligent Systems and Wireless Network Technology and Shenzhen Research Institute of Big Data, Shenzhen, China.
		Q.~Zhou is with the Department of Information Engineering, The Chinese University of Hong Kong, Hong Kong, China.
		L.~Yung is with the Department of Computer Science and Engineering, The Chinese University of Hong Kong, Hong Kong, China.
		K.~Ng is with the Department of Physics, The Chinese University of Hong Kong, Hong Kong, China.
		Emails: \mbox{hfyin@inc.cuhk.edu.hk}, \mbox{shyang@cuhk.edu.cn}, \mbox{zq115@ie.cuhk.edu.hk}, \mbox{lily@link.cuhk.edu.hk}, \mbox{kaheicanaan@link.cuhk.edu.hk}
	}
	\thanks{This work was funded in part by the Shenzhen Science and Technology Innovation Committee (Grant JCYJ20180508162604311, ZDSYS20170725140921348).}
}

\maketitle

\begin{abstract}
Multi-hop networks become popular network topologies in various emerging Internet of things applications.
Batched network coding (BNC) is a solution to reliable communications in such networks with packet loss.
By grouping packets into small batches and restricting recoding to the packets belonging to the same batch, BNC has a much smaller computational and storage requirements at the intermediate nodes compared with a direct application of random linear network coding.
In this paper, we propose a practical recoding scheme called blockwise adaptive recoding (BAR) which learns the latest channel knowledge from short observations so that BAR can adapt to the fluctuation of channel conditions.
We focus on investigating practical concerns such as the design of efficient BAR algorithms.
We also design and investigate feedback schemes for BAR under imperfect feedback systems.
Our numerical evaluations show that BAR has significant throughput gain for small batch size compared with the existing baseline recoding scheme.
More importantly, this gain is insensitive to inaccurate channel knowledge.
This encouraging result suggests that BAR is suitable to be realized in practice as the exact channel model and its parameters could be unknown and subject to change from time to time.
\end{abstract}

\section{Introduction}

Noise, interference and congestion are common causes of packet loss in network communications.
Usually, a packet has to travel through multiple hops before it can arrive at the destination node.
Traditionally, the intermediate nodes apply the store-and-forward strategy.
In order to maintain a reliable communication, retransmission is a common practice.
Feedback mechanism is applied so that a network node can be acknowledged that a packet is lost.
However, due to the delay and the bandwidth consumption of the feedback packets, retransmission schemes come with a cost of degraded system performance.

\emph{Random linear network coding (RLNC)} \cite{random,random2}, which is a simple realization of network coding \cite{flow,alg,linear}, can achieve the capacity of multi-hop networks with packet loss even without the needs of feedback \cite{rate,Lun2008}.
Unfortunately, a direct application of RLNC induces an enormous overhead for the coefficient vectors, and also high computational and storage costs of network coding operations at the intermediate nodes, where those intermediate nodes are usually routers or embedded devices having low computational power and storage space.

\emph{Batched network coding (BNC)} \cite{chou03,Heidarzadeh2010,Mahdaviani12,Silva2009,yang14bats} is a practical variation of RLNC, which resolves the issues of RLNC by encoding the packets for transmission into small \emph{batches} of coded packets, and then applying RLNC on the coded packets belonging to the same batch.
BATS codes \cite{yang14bats,bats_book}, which are a class of BNC, have a close-to-optimal achievable rate where the achievable rate is upper bounded by the expectation of the rank distribution of the \emph{batch transfer matrices} that model the end-to-end network operations (packet erasures, network coding operations, etc.) on the batches \cite{yang11x2}.
This hints that the network coding operations, which also known as \emph{recoding}, have an impact on the throughput of BNC.

\emph{Baseline recoding} is the simplest recoding scheme which generates the same number of recoded packets for every batch.
However, the throughput of baseline recoding is not optimal with finite batch sizes \cite{yang14a}.
The idea of \emph{adaptive recoding}, which aims to outperform baseline recoding by generating different numbers of recoded packets for different batches, was proposed in \cite{yang14a} without truly optimizing the numbers.
Two adaptive recoding optimization models for independent packet loss channels were formulated independently in \cite{scheduling} and the conference version of this paper \cite{adaptive}.
A unified adaptive recoding framework was proposed in \cite{uni} which can subsume both optimization models and support other channel models under certain conditions.

Although adaptive recoding can be applied distributively with local network information, it is a challenge to obtain accurate local information when we deploy adaptive recoding in real-world scenarios.
Adaptive recoding requires two pieces of information: The distribution of the information remained in the received batches and the channel condition of the outgoing link.

The first piece of information may change over time if the channel condition of the incoming link varies.
One reason of the variation is that the link quality can be affected by the interference from the users of other networks around the network node.
A way to adapt to this variation is to group a few batches into a block and observe the distribution from the received batches in this block.
We call this approach \emph{blockwise adaptive recoding (BAR)}.

The second piece of information may also vary from time to time.
In some scenarios such as deep-space and underwater communications, feedback can be expensive or is not available at all so that a feedbackless network is preferred.
Without feedback, we cannot update our knowledge on the channel condition of the outgoing link.
Although we may assume an unchanged channel condition and measure some information such as the packet loss rate of the channel beforehand, this measurement, however, can be inaccurate due to observational errors or precision limits. 

In this paper, we focus on the practical design to apply BAR in real-world applications.
Specifically, we answer the following questions in this paper:
\begin{enumerate}
	\item How does the block size affect the throughput?
	\item Is BAR sensitive to an inaccurate channel condition?
	\item How to calculate the components of BAR and solve the optimization efficiently?
	\item How to make use of link-by-link feedback if it is available?
\end{enumerate}

The first question is related to the trade-off between throughput and delay: A larger block induces a longer delay but gives a higher throughput.
We show by numerical evaluations that a small block size can already give a significant throughput gain compared with baseline recoding.

For the second question, we demonstrate that BAR performs very well with an independent packet loss model on channels with dependent packet loss.
We also show that BAR is insensitive to an inaccurate packet loss rate.
This is an encouraging result as this suggests that it is feasible to apply BAR in real-world applications.

The third question is important in practice as BAR is suppose to run at network nodes which are usually routers or embedded devices having limited computational power but at the same time they have to handle a huge amount of network traffic.
Also, by updating the knowledge of the incoming link from a short observation, we need to recalculate the components of BAR and solve the optimization problem again.
In light of this, we want to reduce the number of computations to improve the reaction time and reduce the stress of congestion.
We answer this question by proposing an on-demand dynamic programming approach to build the components, a efficient greedy algorithm to solve BAR, and an approximation scheme to speed up the greedy algorithm.

Lastly for the fourth question, we consider both a perfect feedback system (e.g., the feedback passes through a side-channel with no packet loss) and a lossy feedback system (e.g., the feedback uses the reverse direction of the lossy channel for data transmission).
We investigate a few ways to estimate the packet loss rate and show that we can further boost the throughput by using feedback.
Also, a rough estimation is sufficient to catch up the variation of the channel condition.
In other words, unless there is another application which requires a more accurate estimation on the packet loss rate, we may consider to use an estimation with low computational cost, e.g., the maximum likelihood estimator.

The paper is organized as follows.
We first formulate BAR and introduce some of its properties in Section~\ref{sec:bar}.
Then, we propose some algorithms to solve BAR efficiently and evaluate the throughput in Section~\ref{sec:algo}.
In Section~\ref{sec:impact+feedback}, we demonstrate that BAR is insensitive to inaccurate channel models and investigate the use of feedback mechanism.
At last, we conclude the paper in Section~\ref{sec:conclusion}.

\section{Blockwise Adaptive Recoding}
\label{sec:bar}

In this section, we briefly introduce batched network coding (BNC) and then formulate blockwise adaptive recoding (BAR).

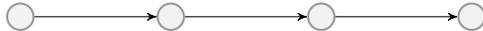
\begin{figure}
	\centering
	\begin{tikzpicture}[font=\footnotesize,dot/.style={circle,draw=gray!80,fill=gray!10,thick,inner
		sep=1pt,minimum size=10pt}]
		\node[dot] (s) at(-2,0) {};
		\node[dot] (a1) at(0,0) {} edge[<-] (s);
		\node[dot] (a2) at(2,0) {} edge[<-] (a1);
		\node[dot] (t) at(4,0) {} edge[<-] (a2);
	\end{tikzpicture}
	\caption{A three-hop line network.
	Network links only exist between two neighboring nodes.
	}
	\label{fig:line}
\end{figure}

We consider line networks in this paper as they are the fundamental building blocks of a general network.
A recoding scheme for line networks can be extended for general unicast networks and certain multicast networks \cite{yang14bats,scheduling}.
A line network is a sequence of network nodes where network links only exist between two neighboring nodes.
An example of a line network is illustrated in Fig.~\ref{fig:line}.

\subsection{Batched Network Coding}

Suppose we want to send a file from a source node to a destination node through a multi-hop line network.
The file is divided into multiple \emph{input packets}, where each packet is regarded as a vector over a fixed finite field.
A batched network code (BNC) has three main components: the encoder, the recoder and the decoder.

An encoder of a BNC is applied at the source node to generate batches from the input packets, where each batch consists of a small number of coded packets.
To generate a batch, the encoder samples a predefined \emph{degree distribution} to obtain a \emph{degree}, where the degree is the number of input packets contributed to the batch.
Depends on the application, there are various ways to formulate the degree distribution \cite{tree,yang18finite,sliding,expanding}.
According to the degree, a set of packets is chosen randomly from the input packets.
Each packet in the batch is formed by taking random linear combinations on the chosen set of packets.
The encoder generates $M$ packets per batch, where $M$ is known as the \emph{batch size}.

Each packet in a batch has a coefficient vector attached to it.
Two packets in a batch are linearly independent of each other if and only if their coefficient vectors are linearly independent of each other.
Right after a batch is generated, the packets in it are defined to be linearly independent of each other.
This can be accomplished by suitably choosing the initial coefficient vectors.

A recoder is applied at each intermediate node, which performs network coding operations to the received batches to generate \emph{recoded packets}.
This procedure is known as \emph{recoding}.
Some packets of a batch may be lost when they pass through a network link.
Each recoded packet of a batch is formed by taking random linear combination on the received packets of this batch.
The number of recoded packets depends on the recoding scheme.
For example, baseline recoding generates the same number of recoded packets for every batch.
Optionally, we can also apply a recoder at the source node so that we can have more than $M$ packets per batch at the beginning. 
After recoding, the recoded packets are sent to the next network node.

At the destination node, a decoder is applied to recover the input packets.
Depends on the specific BNC, we can use different decoding algorithms such as Gaussian elimination, belief propagation and inactivation \cite{Raptormono,inactivation}.

\subsection{Expected Rank Functions} \label{sec:exp_rank}

Define the \emph{rank} of a batch at a network node by the number of linearly independent packets remained in the batch, which is a measure on the amount of information carried by the batch.
Adaptive recoding aims to maximize the sum of the expected value of the rank distribution of each batch arriving at the next network node. 
For simplicity, we called this expected value the \emph{expected rank}.

For a batch $b$, denote the rank of $b$ by $r_b$ and the number of recoded packets to be generated for $b$ by $t_b$.
The expectation of the rank of $b$ at the next network node, denoted by $E(r_b,t_b)$, is known as the \emph{expected rank function}.
We have
\begin{equation} \label{eq:ert_zeta}
	E(r, t) = \sum_{i = 0}^t \Pr(X_t = i) \sum_{j = 0}^{\min\{i,r\}} j \zeta_j^{i,r},
\end{equation}
where $X_t$ is the random variable of the number of packets of a batch received by the next network node when we send $t$ packets for this batch at the current node, and $\zeta_j^{i,r}$ is the probability that a batch of rank $r$ at the current node with $i$ received packets at the next network node has rank $j$ at the next network node.\footnote{%
	Systematic recoding \cite{yang14a,bats_book}, which regards the received packets as recoded packets, can achieve a nearly indistinguishable performance compared with the one which generates all recoded packets by taking random linear combinations \cite{bats_book}.
	So, we can also use \eqref{eq:ert_zeta} to approximate the expected rank functions for systematic recoding accurately.
}
The exact formulation of $\zeta_j^{i, r}$ can be found in \cite{yang14bats}, which is
	$\zeta_j^{i, r} = \frac{\zeta_j^i \zeta_j^r}{\zeta_j^j q^{(i-j)(r-j)}}$,
where $q$ is the field size for the linear algebra operations and
	$\zeta_j^m = \prod_{k = 0}^{j-1} (1-q^{-m+k})$.
%
It is convenience to use $q = 2^8$ in practice as each symbol in this field can be represented by $1$ byte.
For a sufficient large field size, say, $q = 2^8$, $\zeta_j^{i,r}$ is very close to $1$ if $j = \min\{i,r\}$ and is very close to $0$ otherwise.
That is, we can approximate $\zeta_j^{i,r}$ by $\delta_{j,\min\{i,r\}}$ where $\delta_{\cdot, \cdot}$ is the Kronecker delta.
This approximation is also used in literature such as \cite{ge_adaptive,zhou17b}.

\begin{table}
	\tiny
  \centering 
	\caption{Percentage error when approximating expected rank functions}
  \label{tab:pererr}
  \begin{filecontents*}{table.txt}
1	0.39216   0.00153   0.00001   0.00000   0.00000   0.00000   0.00000   0.00000   0.00000   0.00000   0.00000   0.00000   0.00000   0.00000   0.00000   0.00000
2	0.13140   0.15741   0.00061   0.00000   0.00000   0.00000   0.00000   0.00000   0.00000   0.00000   0.00000   0.00000   0.00000   0.00000   0.00000   0.00000
3	0.03841   0.08032   0.08397   0.00033   0.00000   0.00000   0.00000   0.00000   0.00000   0.00000   0.00000   0.00000   0.00000   0.00000   0.00000   0.00000
4	0.01025   0.03091   0.05791   0.05042   0.00020   0.00000   0.00000   0.00000   0.00000   0.00000   0.00000   0.00000   0.00000   0.00000   0.00000   0.00000
5	0.00258   0.01024   0.02761   0.04398   0.03229   0.00013   0.00000   0.00000   0.00000   0.00000   0.00000   0.00000   0.00000   0.00000   0.00000   0.00000
6	0.00062   0.00308   0.01091   0.02502   0.03416   0.02155   0.00008   0.00000   0.00000   0.00000   0.00000   0.00000   0.00000   0.00000   0.00000   0.00000
7	0.00015   0.00087   0.00382   0.01147   0.02258   0.02685   0.01479   0.00006   0.00000   0.00000   0.00000   0.00000   0.00000   0.00000   0.00000   0.00000
8	0.00003   0.00023   0.00122   0.00457   0.01178   0.02023   0.02123   0.01036   0.00004   0.00000   0.00000   0.00000   0.00000   0.00000   0.00000   0.00000
\end{filecontents*}
\pgfplotstabletypeset[
  every head row/.style={
    before row=\toprule,after row=\midrule,},
  every last row/.style={
    after row=\bottomrule},
  every even row/.style={
    before row={\rowcolor[gray]{0.9}}},
  columns={0,1,2,3,4,5,6,7,8},
  columns/0/.style={column name=$t$},
  columns/1/.style={column name={$r=1$},fixed,fixed zerofill,precision=3},
  columns/2/.style={column name={$r=2$},fixed,fixed zerofill,precision=3},
  columns/3/.style={column name={$r=3$},fixed,fixed zerofill,precision=3},
  columns/4/.style={column name={$r=4$},fixed,fixed zerofill,precision=3},
  columns/5/.style={column name={$r=5$},fixed,fixed zerofill,precision=3},
  columns/6/.style={column name={$r=6$},fixed,fixed zerofill,precision=3},
  columns/7/.style={column name={$r=7$},fixed,fixed zerofill,precision=3},
  columns/8/.style={column name={$r=8$},fixed,fixed zerofill,precision=3},
  ]{table.txt}
\end{table}

For the independent packet loss model with packet loss rate $p$, we have $X_t \sim \bin(t, 1-p)$, a binomial distribution.
If $p = 1$, then a store-and-forward technique can guarantee the maximal expected rank.
If $p = 0$, then no matter how many packets we transmitted, the next network node must receive no packet.
So, we assume $0 < p < 1$ in this paper.\footnote{%
	It is easy to prove that the results in this paper are also valid for $p = 0$ or $1$ when we define $0^0 := 1$, which is a convention in combinatorics such that $\bin(t,0)$ and $\bin(t,1)$ are well-defined with a correct interpretation.
}
We demonstrate the accuracy of the approximation $\zeta_j^{i,r} \approx \delta_{j,\min\{i,r\}}$ by showing the percentage error of the expected rank function corrected to $3$ decimal places when $q = 2^8$, $p = 0.2$ and $X_t \sim \bin(t, 1-p)$ in Table~\ref{tab:pererr}.
We can see that only three pairs of $(r,t)$ have percentage errors larger than $0.1\%$, where they occur when $r,t\leq 2$.
For all the other cases, the percentage errors are less than $0.1\%$. 
Therefore, such approximation is accurate enough for practical applications. 
In the remaining text, we assume $\zeta_j^{i,r} = \delta_{j,\min\{i,r\}}$.
That is, for independent packet loss model, we have
\begin{equation} \tag{E-indep} \label{eq:ert}
	\Eindep(r,t) = \sum_{i = 0}^t \binom{t}{i} (1-p)^i p^{t-i} \min\{i, r\}.
\end{equation}

\begin{figure}
	\centering
	\begin{tikzpicture}[scale=0.7,font=\footnotesize,
		state/.style={circle,draw=black,thick,inner sep=1pt,minimum size=20pt,align=center},->,>=stealth',thick,every node/.style={transform shape}]
		\node[state] (G) at (0,0) {\textbf{G}\\$\pg$};
		\node[state] (B) at (3,0) {\textbf{B}\\$\pb$};
		\path (G) edge [bend left] node[below] {$\pgb$} (B);
		\path (B) edge [bend left] node[above] {$\pbg$} (G);
		\path (G) edge [loop left] node[left] {$1-\pgb$} (G);
		\path (B) edge [loop right] node[right] {$1-\pbg$} (B);
	\end{tikzpicture}
	\caption{A Gilbert-Elliott model. 
		In each state, there is an independent event to decide whether a packet is lost or not.
	}
	\label{fig:ge_model}
\end{figure}
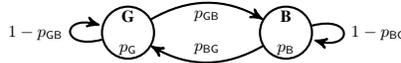

We also consider the expected rank functions for burst packet loss channels modelled by Gilbert-Elliott (GE) models \cite{GilbertBurst,ElliottBurst}, where the GE model was also used in other literature of BNC such as \cite{ge_adaptive,bittt_space}.
A GE model is a $2$-state Markov chain as illustrated in Fig.~\ref{fig:ge_model}.
In each state, there is an independent event to decide whether a packet is lost or not.
Define $f(s, i, t) := \Pr(S_t = s, X_t = i)$, where $S_t$ is the random variable of the state of the GE model after sending $t$ packets of the batch.
By exploiting the structure of the GE model, the computation of $f$ can be done by dynamic programming. 
Then, we have
\begin{equation} \tag{E-GE} \label{eq:ert_ge}
	\EGE(r,t) = \sum_{i = 0}^t (f(\mathbf{G}, i, t) + f(\mathbf{B}, i, t)) \min\{i, r\}.
\end{equation}

It is easy to see that we would take more steps to compute \eqref{eq:ert_ge} than compute \eqref{eq:ert}.
So, a natural question to ask is that for burst packet loss channels, is the throughput gap between adaptive recoding with \eqref{eq:ert} and \eqref{eq:ert_ge} small?
We would demonstrate in Sec.~\ref{sec:inaccurate} that, yes, the gap is small so that we can use \eqref{eq:ert} at all time to get a nice throughput.
Therefore, we mainly focus our investigation on \eqref{eq:ert}.

In the rest of this paper, we refer $E(r,t)$ to $\Eindep(r,t)$ unless otherwise specified.
We first give the recursive formula for $E(r,t)$.
For integers $r \ge 0$ and $t \ge -1$, define
\begin{equation} \label{eq:beta}
	\beta_p(t,r) = \begin{cases}
		1 & \text{if } t \le r-1,\\
		\sum_{i = 0}^{r-1} \binom{t}{i} (1-p)^i p^{t-i} & \text{otherwise}.
	\end{cases}
\end{equation}
When $t \ge 0$, the function $\beta_p(t,r)$ is the partial sum of the probability masses of a binomial distribution $\bin(t, 1-p)$.
The case $t = -1$ will be used in the approximation scheme in Section~\ref{sec:algo} and we will discuss such case in that section.

\begin{lemma} \label{lem:exp_rank2}
	$E(r,t+1) = E(r,t) + (1-p) \beta_p(t,r)$, where $t$ and $r$ are non-negative integers.
\end{lemma}

\begin{IEEEproof}
	Let $Y_i$ be i.i.d. Bernoulli random variables, where $\Pr(Y_i = 1) = 1-p$ for all $i$.
	When $Y_i = 1$, it means that the $i$-th packet is received by the next hop.

	When we transmit one more packet at the current node, $Y_{t+1}$ indicates whether this packet is received by the next network node  or not.
	If $Y_{t+1} = 0$, i.e., the packet is lost, then the expected rank will not change.
	If $Y_{t+1} = 1$, then the packet is linearly independent of all the already received packets at the next network node if the number of received packets at the next network node is less than $r$.
	That is, the rank of this batch at the next network node increases by $1$ if $\sum_{i = 1}^t Y_i < r$.
	So, the increment of $E(r,t)$ is $\Pr(Y_{t+1} = 1, \sum_{i = 1}^t Y_i < r)$.
	Note that $\sum_{i = 1}^t Y_i \sim \bin(t, 1-p)$.
	As $Y_i$ are all i.i.d., we have
		$\Pr \left( Y_{t+1} = 1, \sum_{i = 1}^t Y_i < r \right) = (1-p) \beta_p(t,r)$.
\end{IEEEproof}

The formula shown in Lemma~\ref{lem:exp_rank2} can be interpreted as: A newly received packet is linearly independent of all the already received packets with probability tends to $1$ unless the rank has already reached $r$.
This can also be interpreted as $\zeta_j^{i,r} = \delta_{j,\min\{i,r\}}$ with probability tends to $1$.

\begin{corollary} \label{cor:inc}
	$E(r, t+1) \ge E(r, t)$, where the equality holds if and only if $r = 0$.
\end{corollary}
\begin{IEEEproof}
	It is trivial that $(1-p) \beta_p(t,r) \ge 0$.
	By Lemma~\ref{lem:exp_rank2}, we have $E(r, t+1) \ge E(r, t)$.
	By the definition of $\beta_p$ in \eqref{eq:beta}, we can see that $\beta_p(t,r) = 0$ if and only if $t \ge 0$ and $r = 0$.
\end{IEEEproof}

\subsection{Blockwise Adaptive Recoding} \label{sec:adp}

Let a \emph{block} be a set of batches.
We assume that the blocks at a network node are mutually disjoint.
Blockwise adaptive recoding (BAR) is a recoding scheme which groups the batches into blocks and optimizes jointly the number of recoded packets for each batch in the block.

Fix a network node.
Suppose the node receives a block $\mathcal{L}$.
For each batch $b \in \mathcal{L}$, let $r_b$ and $t_b$ be the rank of $b$ and the number of recoded packets to be generated for $b$ respectively.
A node can only transmit a finite number of packets for a block in practice.
We denote this number by $t^{\mathcal{L}}_\text{max}$, which is an input to the optimization problem.
The following model maximizes the sum of the expected rank of the batches in the block $\mathcal{L}$:
\begin{equation}
\tag{P} \label{eq:P}
	\max_{t_b \in \{0, 1, 2, \ldots\}, \forall b \in \mathcal{L}} \sum_{b \in \mathcal{L}} E(r_b, t_b) \quad \mathrm{s.t.} \quad \sum_{b \in \mathcal{L}} t_b \le t^{\mathcal{L}}_\text{max}.
\end{equation}

The above optimization depends only on the local knowledge at the node.
The batch rank $r_b$ can be known from the coefficient vectors of the received packets of batch $b$.
The value of $t_\text{max}^\mathcal{L}$ can affect the stability of the packet buffer.\footnote{
	For a general network transmission scenario with multiple transmission sessions, the value of $t_\text{max}^\mathcal{L}$ can be determined by optimizing certain local network transmission utility.
	Though we do not discuss such optimizations in this paper, we will solve \eqref{eq:P} for a general value of $t_\text{max}^\mathcal{L}$.
}

\begin{theorem}
\eqref{eq:P} is equivalent to
\begin{equation}
\tag{B} \label{eq:B}
	\max_{t_b \in \{0, 1, 2, \ldots\}, \forall b \in \mathcal{L}} \sum_{b \in \mathcal{L}} E(r_b, t_b) \quad \mathrm{s.t.} \quad \sum_{b \in \mathcal{L}} t_b = t^{\mathcal{L}}_\text{max}.
\end{equation}
\end{theorem}

\begin{IEEEproof}
	Suppose $\{t_b^\ast\}_{b \in \mathcal{L}}$ optimizes \eqref{eq:P} with $\sum_{b \in \mathcal{L}} t_b^\ast < t_\text{max}^{\mathcal{L}}$.
	Let $\alpha_b \ge 0$ be integers such that $\sum_{b \in \mathcal{L}} \alpha_b = t_\text{max}^{\mathcal{L}} - \sum_{b \in \mathcal{L}} t_b^\ast$.
	Corollary~\ref{cor:inc} gives that the objective value with the set $\{t_b^\ast+\alpha_b\}_{b \in \mathcal{L}}$ is no less than the one with the set $\{t_b^\ast\}_{b \in \mathcal{L}}$.
	If the objective value is unchanged, it is no harm to achieve the equality in the constraint in \eqref{eq:P} with $\{t_b^\ast+\alpha_b\}_{b \in \mathcal{L}}$.
	Otherwise, it is a contradiction to the optimality, so we must have to achieve the equality in the constraint.
\end{IEEEproof}

Note that the solution of \eqref{eq:B} may not be unique.
We only need to obtain one of the solutions for recoding purpose.
In general, \eqref{eq:B} is a non-linear integer programming problem.
A linear programming variant of \eqref{eq:B} can be formulated by using a technique in \cite{wang2021smallsample}.
However, such formulation has a huge amount of constraints and requires to calculate the values of $E(r_b,t)$ for all $b \in \mathcal{L}$ and all possible $t$ beforehand.
We defer the discussion of this formulation to Appendix~\ref{sec:linear}.

Now, we formally present BAR. 
It is based on the solution of \eqref{eq:B}.
Similar to baseline recoding, the source node transmits $t_b^{(0)}$ packets for batch $b$. 
Usually we will set an integer $t_b^{(0)} \ge M$, as the first $M$ packets transmitted from the source node must be linearly independent of each other. 
We can also use \eqref{eq:B} to decide the value of $t_b^{(0)}$, i.e., we group some batches generated by the encoder into a block $\mathcal{L}$.
We will discuss in Section~\ref{sec:diff1} that we should have $|t_b^{(0)} - t_{b'}^{(0)}| \le 1$ for all $b, b' \in \mathcal{L}$ at the source node. 

A network node would keep receiving packets until it has received enough batches to form a block $\mathcal{L}$.
A packet buffer is used to store the received packets.
Then, the node solves \eqref{eq:B} and obtains the number of recoded packets for each batch in the block, i.e., $\{t_b\}_{b \in \mathcal{L}}$.
The node then generates and transmits $t_b$ recoded packets for every batch $b \in \mathcal{L}$.
At the same time, the network node keeps receiving new packets.
After all the recoded packets for the block $\mathcal{L}$ are transmitted, the node drops the block from its packet buffer and then repeats the procedure by considering another block.

The size of a block depends on its application.
For example, if an interleaver is applied to $L$ batches, we can group the $L$ batches as a block.
When $|\mathcal{L}| = 1$, the only solution is $t_b = t^\mathcal{L}_\text{max}$, which degenerates into baseline recoding.
Therefore, we need to use a block size at least $2$ in order to enjoy the throughput enhancement of BAR.
Intuitively, it is better to optimize \eqref{eq:B} with a larger block size.
This is formally stated in Theorem~\ref{thm:2blocks} below.
However, the block size is related to the transmission latency as well as the computational and storage burdens at the network nodes.
Note that we cannot conclude the exact rank of each batch in a block until the previous network node finishes sending all the packets of this block.
That is, we need to wait the previous network node for the packets of all the batches in a block until we can solve the optimization problem.
Numerical evaluations in Section~\ref{sec:numerical} show that $|\mathcal{L}| = 2$ already has obvious advantage over $|\mathcal{L}| = 1$, and it may not be necessary to use a block size larger than $8$.

\begin{theorem} \label{thm:2blocks}
	Let $\mathcal{L}$ and $\mathcal{L}'$ be two blocks.
	The sum of objectives of maximizing $\mathcal{L}$ and $\mathcal{L}'$ separately is less than or equal to the objective of maximizing the block $\mathcal{L} \cup \mathcal{L}'$.
\end{theorem}

\begin{IEEEproof}
	See Appendix~\ref{sec:proof:thm:2blocks}.
\end{IEEEproof}

\subsection{Properties of Blockwise Adaptive Recoding} \label{sec:prop}

Due to the non-linear integer programming structure of \eqref{eq:B}, we need to find some properties of the model in order to design efficient algorithms.

Define the probability mass function of the binomial distribution $\bin(t, 1-p)$ by

\begin{equation*}
	B_p(t,i) = \begin{cases}
		\binom{t}{i} (1-p)^i p^{t-i} & \text{if } 0 \le i \le t,\\
		0 & \text{otherwise}.
	\end{cases}
\end{equation*}
We can rewrite \eqref{eq:beta} in terms of $B_p(t,i)$ by
\begin{equation*}
	\beta_p(t,r) = \begin{cases}
		1 & \text{if } t \le r-1,\\
		\sum_{i = 0}^{r-1} B_p(t,i) & \text{otherwise}.
	\end{cases}
\end{equation*}

Due to the fact that $\sum_{i = 0}^t B_p(t,i) = 1$, we have
\begin{equation} \label{eq:beta_range}
	0 \le \beta_p(t,r) \le 1.
\end{equation}
Furthermore, it is easy to see that
\begin{IEEEeqnarray}{rCltrCl}
	\beta_p(t,r) & = & 0 & \text{ if and only if} & r & = & 0 \text{ and } t \ge 0; \label{eq:beta0} \\
	\beta_p(t,r) & = & 1 & \text{ if and only if} & t & \le & r-1. \label{eq:beta1}
\end{IEEEeqnarray}

A tabular form of $\beta_p$ is illustrated in Fig.~\ref{fig:bp_table} after substituting the boundaries with $0$s and $1$s.
\begin{figure}
	\centering
	\begin{tikzpicture}[scale=.73,every node/.style={scale=.73}]
		\matrix (BPSUM) [matrix of math nodes,nodes={text width={width("$B_p(1,0)$")},text height={height("$B_p(1,0)$")},align=center}] at (0,0) {
			1 & 1 & 1 & 1 & 1 \\
			0 & 1 & 1 & 1 & 1 \\
			0 & \beta_p(1,1) & 1 & 1 & 1 \\
			0 & \beta_p(2,1) & \beta_p(2,2) & 1 & 1 \\
			0 & \beta_p(3,1) & \beta_p(3,2) & \beta_p(3,3) & 1 \\
			0 & \beta_p(4,1) & \beta_p(4,2) & \beta_p(4,3) & \beta_p(4,4) \\
		};
		\draw ($(BPSUM-1-1.south)+(-.5,0)$) -- ($(BPSUM-1-5.south)+(.5,0)$);
	\end{tikzpicture}
	\caption{The tabular appearance of the function $\beta_p(t,r)$ after substituting the boundaries with $0$s and $1$s.
		The rows and columns correspond to $t = -1, 0, 1, \ldots$ and $r = 0, 1, 2, \ldots$ respectively.
		The row above the line is $\beta_p(-1,\cdot)$.
	}
	\label{fig:bp_table}
\end{figure}

The regularized incomplete beta function, defined as 
	$I_x(a,b) := \frac{\int_0^x t^{a-1} (1-t)^{b-1}\,dt}{\int_0^1 t^{a-1} (1-t)^{b-1}\,dt}$
\cite[Eq.~8.17.2]{NIST:DLMF},
can be used to express the partial sum of the probability masses of a binomial distribution.
When $t \ge r > 0$, we can apply \cite[Eq.~8.17.5]{NIST:DLMF} and obtain
\begin{equation} \label{eq:betainc0}
	\beta_p(t,r) = \sum_{i = 0}^{r-1} \binom{t}{i} (1-p)^i p^{t-i} = I_p(t-r+1,r).
\end{equation}


\begin{lemma} \label{lem:bp}
	Assume $0 < p < 1$.
	Let $\Lambda$ be an index set.
	\begin{enumerate}[(a)]
		\item $\displaystyle B_p(t+1,i) = (1-p) B_p(t,i-1) + p B_p(t,i)$ for $i = 0, 1, \ldots, t$; \label{lem:bp_rec}
		\item $\displaystyle \beta_p(t+1,r) \le \beta_p(t,r)$ where the equality holds if and only if $t+1 < r$ or $t \ge r = 0$; \label{lem:bp_dec}
		\item $\displaystyle \beta_p(t,r) \le \beta_p(t+1,r+1)$ where the equality holds if and only if $t < r$; \label{lem:bp_diag}
		\item $\displaystyle \beta_p(t,r+1) \ge \beta_p(t,r)$ where the equality holds if and only if $t < r$; \label{lem:bp_inc}
		\item $\displaystyle 1 \ge \max_{b \in \Lambda} \beta_p(t_b,r_b) \ge \beta_p(t_a+s,r_a)$ for all $a \in \Lambda$ and any non-negative integer $s$; \label{lem:bp_max}
		\item $\displaystyle 0 \le \min_{b \in \Lambda} \beta_p(t_b,r_b) \le \beta_p(t_a-s,r_a)$ for all $a \in \Lambda$ and any non-negative integer $s$ such that $t_a-s \ge -1$. \label{lem:bp_min}
	\end{enumerate}
\end{lemma}

\begin{IEEEproof}
	See Appendix~\ref{sec:proof:lem:bp}.
\end{IEEEproof}


\begin{lemma} \label{lem:exp_rank}
	Let $t$ and $r$ are non-negative integers.
	\begin{enumerate}[(a)]
		\item $E(r,t+1) = E(r,t) + (1-p)$ if $t < r$; \label{lem:exp_rank_rec}
		\item $E(r,t) = (1-p) \sum_{j = 0}^{t-1} \beta_p(j,r) = (1-p) \left( \min\{r,t\} + \sum_{j = r}^{t-1} \beta_p(j,r) \right)$. \label{lem:exp_rank_exact}
	\end{enumerate}
\end{lemma}

\begin{IEEEproof}
	See Appendix~\ref{sec:proof:lem:exp_rank}.
\end{IEEEproof}

For a batch of rank $r$ and $t$ recoded packets transmitted, Lemma~\ref{lem:exp_rank2} states that when we transmit one more recoded packet, the expected rank of the batch at the next network node is increased by $(1-p)\beta_p(t,r)$.
Define a multiset $\Omega_r$ which collects the value of $(1-p)\beta_p(t,r)$ for all integers $t \ge 0$, i.e.,
	$\Omega_r = \{(1-p)\beta_p(t,r) \colon t \in \{0,1,2,\ldots\}\}$.
%
The following lemma shows the relationship between $E(r,t)$ and $\Omega_r$.

\begin{lemma} \label{lem:exp_largest_sum}
	$E(r,t)$ equals the sum of the largest $t$ elements in $\Omega_r$.
\end{lemma}

\begin{IEEEproof}
	By Lemma~\ref{lem:exp_rank}\eqref{lem:exp_rank_exact}, $E(r,t) = (1-p) \sum_{j = 0}^{t-1} \beta_p(j,r)$.
	By Lemma~\ref{lem:bp}\eqref{lem:bp_dec}, $\beta_p(t,r)$ is a monotonic decreasing function on $t$.
	By \eqref{eq:beta_range} and \eqref{eq:beta1}, we have $\beta_p(0,r) = 1 \ge \beta_p(t,r)$ for all positive integers $t$.
	So, $E(r,t)$ is the sum of the largest $t$ elements in $\Omega_r$.
\end{IEEEproof}

When $t_\text{max}^\mathcal{L} \le \sum_{b \in \mathcal{L}} r_b$, we can find an optimal solution of \eqref{eq:B} easily by the following lemma.
However, this condition means that the value of $t_\text{max}^\mathcal{L}$ is too small such that the node has just enough or even not enough time to forward the linearly independent packets it received.

\begin{lemma} \label{lem:first_packet}
	If $t_\text{max}^\mathcal{L} \le \sum_{b \in \mathcal{L}} r_b$, then every $\{t_b\}_{b \in \mathcal{L}}$ satisfying $0 \le t_b \le r_b$ and $\sum_{b \in \mathcal{L}} t_b = t_\text{max}^\mathcal{L}$ is a solution to \eqref{eq:B}.
\end{lemma}

\begin{IEEEproof}
	For $t_b = 0, 1, \ldots, r_b-1$, every increment on $t_b$ gains $(1-p)$ to the expected rank by Lemma~\ref{lem:exp_rank}\eqref{lem:exp_rank_rec}, i.e., $E(r_b,t_b+1) = E(r_b,t_b) + (1-p)$.
	For any $t_b' \ge r_b$, Lemma~\ref{lem:exp_rank}\eqref{lem:exp_rank_rec}, \eqref{eq:beta_range} and \eqref{eq:beta1} shows that 	$E(r_b,t_b'+1) = E(r_b,t_b') + (1-p) \beta_p(t_b',r_b) < (1-p)$.
	Therefore, the first $r_b$ packets of each batch can gain the most to the expected rank.
	Note that each of the first $r_b$ packet gains $(1-p)$ regardless of the value of $r_b$.
	This concludes that if $t_\text{max}^{\mathcal{L}} \le \sum_{b \in \mathcal{L}} r_b$, then every assignment with $t_b \le r_b$ and $\sum_{b \in \mathcal{L}} t_b = t_\text{max}^{\mathcal{L}}$ can achieve the largest sum of expected rank, which equals $(1-p)t_\text{max}^{\mathcal{L}}$.
\end{IEEEproof}

Now, we consider the case that we have enough 
time to transmit more than $\sum_{b \in \mathcal{L}} r_b$ recoded packets.
The following theorem states the intuition that we should send more packets for a batch having a higher rank than a batch having a lower rank.

\begin{theorem} \label{thm:adp_recode}
	Let $\mathcal{L}$ be a block where $|\mathcal{L}| \ge 2$.
	If $\{t_b\}_{b \in \mathcal{L}}$ solves \eqref{eq:B} and $t_b \ge r_b$ for all $b \in \mathcal{L}$, then $t_m < t_n$ for all $m, n \in \mathcal{L}$ such that $r_m < r_n$.
\end{theorem}

\begin{IEEEproof}
	See Appendix~\ref{sec:proof:thm:adp_recode}.
\end{IEEEproof}

Theorem~\ref{thm:adp_recode} implies that a network node should not transmit the same number of packets for the batches having different ranks.
Baseline recoding has $t_b = M$ for all $b \in \mathcal{L}$ so it violates this condition when not all the batches have the same rank.
This means that, baseline recoding is not optimal, which is consistent with the result from \cite{yang14a}.
Next, we give the following theorem which shows the necessary and sufficient conditions for a non-optimal solution of \eqref{eq:B}.

\begin{theorem} \label{thm:not_opt}
	Let $\{t_b\}_{b \in \mathcal{L}}$ where $\sum_{b \in \mathcal{L}} t_b = t_\text{max}^{\mathcal{L}}$ be a feasible solution of \eqref{eq:B}.
	$\{t_b\}_{b \in \mathcal{L}}$ is not an optimal solution of \eqref{eq:B} if and only if there exists two distinct batches $\kappa, \rho$ with $t_\rho \ge 1$ such that
		$(1-p) \beta_p(t_\kappa,r_\kappa) > (1-p) \beta_p(t_\rho-1,r_\rho)$.
\end{theorem}

\begin{IEEEproof}
	See Appendix~\ref{sec:proof:thm:not_opt}.
\end{IEEEproof}

We can take contrapositive to obtain the necessary and sufficient conditions for an optimal solution.
Finally, the following corollary gives a contrast to Lemma~\ref{lem:first_packet}, which is also an assumption required by Theorem~\ref{thm:adp_recode}.

\begin{corollary} \label{cor:first_packet}
	Let $\{t_b\}_{b \in \mathcal{L}}$ be a solution of \eqref{eq:B}.
	If $t_\text{max}^\mathcal{L} > \sum_{b \in \mathcal{L}} r_b$, then $t_b \ge r_b$ for all $b \in \mathcal{L}$.
\end{corollary}

\begin{IEEEproof}
	Suppose there exists some $b \in \mathcal{L}$ such that $t_b < r_b$.
	The constraint of \eqref{eq:B}, i.e., $\sum_{b \in \mathcal{L}} t_b = t_\text{max}^\mathcal{L}$, suggests that we must have at least one $b' \in \mathcal{L}$ such that $t_{b'} > r_{b'}$.
	By \eqref{eq:beta1}, we know that $\beta_p(t_b, r_b) = 1$ and $\beta_p(t_{b'}, r_{b'}) \neq 1$.
	Further by \eqref{eq:beta_range}, we have $\beta_p(t_{b'}, r_{b'}) \le 1$.
	That is, we have $(1-p)\beta_p(t_b, r_b) > (1-p)\beta_p(t_{b'}, r_{b'})$ as $0 < p < 1$.
	By Theorem~\ref{thm:not_opt}, $\{t_b\}_{b \in \mathcal{L}}$ cannot solve \eqref{eq:B}, which is a contradiction.
	So, we must have $t_b \ge r_b$ for all $b \in \mathcal{L}$.
\end{IEEEproof}

\section{Algorithms for Blockwise Adaptive Recoding} \label{sec:algo}

In this section, we first propose a greedy algorithm to solve \eqref{eq:B} efficiently.
This algorithm gives an insight to the characteristic of the solution, which will be discussed in Section~\ref{sec:diff1}.
We also propose an approximation scheme based on Theorem~\ref{thm:adp_recode} to speed up the solver for practical implementations.
We defer a discussion on connecting BAR with other adaptive recoding formulations which assume that the incoming link condition is known in advance to Appendix~\ref{sec:longrun}.

The algorithms we propose in this paper frequently query and compare the values of $(1-p)\beta_p(t,r)$ for different $t \in \{-1, 0, 1, \ldots, t_\text{max}^\mathcal{L}\}$ and $r \in \{0, 1, 2, \ldots, M\}$.
We suppose a lookup table is constructed so that the queries can be done in $\mathcal{O}(1)$ time.
The table is reusable if the packet loss rate of the outgoing link is unchanged.
We only consider the subset $\{-1, 0, 1, \ldots, t_\text{max}^\mathcal{L}\} \times \{0, 1, 2, \ldots, M\}$ of the domain of $\beta_p$ because
\begin{enumerate}
	\item the maximum rank of a batch is $M$; and
	\item any $t_b$ cannot excess $t_\text{max}^\mathcal{L}$ as $\sum_{b \in \mathcal{L}} t_b = t_\text{max}^\mathcal{L}$.
\end{enumerate}
The case $t = -1$ will be used by our approximation scheme so we keep it in the lookup table.
We can build the table on-demand by dynamic programming, which will be discussed in Section~\ref{sec:table}.

\subsection{Greedy Algorithm} \label{sec:greedy}

We have discussed the case $t_\text{max}^\mathcal{L} \le \sum_{b \in \mathcal{L}} r_b$ in Lemma~\ref{lem:first_packet}.
Now, we consider $t_\text{max}^\mathcal{L} > \sum_{b \in \mathcal{L}} r_b$. 
We first define the subproblems \eqref{eq:Bk} of \eqref{eq:B} for $k \in \{0, 1, \ldots, t_\text{max}^\mathcal{L}\}$ so that we can investigate the optimal substructure for our greedy algorithm:
\begin{equation}
	\tag{B${}^{(k)}$} \label{eq:Bk}
		\max_{\substack{t_b \in \{0, 1, 2, \ldots\}\\\forall b \in \mathcal{L}}} \sum_{b \in \mathcal{L}} E(r_b, t_b) \quad \mathrm{s.t.} \quad \sum_{b \in \mathcal{L}} t_b = k.
\end{equation}

Fix a block $\mathcal{L}$.
We define a multiset 
	$\Omega := \biguplus_{b \in \mathcal{L}} \Omega_{r_b} = \{(1-p)\beta_p(t,r_b) \colon t \in \{0,1,2,\ldots\}, b \in \mathcal{L}\}$.
If two batches $a, b \in \mathcal{L}$ have the same rank, i.e., $r_a = r_b$, then for all $t$, we have $(1-p)\beta_p(t,r_a) = (1-p)\beta_p(t,r_b)$.
As $\Omega$ is a multiset, the duplicated values are not eliminated.

We have shown the relationship between $E(r,t)$ and $\Omega_r$ in Lemma~\ref{lem:exp_largest_sum}.
Now, we have the following lemma to connect \eqref{eq:Bk} and $\Omega$.

\begin{lemma} \label{lem:largest_sum}
	The optimal value of \eqref{eq:Bk} is the sum of the largest $k$ elements in $\Omega$.
\end{lemma}

\begin{IEEEproof}
	Let $\{t_b\}_{b \in \mathcal{L}}$ solves \eqref{eq:Bk}.
	Suppose the optimal value is not the sum of the largest $k$ elements in $\Omega$.
	However, Lemma~\ref{lem:exp_largest_sum} states that $E(r_b,t_b)$ equals the sum of the largest $t_b$ elements in $\Omega_{r_b}$ for all $b \in \mathcal{L}$.
	This means that there exists two distinct batches $\kappa, \rho \in \mathcal{L}$ with $t_\rho \ge 1$ such that
		$(1-p) \beta_p(t_\kappa,r_\kappa) > (1-p) \beta_p(t_\rho-1,r_\rho)$.

	By setting $t_\text{max}^\mathcal{L} = k$, we can apply Theorem~\ref{thm:not_opt} which gives that $\{t_b\}_{b \in \mathcal{L}}$ is not an optimal solution of \eqref{eq:Bk}.
	The proof is done by contradiction.
\end{IEEEproof}

The largest $k+1$ elements in $\Omega$ must contain the largest $k$ elements in $\Omega$ in terms of the values.
That is, Lemma~\ref{lem:largest_sum} shows the optimal substructure of \eqref{eq:B}.

\begin{figure}
\removelatexerror
\begin{algorithm}[H]
	\footnotesize
	\caption{Solver of BAR}
	\label{alg:opt1}
	\KwData{$t_\text{max}^\mathcal{L}; \{r_b\}_{b \in \mathcal{L}}$}
	\KwResult{An assignment $\{t_b^\ast\}_{b \in \mathcal{L}}$ solving \eqref{eq:B}}
	$t \leftarrow t_\text{max}^\mathcal{L}$ ; $t_b \leftarrow 0, \forall b \in \mathcal{L}$ \;
	\ForEach{$b \in \mathcal{L}$}{
		\lIf{$r_b \ge t$}{
			$t_b \leftarrow t$ ; \Return The assignment $\{t_b\}_{b \in \mathcal{L}}$
		}\lElse{
			$t_b \leftarrow r_b$ ; $t \leftarrow t - r_b$
		}
	}
	\lWhile{$t > 0$}{
		$b \leftarrow$ an element in $\argmax_{b \in \mathcal{L}} \beta_p(t_b,r_b)$ ; $t_b \leftarrow t_b + 1$ ; $t \leftarrow t - 1$ 
	}
	\Return The assignment $\{t_b\}_{b \in \mathcal{L}}$ \;
\end{algorithm}
\end{figure}

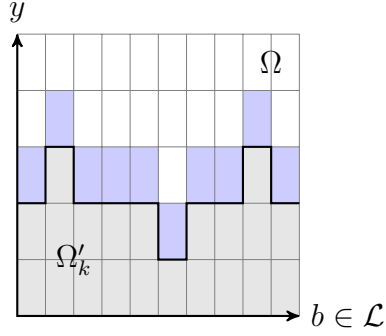
\begin{figure}
	\centering
	\begin{tikzpicture}[scale=.75]
		\fill[blue!20] (0,2) rectangle ++(.5,1);
		\fill[blue!20] (.5,3) rectangle ++(.5,1);
		\fill[blue!20] (1,2) rectangle ++(1.5,1);
		\fill[blue!20] (2.5,1) rectangle ++(.5,1);
		\fill[blue!20] (3,2) rectangle ++(1,1);
		\fill[blue!20] (4,3) rectangle ++(.5,1);
		\fill[blue!20] (4.5,2) rectangle ++(.5,1);
		\fill[gray!20] (0,2) -- ++(.5,0) -- ++(0,1) -- ++(.5,0) -- ++(0,-1) -- ++(1.5,0) -- ++(0,-1) -- ++(.5,0) -- ++(0,1) -- ++(1,0) -- ++(0,1) -- ++(.5,0) -- ++(0,-1) -- ++(.5,0) -- (5,0) -- (0,0) -- cycle;
		\draw[xscale=.5,step=1,gray,very thin] (0,0) grid (10,5);
		\draw[thick] (0,2) -- ++(.5,0) -- ++(0,1) -- ++(.5,0) -- ++(0,-1) -- ++(1.5,0) -- ++(0,-1) -- ++(.5,0) -- ++(0,1) -- ++(1,0) -- ++(0,1) -- ++(.5,0) -- ++(0,-1) -- ++(.5,0);
		\draw[<->,thick] (0,5) node (yaxis) [above] {$y$} |- (5,0) node (xaxis) [right] {$b \in \mathcal{L}$};
		\node at (4.5,4.5) {$\Omega$};
		\node at (1,1) {$\Omega_k'$};
	\end{tikzpicture}
	\caption{
		Each grid cell represents a value of $(1-p)\beta_p(y,r_b)$.
		The universe is the multiset $\Omega$.
		The grey region $\Omega_k'$ contains the largest $k$ elements in $\Omega$, which represents an optimal solution of \eqref{eq:Bk}.
		To achieve an optimal solution of $(\text{B}^{(k+1)})$, we need to find the $(k+1)$-th largest element in $\Omega$, which is the largest element above the solid line.
		By Lemma~\ref{lem:bp}\eqref{lem:bp_dec}, we have $\beta_p(y,r_b) \ge \beta_p(y+1,r_b)$ for all $b \in \mathcal{L}$, which implies that the largest element above the solid line is the largest one among the blue cells.
	}
	\label{fig:alg1}
\end{figure}

Algorithm~\ref{alg:opt1} is a greedy algorithm which makes use of this optimal substructure.
After an initialization, the algorithm repeatedly finds the batch $b$ which has the largest $(1-p)\beta_p(t_b,r_b)$ and increases the corresponding $t_b$ by one, i.e., we are solving \eqref{eq:Bk} for different $k$ incrementally.
Fig.~\ref{fig:alg1} illustrates an iteration of the algorithm.

\begin{theorem} \label{thm:alg:opt1}
	Algorithm~\ref{alg:opt1} solves \eqref{eq:B} in $\mathcal{O}(|\mathcal{L}|+\max\{0, t_\text{max}^\mathcal{L}-\sum_{b \in \mathcal{L}} r_b\} \log |\mathcal{L}|)$ time.
\end{theorem}

\begin{IEEEproof}
	Note that the variable $t$ in Algorithm~\ref{alg:opt1} represents the number of unassigned packets.
	The algorithm stops when $t = 0$, i.e., $\sum_{b \in \mathcal{L}} t_b = t_\text{max}^{\mathcal{L}}$, which is in the feasible region.

	If $t_\text{max}^{\mathcal{L}} \le \sum_{b \in \mathcal{L}} r_b$, the algorithm returns $\{t_b\}_{b \in \mathcal{L}}$ such that $t_b \le r_b$ for all $b \in \mathcal{L}$.
	Its optimality follows Lemma~\ref{lem:first_packet}.
	The algorithm takes $\mathcal{O}(|\mathcal{L}|)$ time in this case.

	Now, we consider $t_\text{max}^\mathcal{L} > \sum_{b \in \mathcal{L}} r_b$.
	We prove the correctness of Algorithm~\ref{alg:opt1} by induction.
	First, the \textbf{foreach} loop initializes $t_b = r_b$ for all $b \in \mathcal{L}$ in $\mathcal{O}(|\mathcal{L}|)$ time.
	By Lemma~\ref{lem:first_packet}, this initialization solves the subproblem $(\text{B}^{(k_0)})$ where $k_0 = \sum_{b \in \mathcal{L}} r_b$.

	Assume $\{t_b^{(k)}\}_{b \in \mathcal{L}}$ solves \eqref{eq:Bk} where $k = k_0, k_0+1, \ldots, t_\text{max}^{\mathcal{L}}-1$.
	Define a multiset
		$\Omega_k' = \biguplus_{b \in \mathcal{L}} \{(1-p)\beta_p(t,r_b) \colon t \in \{0, 1, \ldots, t_b^{(k)}-1\}\} \subset \Omega$.
%
	By Lemma~\ref{lem:largest_sum}, $\sum_{\omega \in \Omega_k'} \omega$ is the optimal value of \eqref{eq:Bk}, and $\Omega_k'$ is the collection of the largest $k$ elements in $\Omega$.
	By Lemma~\ref{lem:bp}\eqref{lem:bp_max}, we have
		$\max_{b \in \mathcal{L}} (1-p)\beta_p(t_b^{(k)},r_b) = \max \left( \Omega \setminus \Omega_k' \right)$,
	i.e., $\max_{b \in \mathcal{L}} (1-p)\beta_p(t_b^{(k)},r_b)$ is the $(k+1)$-th largest element in $\Omega$.
	The algorithm chooses a batch $\nu \in \mathcal{L}$ where
		$\nu \in \argmax_{b \in \mathcal{L}} \beta_p(t_b^{(k)},r_b) = \argmax_{b \in \mathcal{L}} (1-p) \beta_p(t_b^{(k)},r_b)$,
	and	gives $\{t_b^{(k+1)}\}_{b \in \mathcal{L}}$ by
	\begin{equation*}
		t_b^{(k+1)} = \begin{cases}
			t_b^{(k)}+1 & \text{if } b = \nu,\\
			t_b^{(k)} & \text{otherwise}
		\end{cases}
	\end{equation*}
	for all $b \in \mathcal{L}$.
	By Lemma~\ref{lem:exp_largest_sum}, we can express the objective value of $(\text{B}^{(k+1)})$ by
		$\sum_{b \in \mathcal{L}} E\left(r_b, t_b^{(k+1)}\right) = \sum_{\omega \in \Omega'_{k+1}} \omega$,
	where $\Omega_{k+1}'$ is the multiset defined by
		$\Omega_{k+1}' = \Omega_k' \uplus \{(1-p)\beta_p(t_\nu^{(k)},r_\nu)\} \subset \Omega$.
%
	The multiset $\Omega_{k+1}'$ contains the largest $k+1$ elements in $\Omega$.
	So, by Lemma~\ref{lem:largest_sum}, $\{t_b^{(k+1)}\}_{b \in \mathcal{L}}$ solves $(\text{B}^{(k+1)})$.
	The correctness of the algorithm is proved by induction.

	Now we calculate the time complexity of the algorithm.
	There are totally $\max\{0, t_\text{max}^{\mathcal{L}} - \sum_{b \in \mathcal{L}} r_b\}$ iterations for the \textbf{while} loop.
	The query of $\mu = \argmax_{b \in \mathcal{L}} \beta_p(t_b,r_b)$ can be implemented by using a binary heap.
	The initialization of the heap, i.e., heapify, takes $\mathcal{O}(|\mathcal{L}|)$ time, which can be done outside the loop.
	Each query in the loop takes $\mathcal{O}(1)$ time.
	The update after changing $\beta_p(t_\mu,r_\mu)$ into $\beta_p(t_\mu+1,r_\mu)$ takes $\mathcal{O}(\log |\mathcal{L}|)$ time. 
		Note that by Lemma~\ref{lem:bp}\eqref{lem:bp_dec}, $\beta_p(t_\mu,r_\mu) \ge \beta_p(t_\mu+1,r_\mu)$, so the update is a decrease key operation in a max-heap.
		A Fibonacci heap \cite{fibonacci} cannot benefit this operation here.
	Therefore, the overall time complexity is $\mathcal{O}(|\mathcal{L}| + \max\{0, t_\text{max}^{\mathcal{L}}-\sum_{b \in \mathcal{L}} r_b\}\log |\mathcal{L}|)$.
\end{IEEEproof}

\subsection{Equal Opportunity Approximation Scheme}

Algorithm~\ref{alg:opt1} increases $t_b$ step-by-step.
In the geometric point of view, the algorithm finds a path from the interior of a compact convex polytope $\mathcal{P}$ to the facet $\mathcal{H} \colon \sum_{b \in \mathcal{L}} t_b = t_\text{max}^\mathcal{L}$, where the half-space representation of $\mathcal{P}$ is the system of linear inequalities
\begin{equation*}
	\begin{cases}
		\sum_{b \in \mathcal{L}} t_b \le t_\text{max}^\mathcal{L}&\\
		t_b \ge 0, & \forall b \in \mathcal{L}.
	\end{cases}
\end{equation*}
Equivalently, $\mathcal{P}$ is the convex hull of the points $(t_\text{max}^\mathcal{L}, 0, \ldots, 0)$, $(0, t_\text{max}^\mathcal{L}, \ldots, 0)$, $\ldots$, $(0, 0, \ldots, t_\text{max}^\mathcal{L})$ and the origin.

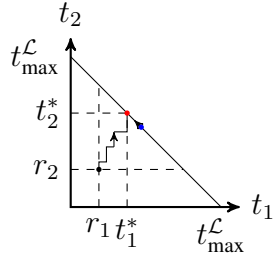
\begin{figure}
	\centering
	\begin{tikzpicture}[scale=.5]
		\draw [<->,thick] (0,4.5) node (yaxis) [above] {$t_2$} |- (4.5,0) node (xaxis) [right] {$t_1$};
		\draw (0,4) coordinate (ymax) -- (.75,3.25) coordinate (r1);
		\draw (4,0) coordinate (xmax) -- (3,1) coordinate (r2);
		\node (nr1) at (r1) {};
		\node (nr2) at (r2) {};
		\draw (r1) -- (1.5,2.5) coordinate (opt);
		\draw (r2) -- ($.5*(nr1)+.5*(nr2)$) coordinate (mid);
		\draw[->-] (mid) -- (opt);
		\node at (ymax) [left] {$t_\text{max}^\mathcal{L}$};
		\node at (xmax) [below] {$t_\text{max}^\mathcal{L}$};
		\draw[dashed] (yaxis |- opt) node[left] {$t_2^\ast$} -| (xaxis -| opt) node[below] {$t_1^\ast$};
		\draw[dashed] (yaxis |- r2) node[left] (lr2) {$r_2$} -- (r2);
		\draw[dashed] (xaxis -| r1) node[below] (lr1) {$r_1$} -- (r1);
		\draw (intersection of lr2--nr2 and lr1--nr1) coordinate (ori) -- ++(0,.2) -- ++(.2,0) -- ++(0,.4) -- ++(.2,0) coordinate (a);
		\draw[->] (a) -- ++(0,.4) coordinate (b);
		\draw (b) -| (opt);
		\fill[red] (opt) circle (2pt);
		\fill[blue] (mid) circle (2pt);
		\fill[black] (ori) circle (2pt);
	\end{tikzpicture}
	\caption{This figure illustrates the idea of modifying the output of an approximation scheme with two batches, where $t_\text{max}^\mathcal{L} \ge r_1+r_2$.
	The red and blue dots represent the optimal solution and the approximate solution on the facet $t_1+t_2 = t_\text{max}^\mathcal{L}$ respectively.
	Algorithm~\ref{alg:opt1} starts the search from an interior point $(r_1,r_2)$, while a modification approach starts the search from the blue dot.
	}
	\label{fig:opt_plot}
\end{figure}

If we have a method to move a non-optimal feasible point on $\mathcal{H}$ towards an optimal point, together with a fast and accurate enough approximation to \eqref{eq:B}, then we can combine them to solve \eqref{eq:B} faster than using Algorithm~\ref{alg:opt1} directly.
This idea is illustrated in Fig~\ref{fig:opt_plot}.

We first give an approximation scheme in this subsection.
%
As we cannot generate any linearly independent packet for a batch of rank $0$, we have $E(0, \cdot) = 0$.
So, we can exclude those batches having rank $0$ from $\mathcal{L}$ before we start the approximation.

Define $\mathfrak{L} = \{b \in \mathcal{L} \colon r_b > 0 \} \subseteq \mathcal{L}$.
When $t_\text{max}^\mathcal{L} > \sum_{b \in \mathcal{L}} r_b$, Corollary~\ref{cor:first_packet} shows that we should have $t_b \ge r_b$ for all $b \in \mathcal{L}$.
An easy way to give an approximation is to assign $\{t_b\}_{b \in \mathcal{L}}$ following the guideline given in Theorem~\ref{thm:adp_recode} by:
		$t_b = 0$ for all $b \in \mathcal{L} \setminus \mathfrak{L}$; and
		$t_b = r_b + \ell$ for all $b \in \mathfrak{L}$,
where $\ell = (t_\text{max}^\mathcal{L}-\sum_{b \in \mathcal{L}} r_b)/|\mathfrak{L}|$.
In case $\ell$ is not an integer, we can round it up for the batches having higher ranks and round it down for those having lower ranks.

\begin{figure}
\removelatexerror
\begin{algorithm}[H]
	\footnotesize
	\caption{Equal Opportunity Approximation Scheme}
	\label{alg:approx}
	\KwData{$t_\text{max}^\mathcal{L}; \{r_b\}_{b \in \mathcal{L}}$}
	\KwResult{An assignment $\{t_b\}_{b \in \mathcal{L}}$ approximating \eqref{eq:B}}
	$\ell \leftarrow t_\text{max}^\mathcal{L}$ ; $t_b \leftarrow 0, \forall b \in \mathcal{L}$ ; $L \leftarrow 0$ \;
	\ForEach{$b \in \mathcal{L}$}{
		\lIf{$r_b \ge \ell$}{
			$t_b \leftarrow \ell$ ; \Return The assignment $\{t_b\}_{b \in \mathcal{L}}$ 
		}\lElseIf{$r_b > 0$}{
			$t_b \leftarrow r_b$ ; $\ell \leftarrow \ell - r_b$ ; $L \leftarrow L + 1$ 
		}
	}
	\lIf{$L = 0$}{
		\Return An arbitrary feasible solution $\{t_b\}_{b \in \mathcal{L}}$ 
	}
	$r \leftarrow \ell \bmod L$ ; $\ell \leftarrow \left\lfloor \ell / L \right\rfloor$ \;
	\lForEach{$b \in \mathcal{L}$ s.t. $r_b > 0$}{
		$t_b \leftarrow r_b + \ell$ 
	}
	\lFor{the $r$ elements which have the largest $r_b, b \in \mathcal{L}$}{
		$t_b \leftarrow t_b+1$ 
	}
	\Return The assignment $\{t_b\}_{b \in \mathcal{L}}$ \;
\end{algorithm}
\end{figure}

\begin{figure}
	\centering
	\begin{tikzpicture}[scale=.6]
		\fill[green!20] (0,2) rectangle ++(.5,2);
		\fill[yellow!20] (0,4) rectangle ++(.5,1);
		\fill[green!20] (.5,3) rectangle ++(.5,2);
		\fill[yellow!20] (.5,5) rectangle ++(.5,1);
		\fill[green!20] (1,2) rectangle ++(1.5,2);
		\fill[yellow!20] (1,4) rectangle ++(.5,1);
		\fill[green!20] (2.5,1) rectangle ++(.5,2);
		\fill[green!20] (3,2) rectangle ++(1,2);
		\fill[green!20] (4,3) rectangle ++(.5,2);
		\fill[yellow!20] (4,5) rectangle ++(.5,1);
		\fill[green!20] (4.5,2) rectangle ++(.5,2);
		\fill[gray!20] (0,2) -- ++(.5,0) -- ++(0,1) -- ++(.5,0) -- ++(0,-1) -- ++(1.5,0) -- ++(0,-1) -- ++(.5,0) -- ++(0,1) -- ++(1,0) -- ++(0,1) -- ++(.5,0) -- ++(0,-1) -- ++(.5,0) -- (5,0) -- (0,0) -- cycle;
		\draw[xscale=.5,step=1,gray,very thin] (0,0) grid (10,6);
		\draw[<->,thick] (0,6) node (yaxis) [above] {$y$} |- (5,0) node (xaxis) [right] {$b \in \mathcal{L}$};
	\end{tikzpicture}
	\caption{
		The equal opportunity approximation scheme first assigns the grey cells, where the number of grey cells for each batch $b \in \mathcal{L}$ is equal to its rank $r_b$.
		Next, the scheme assigns the same number of green cells for all the batches which have non-zero rank.
		When $\ell$ is not an integer, the remaining packets are assigned to the batches having the highest ranks, where only one extra packet is allowed for each batch.
		This final step is shown as the yellow cells.
	}
	\label{fig:alg2}
\end{figure}
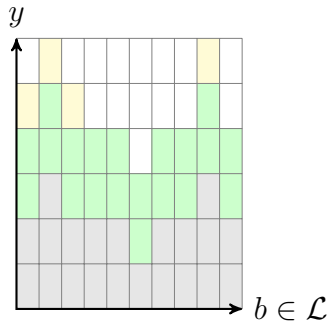

The above rules allocate the unassigned packets to the batches equally after $r_b$ packets are assigned to each batch $b$.
So, we call this approach the \emph{equal opportunity approximation scheme}.
The steps of this scheme is summarized in Algorithm~\ref{alg:approx} and illustrated in Fig.~\ref{fig:alg2}.

Note that we do not need to know the packet loss rate $p$ to apply this approximation.
This is, if we do not know the value of $p$, 
we can still apply this approximation to outperform baseline recoding.

\begin{theorem} \label{thm:approx}
	Algorithm~\ref{alg:approx} approximates \eqref{eq:B} in $\mathcal{O}(|\mathcal{L}|)$ time.
	If $t_\text{max}^\mathcal{L} \le \sum_{b \in \mathcal{L}} r_b$, then the algorithm solves \eqref{eq:B}.
\end{theorem}

\begin{IEEEproof}
	It is easy to see that Algorithm~\ref{alg:approx} outputs $\{t_b\}_{b \in \mathcal{L}}$ which satisfies $\sum_{b \in \mathcal{L}} t_b = t_\text{max}^\mathcal{L}$.
	That is, the output is a feasible solution of \eqref{eq:B}.
	Note that $|\mathfrak{L}| \le |\mathcal{L}|$, 
	so the assignments and branches before the last \textbf{for} loop totally take $\mathcal{O}(|\mathcal{L}|)$ time.
	The variable $L$ after the first \textbf{foreach} loop equals $|\mathfrak{L}|$.
	Adding one to the number of recoded packets for those $r = \ell \bmod L$ batches having the highest ranks can be done in $\mathcal{O}(|\mathcal{L}|)$ time (See Appendix~\ref{sec:approx:linear}).
	So, the overall running time is $\mathcal{O}(|\mathcal{L}|)$.

	If $\mathfrak{L} = \emptyset$, i.e., the whole block is lost, then any feasible $\{t_b\}_{b \in \mathcal{L}}$ is a solution, and the optimal objective value is $0$.
	If $t_\text{max}^\mathcal{L} \le \sum_{b \in \mathcal{L}} r_b$, then the algorithm terminates with an output satisfying $t_b \le r_b$ for all $b \in \mathcal{L}$.
	By Lemma~\ref{lem:first_packet}, such solution is optimal.
\end{IEEEproof}

In practice, the batch size $M$ is small.
We can search those $r$ batches having the highest ranks in $\mathcal{O}(|\mathcal{L}|+M)$ time by a counting technique (see Appendix~\ref{sec:approx:count}) 
as an efficient alternative.

Algorithm~\ref{alg:approx} is a $(1-p)$-approximation algorithm, although the relative performance guarantee factor $1-p$ is not tight in general.
However, it suggests that the smaller the packet loss rate $p$, the more accurate output the algorithm gives.
We leave the discussion regarding this approximation in Appendix~\ref{sec:proof:thm:approx2}.

\subsection{Speed-up via Approximation}

In this subsection, we investigate a method which corrects an approximate solution to an optimal solution.

Note that all integral points on the facet $\mathcal{H}$ are feasible solutions to \eqref{eq:B}.
When we move a point on $\mathcal{H}$, 
the minimal change is to increase $t_b$ by $1$ and decrease $t_{b'}$ by $1$, where $b, b' \in \mathcal{L}$ and $b \neq b'$.
We call such minimal change a \emph{step}.
A sequence of steps is called a \emph{path}.

\begin{lemma} \label{lem:path}
	Suppose $\sum_{b \in \mathcal{L}} r_b < t_\text{max}^\mathcal{L}$.
	For any non-optimal point $\mathcal{T}$ on the facet $\mathcal{H}$, there exists a path of finite steps from $\mathcal{T}$ to an optimal point, where the objective value is strictly increasing along the path.
\end{lemma}

\begin{IEEEproof}
		If $t_\text{max}^\mathcal{L} = 0$, then the facet $\mathcal{H}$ degenerates into a single point, which is the origin.
	The only feasible point is the optimal point, so we cannot have a non-optimal solution.

	Consider $t_\text{max}^\mathcal{L} > 0$.
	Define a multiset $\Omega_{t_\text{max}^\mathcal{L}}'$ which is a collection of the largest $t_\text{max}^\mathcal{L}$ elements in $\Omega$.
	By Lemma~\ref{lem:largest_sum}, $\sum_{\omega \in \Omega_{t_\text{max}^\mathcal{L}}'} \omega$ is the optimal value of \eqref{eq:B}.
	For any non-optimal point on $\mathcal{H}$ locating at $(t_b^{(k)})_{b \in \mathcal{L}}$, we define a multiset
		$\mho_k := \{(1-p) \beta_p(t,r_b) \colon t \in \{0,1,\ldots,t_b^{(k)}-1\}, b \in \mathcal{L}\}$,
	where the value $k \in \{0, 1, \ldots, t_\text{max}^\mathcal{L} - 1\}$ is the number of elements in $\Omega'_{t_\text{max}^\mathcal{L}}$ which are also contained in $\mho_k$.
	We know that there exists a $m$ such that the given non-optimal point $\mathcal{T}$ is located at $(t_b^{(m)})$.

	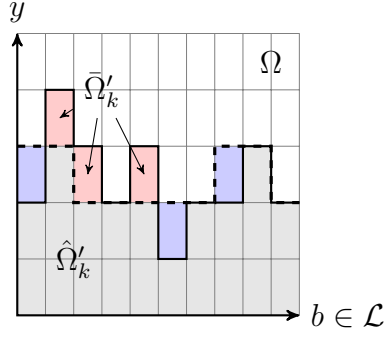
\begin{figure}
		\centering
		\begin{tikzpicture}[scale=.75]
			\fill[blue!20] (0,2) rectangle ++(.5,1);
			\fill[blue!20] (2.5,1) rectangle ++(.5,1);
			\fill[blue!20] (3.5,2) rectangle ++(.5,1);
			\fill[red!20] (.5,3) rectangle ++(.5,1);
			\fill[red!20] (1,2) rectangle ++(.5,1);
			\fill[red!20] (2,2) rectangle ++(.5,1);
			\fill[gray!20] (0,2) -- ++(.5,0) -- ++(0,1) -- ++(.5,0) -- ++(0,-1) -- ++(1.5,0) -- ++(0,-1) -- ++(.5,0) -- ++(0,1) -- ++(1,0) -- ++(0,1) -- ++(.5,0) -- ++(0,-1) -- ++(.5,0) -- (5,0) -- (0,0) -- cycle;
			\draw[xscale=.5,step=1,gray,very thin] (0,0) grid (10,5);
			\draw[dashed,very thick] (0,3) -- ++(1,0) -- ++(0,-1) -- ++(2.5,0) -- ++(0,1) -- ++(1,0) -- ++(0,-1) -- ++(.5,0);
			\draw[thick] (0,2) -- ++(.5,0) -- ++(0,2) -- ++(.5,0) -- ++(0,-1) -- ++(.5,0) -- ++(0,-1) -- ++(.5,0) -- ++(0,1) -- ++(.5,0) -- ++(0,-2) -- ++(.5,0) -- ++(0,1) -- ++(1,0) -- ++(0,1) -- ++(.5,0) -- ++(0,-1) -- ++(.5,0);
			\draw[<->,thick] (0,5) node (yaxis) [above] {$y$} |- (5,0) node (xaxis) [right] {$b \in \mathcal{L}$};
			\node (bar) at (1.5,4) {$\bar\Omega_k'$};
			\draw[->] (bar) -- (.75,3.5);
			\draw[->] (bar) -- (1.25,2.5);
			\draw[->] (bar) -- (2.25,2.5);
			\node at (4.5,4.5) {$\Omega$};
			\node at (1,1) {$\hat\Omega_k'$};
		\end{tikzpicture}
		\caption[Caption for LOF]{
			The region under the solid line, i.e., the red and grey regions, is the multiset $\mho_k$, which represents the current non-optimal solution.
			The region under the dashed line, i.e., the blue and grey regions, is the multiset $\Omega_{t_\text{max}^\mathcal{L}}'$, which represents the optimal solution.
			Note that the graphical representation of $\Omega_{t_\text{max}^\mathcal{L}}'$ may not be unique.
			By the definition of $\Omega_{t_\text{max}^\mathcal{L}}'$, we can swap a cell under the dashed line with a cell above it if they have the same value.
			We can always obtain the largest grey region due to this flexibility.
			The intersection of the current non-optimal and optimal solutions, i.e., the multiset $\hat\Omega_k'$, is the grey region.
			The red region is denoted by the multiset $\bar\Omega_k'$, which contains the non-optimal values in $\mho_k$ which should be removed.
			The blue region contains the values which are part of the optimal solution but not in $\mho_k$ yet.
		}
		\label{fig:lem:path}
	\end{figure}

	We provide a constructive proof for this lemma.
	Given any $k \in \{m, m+1, \ldots, t_\text{max}^\mathcal{L} - 1\}$, we are going to show that there exists a step to obtain a point locating at $(t_b^{(k+1)})_{b \in \mathcal{L}}$ which has a larger objective value.
	Suppose we have a non-optimal point located at $(t_b^{(k)})_{b \in \mathcal{L}}$.
	Define
	$\hat\Omega_k' = \Omega_{t_\text{max}^\mathcal{L}}' \cap \mho_k$ and $\bar\Omega_k' = \mho_k \setminus \hat\Omega_k'$.
	By definition, $|\hat\Omega_k'| = k$.
	$\hat\Omega_k'$ contains the elements in $\mho_k$ which are part of the optimal solution.
	On the other side, $\bar\Omega_k'$ contains the elements in $\mho_k$ which are not the largest $t_\text{max}^\mathcal{L}$ elements in $\Omega$.
	By Lemma~\ref{lem:largest_sum}, the elements in $\bar\Omega_k'$ are the cause of making the solution non-optimal.
	It is clear that for all $\omega_1 \in \hat\Omega_k', \omega_2 \in \bar\Omega_k'$, we have $\omega_1 > \omega_2$.
	The relationship between $\mho_k, \Omega_{t_\text{max}^\mathcal{L}}', \hat\Omega_k'$ and $\bar\Omega_k'$ is illustrated in Fig.~\ref{fig:lem:path}.

	As $\mho_k \neq \Omega_{t_\text{max}^\mathcal{L}}'$, i.e., the current point is not optimal, we can apply Theorem~\ref{thm:not_opt} and know that there exists two batches $\kappa, \rho \in \mathcal{L}$ with $t_\rho \ge 1$ such that
	\begin{equation} \label{eq:lem:path0}
		(1-p) \beta_p(t_\kappa,r_\kappa) > (1-p) \beta_p(t_\rho-1,r_\rho).
	\end{equation}
	Without loss of generality, take any 
	$\kappa \in \argmax_{b \in \mathcal{L}} \beta_p(t_b,r_b)$ and $\rho \in \argmin_{b \in \mathcal{L}} \beta_p(t_b-1,r_b)$.
%
	Note that $\max (\Omega\setminus\mho_k) \in \Omega_{t_\text{max}^\mathcal{L}}'$, and
		$\max (\Omega\setminus\mho_k) > \min (\bar\Omega_k') = \min (\mho_k)$.
%
	By Lemma~\ref{lem:bp}\eqref{lem:bp_max} and \eqref{lem:bp_min}, we have
	$\max (\Omega\setminus\mho_k) = (1-p) \beta_p(t_\kappa,r_\kappa)$, and 
	$\min (\mho_k) = (1-p) \beta_p(t_\rho-1,r_\rho)$.

	We construct a step by removing $\min\{\mho_k\}$ from $\mho_k$ and inserting $\max\{\Omega \setminus \mho_k\}$ into it.
	Mathematically, we are creating the following multisets:
	$\hat\Omega_{k+1}' = \hat\Omega_k' \uplus \{(1-p) \beta_p(t_\kappa,r_\kappa)\}$, 
	$\bar\Omega_{k+1}' = \bar\Omega_k' \setminus \{(1-p) \beta_p(t_\rho-1,r_\rho)\}$, and
	$\mho_{k+1} = \hat\Omega_{k+1}' \uplus \bar\Omega_{k+1}'$.
	The corresponding new point on $\mathcal{H}$ is located at $(t_b^{(k+1)})_{b \in \mathcal{L}}$, where
	\begin{equation*}
		t_b^{(k+1)} = \begin{cases}
			t_\kappa^{(k)} + 1 & \text{if } b = \kappa,\\
			t_\rho^{(k)} - 1 & \text{if } b = \rho,\\
			t_b^{(k)} & \text{otherwise.}
		\end{cases}
	\end{equation*}

	On the other hand, consider
	\begin{IEEEeqnarray*}{rCl}
		\sum_{\omega \in \mho_{k+1}} \omega & = & \sum_{\omega_0 \in \hat\Omega_{k+1}'} \omega_0 + \sum_{\omega_1 \in \bar\Omega_{k+1}'} \omega_1\\
		& = & \left( \sum_{\omega_0 \in \hat\Omega_k'} \omega_0 + (1-p) \beta_p(t_\kappa,r_\kappa) \right) + \left( \sum_{\omega_1 \in \bar\Omega_k'} \omega_1 - (1-p) \beta_p(t_\rho-1,r_\rho) \right)\\
		& > & \sum_{\omega_0 \in \hat\Omega_k'} \omega_0 + \sum_{\omega_1 \in \bar\Omega_k'} \omega_1 \yesnumber \label{eq:lem:path1} 
		 =  \sum_{\omega \in \mho_k} \omega,
	\end{IEEEeqnarray*}
	where \eqref{eq:lem:path1} follows \eqref{eq:lem:path0}.
	Lastly, by Lemma~\ref{lem:exp_rank}\eqref{lem:exp_rank_exact}, we have
		$\sum_{b \in \mathcal{L}} E(r_b,t_b^{(k+1)}) = \sum_{\omega \in \mho_{k+1}} \omega > \sum_{\omega \in \mho_k} \omega = \sum_{b \in \mathcal{L}} E(r_b,t_b^{(k)})$,
	which is strictly increasing.
	By induction, the existence of such path is shown.

\end{IEEEproof}

\begin{figure}
\removelatexerror
\begin{algorithm}[H]
	\footnotesize
	\caption{Solver of BAR via Approximation}
	\label{alg:opt2}
	\KwData{$t_\text{max}^\mathcal{L}; r_b, b \in \mathcal{L}$}
	\KwResult{An assignment $\{t_b^\ast\}_{b \in \mathcal{L}}$ solving \eqref{eq:B}}
	$t_b \leftarrow 0, \forall b \in \mathcal{L}$ \;
	Run an approximation to get $t_b, b \in \{a \in \mathcal{L} \colon r_a > 0\}$ \;
	\While{$\min_{a \in \mathcal{L}} \beta_p(t_a-1,r_a) < \max_{b \in \mathcal{L}} \beta_p(t_b,r_b)$}{
		$a \leftarrow$ an element in $\argmin_{a \in \mathcal{L}} \beta_p(t_a-1,r_a)$ \;
		$b \leftarrow$ an element in $\argmax_{b \in \mathcal{L}} \beta_p(t_b,r_b)$ \;
		$t_a \leftarrow t_a - 1$ ; $t_b \leftarrow t_b + 1$ \;
	}
	\Return The assignment $\{t_b\}_{b \in \mathcal{L}}$ \;
\end{algorithm}
\end{figure}

\begin{figure}
	\centering
	\begin{tikzpicture}[scale=.75]
		\fill[blue!20] (0,2) rectangle ++(.5,1);
		\fill[blue!20] (.5,4) rectangle ++(.5,1);
		\fill[blue!20] (1,3) rectangle ++(.5,1);
		\fill[blue!20] (1.5,2) rectangle ++(.5,1);
		\fill[blue!20] (2,3) rectangle ++(.5,1);
		\fill[blue!20] (2.5,1) rectangle ++(.5,1);
		\fill[blue!20] (3,2) rectangle ++(1,1);
		\fill[blue!20] (4,3) rectangle ++(.5,1);
		\fill[blue!20] (4.5,2) rectangle ++(.5,1);
		\fill[gray!20] (0,2) -- ++(.5,0) -- ++(0,2) -- ++(.5,0) -- ++(0,-1) -- ++(.5,0) -- ++(0,-1) -- ++(.5,0) -- ++(0,1) -- ++(.5,0) -- ++(0,-2) -- ++(.5,0) -- ++(0,1) -- ++(1,0) -- ++(0,1) -- ++(.5,0) -- ++(0,-1) -- ++(.5,0) -- (5,0) -- (0,0) -- cycle;
		\fill[red!20] (0,1) rectangle ++(.5,1);
		\fill[red!20] (.5,3) rectangle ++(.5,1);
		\fill[red!20] (1,2) rectangle ++(.5,1);
		\fill[red!20] (1.5,1) rectangle ++(.5,1);
		\fill[red!20] (2,2) rectangle ++(.5,1);
		\fill[red!20] (2.5,0) rectangle ++(.5,1);
		\fill[red!20] (3,1) rectangle ++(1,1);
		\fill[red!20] (4,2) rectangle ++(.5,1);
		\fill[red!20] (4.5,1) rectangle ++(.5,1);
		\draw[xscale=.5,step=1,gray,very thin] (0,0) grid (10,5);
		\draw[thick] (0,2) -- ++(.5,0) -- ++(0,2) -- ++(.5,0) -- ++(0,-1) -- ++(.5,0) -- ++(0,-1) -- ++(.5,0) -- ++(0,1) -- ++(.5,0) -- ++(0,-2) -- ++(.5,0) -- ++(0,1) -- ++(1,0) -- ++(0,1) -- ++(.5,0) -- ++(0,-1) -- ++(.5,0);
		\draw[<->,thick] (0,5) node (yaxis) [above] {$y$} |- (5,0) node (xaxis) [right] {$b \in \mathcal{L}$};
	\end{tikzpicture}
	\caption{
		When the current solution is not optimal, Theorem~\ref{thm:not_opt} states that one of the blue cells has a value strictly larger than the value of one of the red cells.
		We can choose a pair of cells by finding the largest and smallest one among the blue and red cells respectively.
		Then, we can modify the solution by raising and lowering the solid line to include the selected blue cell and exclude the selected red cell.
	}
	\label{fig:alg3}
\end{figure}

Algorithm~\ref{alg:opt2} is a greedy algorithm which uses any point $\mathcal{T}$ on the facet $\mathcal{H}$ as a starting point.
Then, it follows the path constructed in the proof of Lemma~\ref{lem:path} to obtain an optimal solution.
An iteration to modify the solution is illustrated in Fig.~\ref{fig:alg3}.
Note that the algorithm may query $\beta_p(t_a-1,r_a)$ for $a \in \mathcal{L}$.
If $t_a = 0$, then it is accessing the value $\beta_p(-1,r_a)$.
Recall that we have defined $\beta_p(-1, \cdot) = 1$, which is the upper bound of $\beta_p(\cdot,\cdot)$ by \eqref{eq:beta_range}.
So, these values act as barriers to prevent outputing negative number of recoded packets.

\begin{theorem} \label{thm:alg:opt2}
	Let $\{t_b\}_{b \in \mathcal{L}}$ be an approximate solution of \eqref{eq:B} computed in $\mathcal{O}(T_\text{approx})$ time.
	Algorithm~\ref{alg:opt2} outputs $\{t_b^\ast\}_{b \in \mathcal{L}}$ which solves \eqref{eq:B} in $\mathcal{O}(T_\text{approx} + |\mathcal{L}| + \sum_{b \in \mathcal{L}} |t_b^\ast-t_b| \log |\mathcal{L}|)$ time. 
\end{theorem}

\begin{IEEEproof}
	Theorem~\ref{thm:not_opt} suggests that if the current feasible solution is not optimal, then we must have two distinct batches $a, b \in \mathcal{L}$ with $a \ge 1$ such that
	\begin{equation} \label{eq:thm:opt2:not_opt}
		(1-p)\beta_p(t_a-1,r_a) < (1-p)\beta_p(t_b,r_b).
	\end{equation}

	Suppose $t_a = 0$ for some $a \in \mathcal{L}$, then by \eqref{eq:beta1}, we have $\beta_p(t_a-1,r_a) = \beta_p(-1,r_a) = 1$.
	By \eqref{eq:beta_range}, we know that it is not possible to have a $b \in \mathcal{L}$ satisfying \eqref{eq:thm:opt2:not_opt}.
	That is, we do not need to check whether $t_a = 0$ or not in the algorithm: The value $\beta_p(-1,r_a)$ does this purpose.

	By Lemma~\ref{lem:bp}\eqref{lem:bp_max} and \eqref{lem:bp_min}, we have
	$\beta_p(t_\rho-1,r_\rho) \ge \min_{a \in \mathcal{L}} \beta_p(t_a-1,r_a)$ and 
	$\max_{b \in \mathcal{L}} \beta_p(t_b,r_b) \ge \beta_p(t_\kappa,r_\kappa)$
	for all batches $\rho, \kappa \in \mathcal{L}$.
	That is, we can rewrite \eqref{eq:thm:opt2:not_opt} as
		$\min_{a \in \mathcal{L}} \beta_p(t_a-1,r_a) < \max_{b \in \mathcal{L}} \beta_p(t_b,r_b)$
	as the condition to continue the iterations.
	Each iteration in the algorithm takes the next step following the path constructed in the proof of Lemma~\ref{lem:path} exactly.
	The correctness is implied directly.

	Now we calculate the time complexity of the algorithm.
	The assignments before the \textbf{while} loop takes $\mathcal{O}(T_\text{approx}+|\mathcal{L}|)$ time.
	There are totally $\sum_{b \in \mathcal{L}} |t_b^\ast - t_b|/2$ iterations for the \textbf{while} loop.
	The queries for the minimum and maximum values can be implemented by using a min-heap and a max-heap respectively.
	Similar to Algorithm~\ref{alg:opt1}, we can use binary heaps, which takes $\mathcal{O}(|\mathcal{L}|)$ initialization time, $\mathcal{O}(1)$ query time, and $\mathcal{O}(\log |\mathcal{L}|)$ update time.
	Each iteration contains two heap queries and two heap updates (see Appendix~\ref{sec:corrupted_heap}). 
	The overall time complexity is then $\mathcal{O}(T_\text{approx} + |\mathcal{L}| + \sum_{b \in \mathcal{L}} |t_b^\ast - t_b| \log |\mathcal{L}|)$.
\end{IEEEproof}

\subsection{Solution Characteristic} \label{sec:diff1}

In this subsection, we discuss an observation inspired by Algorithm~\ref{alg:opt1}.
Consider a block $\mathcal{L}$.
Let $\mathcal{B}_r = \{b \in \mathcal{L} \colon r_b = r\}$ be a set containing all batches with rank $r$ in the block.
It is easy to see that $\bigcup_{r = 0}^M \mathcal{B}_r = \mathcal{L}$ and $\mathcal{B}_\alpha \cap \mathcal{B}_\beta = \emptyset$ for all distinct $\alpha, \beta \in \{0, 1, \ldots, M\}$.

Suppose after some iterations, we have $t_b = t_{b'}$ for all $\mathcal{B}_r$, $r = 0, 1, \ldots, M$.
This condition holds for the subproblem $(\text{B}^{(k_0)})$ where $k_0 = \sum_{b \in \mathcal{L}} r_b$.
The current iteration selects a batch $b \in \argmax_{b \in \mathcal{L}} \beta_p(t_b, r_b)$.
Observe that actually all batches in $\mathcal{B}_{r_b}$ are in $\argmax_{b \in \mathcal{L}} \beta_p(t_b, r_b)$.
That is, the algorithm can select another batch in $\mathcal{B}_{r_b}$ in the next iteration, so on and so forth.
This suggests that the difference between every pair of $t_b, t_{b'}$, where $b, b' \in \mathcal{B}_r$, is at most $1$ for all $r = 0, 1, \ldots, M$.
Also, we can select all the batches in $\mathcal{B}_r$ before we consider another rank, which suggests that at most one $r$ we have a difference $1$ mentioned above.
The following theorem formally states this observation.

\begin{theorem} \label{thm:diff1}
	There exists a set of $\{t_b\}_{b \in \mathcal{L}}$ solving \eqref{eq:B} where $|t_b - t_{b'}| \le 1$ for all $b, b' \in \mathcal{B}_r$, $r = 0, 1, \ldots, M$.
	Furthermore, there is at most one $r$ which can achieve the equality.
\end{theorem}

\begin{IEEEproof}
	See Appendix~\ref{sec:proof:thm:diff1}.
\end{IEEEproof}

Corollary~\ref{cor:diff1} below is a response to the discussion in Section~\ref{sec:adp}.

\begin{corollary} \label{cor:diff1}
	There exists a set of $\{t_b\}_{b \in \mathcal{L}}$ solving \eqref{eq:B} at the source node where $|t_b - t_{b'}| \le 1$ for all $b, b' \in \mathcal{L}$.
\end{corollary}

\begin{IEEEproof}
	All the batches generated by the encoder have rank $M$, i.e., $\mathcal{B}_M = \mathcal{L}$ and $\mathcal{B}_r = \emptyset$ for all $r = 0, 1, \ldots, M-1$.
	The proof is done by applying Theorem~\ref{thm:diff1} directly.
\end{IEEEproof}

\subsection{Construction of the Lookup Table}
\label{sec:table}

In the algorithms, we assume that we have a lookup table for the function $\beta_p(\cdot,\cdot)$ so that we can query its values quickly.
As shown in \eqref{eq:betainc0}, we can consider our problem as accessing the value of a regularized incomplete beta function $I_p(\cdot, \cdot)$. 
Most available implementations consider non-negative real parameters and calculate different queries independently.
This consideration is too general for our application, as we only need to query the integral points efficiently.
In this subsection, we propose an on-demand approach to construct the lookup table. 

Being a dynamic programming approach, we need the following recursive relations:
\begin{IEEEeqnarray}{rClL}
	B_p(t+1,r) & = & (1-p) B_p(t,r-1) + p B_p(t,r) & \quad \text{for } 0 \le r \le t; \label{eq:bp_rec}\\
	B_p(r,r) & = & (1-p) B_p(r-1,r-1) & \quad \text{for } r > 0; \label{eq:bp_rr}\\
	\beta_p(t,r) & = & \beta_p(t,r-1) + B_p(t,r-1) & \quad \text{for } 1 < r \le t+1, \label{eq:beta_rec}
\end{IEEEeqnarray}
where
		\eqref{eq:bp_rec} is stated in Lemma~\ref{lem:bp}\eqref{lem:bp_rec}; and
		\eqref{eq:bp_rr} and \eqref{eq:beta_rec} are by the definitions of $B_p(t,r)$ and $\beta_p(t,r)$ 
		respectively.
%
The boundary conditions are
$B_p(0,0) = 1$, $B_p(i,-1) = 0$, and $\beta_p(i,1) = B_p(i,0)$ for $i = 0, 1, \ldots$.
%
The table can be built in-place in two stages.
The first stage fills in $B_p(y,x-1)$ at the $(y,x)$ position of the table.
The second stage finishes the table by using \eqref{eq:beta_rec}.
Fig.~\ref{fig:alg:bp} illustrates the two stages where the arrows represent the recursive relations \eqref{eq:bp_rec}-\eqref{eq:beta_rec}.
As $\beta_p(0,1) = \beta_p(1,2) = \ldots = 1$, the corresponding entries can be substituted directly. 

\begin{figure}
	\centering
	\begin{subfigure}{0.45\textwidth}
		\centering
		\begin{tikzpicture}[scale=.73,every node/.style={scale=.73}]
			\matrix (BP) [matrix of math nodes,nodes={text width={width("$B_p(1,0)$")},text height={height("$B_p(1,0)$")},align=center}] at (0,0) {
				1 & 1 & 1 & 1 & 1 \\
				0 & B_p(0,0) & 1 & 1 & 1 \\
				0 & B_p(1,0) & B_p(1,1) & 1 & 1 \\
				0 & B_p(2,0) & B_p(2,1) & B_p(2,2) & 1 \\
				0 & B_p(3,0) & B_p(3,1) & B_p(3,2) & B_p(3,3) \\
				0 & B_p(4,0) & B_p(4,1) & B_p(4,2) & B_p(4,3) \\
			};
			\foreach \y in {3,...,6}{
				\draw[->] ($(BP-\y-2.north west)+(-.5,.15)$) -- ($(BP-\y-2.north west)+(.15,-.15)$);
			}
			\foreach \x in {2,3,4}{
				\pgfmathtruncatemacro{\xx}{\x+1}
				\draw[dashed,->] ($(BP-\x-\x.south east)+(-.15,.15)$) -- ($(BP-\xx-\xx.north west)+(.15,-.15)$);
				\foreach \y in {\xx,...,5}{
					\pgfmathtruncatemacro{\yy}{\y+1}
					\draw[->] ($(BP-\y-\x.south east)+(-.15,.15)$) -- ($(BP-\yy-\xx.north west)+(.15,-.15)$);
				}
			}
			\foreach \x in {2,3,4,5}{
				\foreach \y in {\x,...,5}{
					\pgfmathtruncatemacro{\yy}{\y+1}
					\draw[->] ($(BP-\y-\x.south)+(0,.1)$) -- ($(BP-\yy-\x.north)+(0,-.15)$);
				}
			}
			\draw ($(BP-1-1.south)+(-.5,0)$) -- ($(BP-1-5.south)+(.5,0)$);
		\end{tikzpicture}
		\caption{The first stage of the table generation.
			The $1$s and $0$s paddings are generated first.
			The solid and dashed arrows represent \eqref{eq:bp_rec} and \eqref{eq:bp_rr} respectively.
	}
	\end{subfigure}~
	\begin{subfigure}{0.45\textwidth}
		\centering
		\begin{tikzpicture}[scale=.73,every node/.style={scale=.73}]
			\matrix (BPSUM) [matrix of math nodes,nodes={text width={width("$B_p(1,0)$")},text height={height("$B_p(1,0)$")},align=center}] at (0,0) {
				1 & 1 & 1 & 1 & 1 \\
				0 & \beta_p(0,1) & 1 & 1 & 1 \\
				0 & \beta_p(1,1) & \beta_p(1,2) & 1 & 1 \\
				0 & \beta_p(2,1) & \beta_p(2,2) & \beta_p(2,3) & 1 \\
				0 & \beta_p(3,1) & \beta_p(3,2) & \beta_p(3,3) & \beta_p(3,4) \\
				0 & \beta_p(4,1) & \beta_p(4,2) & \beta_p(4,3) & \beta_p(4,4) \\
			};
			\foreach \x in {2,3,4}{
				\pgfmathtruncatemacro{\xx}{\x+1}
				\foreach \y in {\xx,...,6}{
					\draw[->] ($(BPSUM-\y-\x.east)+(-.15,0)$) -- ($(BPSUM-\y-\xx.west)+(.15,0)$);
				}
			}
			\draw ($(BPSUM-1-1.south)+(-.5,0)$) -- ($(BPSUM-1-5.south)+(.5,0)$);
		\end{tikzpicture}
		\caption{The second stage of the table generation.
			The $1$s and $0$s paddings are kept.
			The arrows represent the recursive relation \eqref{eq:beta_rec} with the $B_p$ function in-place.
		}
	\end{subfigure}
	\caption{The figures illustrate the two stages of the table generation.
		The indices start from $(-1,0)$.
		The first row has the index $y = -1$, which is the row above the line.
		Comparing to Fig.~\ref{fig:bp_table}, $\beta_p(0,1) = \beta_p(1,2) = \ldots = 1$ can be substituted directly without using the relation \eqref{eq:beta_rec}.
	}
	\label{fig:alg:bp}
\end{figure}

\begin{figure}
	\centering
	\begin{tikzpicture}[scale=.73,every node/.style={scale=.73}]
		\matrix (BPSUM) [matrix of math nodes,nodes={text width={width("$B_p(1,0)$")},text height={height("$B_p(1,0)$")},align=center}] at (0,0) {
			1 & 1 & 1 & 1 & 1 \\
			0 &[-1em] \beta_p(0,1) & 1 & 1 & 1 \\
			0 & \beta_p(1,1) & \beta_p(1,2) & 1 & 1 \\
			0 & \beta_p(2,1) & \beta_p(2,2) & \beta_p(2,3) & 1 \\
			0 & B_p(3,0) & B_p(3,1) & B_p(3,2) & B_p(3,3) \\
			\vphantom{\beta_p(0,1)}\ast & \ast & \ast & \ast & \ast \\
		};
		\draw ($(BPSUM-1-1.south)+(-.5,0)$) -- ($(BPSUM-1-5.south)+(.5,0)$);
	\end{tikzpicture}
	\caption{The values we have prepared when $t_{b'} = 2$.
		The asterisks represent the values that are not initialized yet.
	}
	\label{fig:alg:bp3}
\end{figure}

We can compute the values in the table on demand.
Suppose we have $\{t_b\}_{b \in \mathcal{L}}$ in an iteration of Algorithm~\ref{alg:opt1} so that we need the values of $\beta_p(t_b,r_b)$ for $b \in \mathcal{L}$.
Let $b'$ be an element in $\argmax_{b \in \mathcal{L}} r_b$.
The table has 
$r_{b'}+1$ columns.
By the criteria of selecting $b$ in Algorithm~\ref{alg:opt1} and also by Lemmas~\ref{lem:bp}(\ref{lem:bp_diag}) and \ref{lem:bp}(\ref{lem:bp_inc}), we have $\max_{b \in \mathcal{L}} t_b = t_{b'}$.
From Fig.~\ref{fig:alg:bp}, we know that we have to calculate all rows of $\beta(t, r)$ for $t \le t_{b'}$.
Also, the recursive relations on a row only depend on the previous row, we need to prepare ahead the values of $B_p$ in the next row so that we have the values to compute $\beta_p$ in the next row.
As an example, Fig.~\ref{fig:alg:bp3} illustrates the values we have prepared when $t_{b'} = 2$.

Each entry in the table is modified at most twice during the two stages.
Each assignment takes $\mathcal{O}(1)$ time.
Therefore, the time and space complexities for building the table are both $\mathcal{O}(MR)$, where $R$ is the number of rows we want to construct.
As restricted by the block size, we know that $R \le t_\text{max}^\mathcal{L}$.
The worst case is that we only receive one rank-$M$ batch for the whole block, which is unlikely to occur.
In this case, we have the worst case complexity $\mathcal{O}(M\tmax)$.

Note that we can use fixed-point numbers instead of floating point numbers for a more efficient table construction. 
Also, the numerical values in the table are not important as long as the orders of any pair of values in the table are the same.

\subsection{Throughput Evaluations} \label{sec:numerical}

We now evaluate the performance of BAR in a feedbackless multi-hop network.
Note that baseline recoding is a special case of BAR with block size $1$.
Let $(h_0, h_1, \ldots, h_M)$ be the rank distribution of the batches arriving at a network node.
The \emph{normalized throughput} at a network node is defined to be the average rank of the received batches divided by the batch size, i.e., 
$\sum_{i = 0}^M ih_i / M$.
In our evaluations in this subsection, we set $t_\text{max}^\mathcal{L} = M|\mathcal{L}|$ for every block $\mathcal{L}$.
That is, the source node transmits $M$ packets per batch.
We assume that every link in the line network has independent packet loss with the same packet loss rate.

\input{arbr}

We first evaluate the normalized throughput with different batch sizes and packet loss rates.
Fig.~\ref{fig:tp} compares adaptive recoding (AR) and baseline recoding (BR) when we know the rank distribution of the batches arriving at each network node before the node applies BAR.
In other words, Fig.~\ref{fig:tp} shows the best possible throughput of adaptive recoding.
We will compare the effect of block sizes later.
We observe that
\begin{enumerate}
\item adaptive recoding has a higher throughput than baseline recoding under the same setting; 
\item the difference in throughput between adaptive recoding and baseline recoding is larger when
the batch size is smaller, the packet loss probability is larger, or the length of the line network is longer.
\end{enumerate}
In terms of throughput, the percentage gains of adaptive recoding over baseline recoding using $M = 4$ and $p = 0.2$ are $23.3\%$ and $33.7\%$ at the $20$-th and $40$-th hops respectively.
They become $43.8\%$ and $70.3\%$ respectively when $p = 0.3$.

\input{Lplot}

Now, we consider the effect of different block sizes.
Fig.~\ref{fig:L} shows the normalized throughput of different $|\mathcal{L}|$ and $p$ with $M = 8$.
We observe that 
\begin{enumerate}
	\item a larger $|\mathcal{L}|$ results a better throughput;
	\item using $|\mathcal{L}|=2$ already gives much larger throughput than using $|\mathcal{L}|=1$; and
	\item using $|\mathcal{L}|>8$ gives little extra gain in terms of throughput.
\end{enumerate}

\input{aras}

Next, we show the performance of the equal opportunity approximation scheme.
Fig.~\ref{fig:approx} compares the normalized throughput achieved by
Algorithm~\ref{alg:approx} (AS) and the true optimal throughput (AR).
We also compare the best possible throughput of adaptive recoding here.
We observe that 
\begin{enumerate}
	\item the approximation is close to the optimal solution; and
	\item the gap in normalized throughput is smaller when the batch size is larger, the packet loss probability is smaller, or the length of the line network is shorter.
\end{enumerate}

\section{Impact of Inaccurate Channel Models}
\label{sec:impact+feedback}

In this section, we first demonstrate that the throughput of BAR is insensitive to inaccurate channel models and inaccurate packet loss rates.
Then, we investigate the feedback design and show that feedback can enhance the throughput a little bit.

\subsection{Sensitivity of $\beta_p(r,t)$}

\begin{figure}
	\begin{subfigure}{.30\textwidth}
	\centering
	\begin{tikzpicture}[scale=.65,every node/.style={scale=.65}]
		\matrix (X) [matrix of math nodes,nodes={align=center}] at (0,0) {
			0.1000&	1&		1&		1\\
			0.0100&	0.1900&	1&		1\\
			0.0010&	0.0280&	0.2710&	1\\
			0.0001&	0.0037&	0.0523&	0.3439\\
			0.0000&	0.0005&	0.0086&	0.0815\\
			0.0000&	0.0001&	0.0013&	0.0159\\
			0.0000&	0.0000&	0.0002&	0.0027\\
		};
		\begin{scope}[on background layer]
			\node[fill=red!30,inner xsep=1mm,inner ysep=.6mm, fit=(X-1-1) (X-1-1) (X-1-1) (X-1-1)] {};
			\node[fill=red!30,inner xsep=1mm,inner ysep=.6mm, fit=(X-2-2) (X-2-2) (X-2-2) (X-2-2)] {};
			\node[fill=red!30,inner xsep=1mm,inner ysep=.6mm, fit=(X-3-3) (X-3-3) (X-3-3) (X-3-3)] {};
			\node[fill=red!30,inner xsep=1mm,inner ysep=.6mm, fit=(X-4-4) (X-4-4) (X-4-4) (X-4-4)] {};
			\node[fill=red!30,inner xsep=1mm,inner ysep=.6mm, fit=(X-3-2) (X-3-2) (X-3-2) (X-3-2)] {};
			\node[fill=red!30,inner xsep=1mm,inner ysep=.6mm, fit=(X-4-3) (X-4-3) (X-4-3) (X-4-3)] {};
			\node[fill=red!30,inner xsep=1mm,inner ysep=.6mm, fit=(X-5-4) (X-5-4) (X-5-4) (X-5-4)] {};
			\node[fill=red!30,inner xsep=1mm,inner ysep=.6mm, fit=(X-6-4) (X-6-4) (X-6-4) (X-6-4)] {};
		\end{scope}
	\end{tikzpicture}
	\caption{$p = 0.1$.}
	\end{subfigure}~~~~~\vrule
	\begin{subfigure}{.30\textwidth}
	\centering
	\begin{tikzpicture}[scale=.65,every node/.style={scale=.65}]
		\matrix (X) [matrix of math nodes,nodes={align=center}] at (0,0) {
			0.1010&	1&		1&		1\\
			0.0102&	0.1918&	1&		1\\
			0.0010&	0.0285&	0.2734&	1\\
			0.0001&	0.0038&	0.0533&	0.3468\\
			0.0000&	0.0005&	0.0088&	0.0829\\
			0.0000&	0.0001&	0.0013&	0.0163\\
			0.0000&	0.0000&	0.0002&	0.0028\\
		};
		\begin{scope}[on background layer]
			\node[fill=red!30,inner xsep=1mm,inner ysep=.6mm, fit=(X-1-1) (X-1-1) (X-1-1) (X-1-1)] {};
			\node[fill=red!30,inner xsep=1mm,inner ysep=.6mm, fit=(X-2-2) (X-2-2) (X-2-2) (X-2-2)] {};
			\node[fill=red!30,inner xsep=1mm,inner ysep=.6mm, fit=(X-3-3) (X-3-3) (X-3-3) (X-3-3)] {};
			\node[fill=red!30,inner xsep=1mm,inner ysep=.6mm, fit=(X-4-4) (X-4-4) (X-4-4) (X-4-4)] {};
			\node[fill=red!30,inner xsep=1mm,inner ysep=.6mm, fit=(X-3-2) (X-3-2) (X-3-2) (X-3-2)] {};
			\node[fill=red!30,inner xsep=1mm,inner ysep=.6mm, fit=(X-4-3) (X-4-3) (X-4-3) (X-4-3)] {};
			\node[fill=red!30,inner xsep=1mm,inner ysep=.6mm, fit=(X-5-4) (X-5-4) (X-5-4) (X-5-4)] {};
			\node[fill=red!30,inner xsep=1mm,inner ysep=.6mm, fit=(X-6-4) (X-6-4) (X-6-4) (X-6-4)] {};
		\end{scope}
	\end{tikzpicture}
	\caption{$p = 0.101$.}
	\end{subfigure}~~~~\vrule
	\begin{subfigure}{.30\textwidth}
	\centering
	\begin{tikzpicture}[scale=.65,every node/.style={scale=.65}]
		\matrix (X) [matrix of math nodes,nodes={align=center}] at (0,0) {
			0.0990&	1&		1&		1\\
			0.0098&	0.1882&	1&		1\\
			0.0010&	0.0275&	0.2686&	1\\
			0.0001&	0.0036&	0.0513&	0.3410\\
			0.0000&	0.0004&	0.0083&	0.0800\\
			0.0000&	0.0001&	0.0012&	0.0154\\
			0.0000&	0.0000&	0.0002&	0.0026\\
		};
		\begin{scope}[on background layer]
			\node[fill=red!30,inner xsep=1mm,inner ysep=.6mm, fit=(X-1-1) (X-1-1) (X-1-1) (X-1-1)] {};
			\node[fill=red!30,inner xsep=1mm,inner ysep=.6mm, fit=(X-2-2) (X-2-2) (X-2-2) (X-2-2)] {};
			\node[fill=red!30,inner xsep=1mm,inner ysep=.6mm, fit=(X-3-3) (X-3-3) (X-3-3) (X-3-3)] {};
			\node[fill=red!30,inner xsep=1mm,inner ysep=.6mm, fit=(X-4-4) (X-4-4) (X-4-4) (X-4-4)] {};
			\node[fill=red!30,inner xsep=1mm,inner ysep=.6mm, fit=(X-3-2) (X-3-2) (X-3-2) (X-3-2)] {};
			\node[fill=red!30,inner xsep=1mm,inner ysep=.6mm, fit=(X-4-3) (X-4-3) (X-4-3) (X-4-3)] {};
			\node[fill=red!30,inner xsep=1mm,inner ysep=.6mm, fit=(X-5-4) (X-5-4) (X-5-4) (X-5-4)] {};
			\node[fill=red!30,inner xsep=1mm,inner ysep=.6mm, fit=(X-6-4) (X-6-4) (X-6-4) (X-6-4)] {};
		\end{scope}
	\end{tikzpicture}
	\caption{$p = 0.099$.}
	\end{subfigure}

	\begin{subfigure}{.30\textwidth}
	\centering
	\begin{tikzpicture}[scale=.65,every node/.style={scale=.65}]
		\matrix (X) [matrix of math nodes,nodes={align=center}] at (0,0) {
			0.4500&	1&		1&		1\\
			0.2025&	0.6975&	1&		1\\
			0.0911&	0.4252&	0.8336&	1\\
			0.0410&	0.2415&	0.6090&	0.9085\\
			0.0185&	0.1312&	0.4069&	0.7438\\
			0.0083&	0.0692&	0.2553&	0.5585\\
			0.0037&	0.0357&	0.1529&	0.3917\\
		};
		\begin{scope}[on background layer]
			\node[fill=red!30,inner xsep=1mm,inner ysep=.6mm, fit=(X-2-2) (X-2-2) (X-2-2) (X-2-2)] {};
			\node[fill=red!30,inner xsep=1mm,inner ysep=.6mm, fit=(X-3-3) (X-3-3) (X-3-3) (X-3-3)] {};
			\node[fill=red!30,inner xsep=1mm,inner ysep=.6mm, fit=(X-4-4) (X-4-4) (X-4-4) (X-4-4)] {};
			\node[fill=red!30,inner xsep=1mm,inner ysep=.6mm, fit=(X-5-4) (X-5-4) (X-5-4) (X-5-4)] {};
			\node[fill=red!30,inner xsep=1mm,inner ysep=.6mm, fit=(X-3-2) (X-3-2) (X-3-2) (X-3-2)] {};
			\node[fill=red!30,inner xsep=1mm,inner ysep=.6mm, fit=(X-4-3) (X-4-3) (X-4-3) (X-4-3)] {};
			\node[fill=red!30,inner xsep=1mm,inner ysep=.6mm, fit=(X-6-4) (X-6-4) (X-6-4) (X-6-4)] {};
			\node[fill=red!30,inner xsep=1mm,inner ysep=.6mm, fit=(X-1-1) (X-1-1) (X-1-1) (X-1-1)] {};
		\end{scope}
	\end{tikzpicture}
	\caption{$p = 0.45$.}
	\end{subfigure}~~~~~\vrule
	\begin{subfigure}{.30\textwidth}
	\centering
	\begin{tikzpicture}[scale=.65,every node/.style={scale=.65}]
		\matrix (X) [matrix of math nodes,nodes={align=center}] at (0,0) {
			0.4545&	1&		1&		1\\
			0.2066&	0.7024&	1&		1\\
			0.0939&	0.4319&	0.8377&	1\\
			0.0427&	0.2475&	0.6163&	0.9115\\
			0.0194&	0.1358&	0.4152&	0.7505\\
			0.0088&	0.0723&	0.2628&	0.5676\\
			0.0040&	0.0377&	0.1589&	0.4013\\
		};
		\begin{scope}[on background layer]
			\node[fill=red!30,inner xsep=1mm,inner ysep=.6mm, fit=(X-2-2) (X-2-2) (X-2-2) (X-2-2)] {};
			\node[fill=red!30,inner xsep=1mm,inner ysep=.6mm, fit=(X-3-3) (X-3-3) (X-3-3) (X-3-3)] {};
			\node[fill=red!30,inner xsep=1mm,inner ysep=.6mm, fit=(X-4-4) (X-4-4) (X-4-4) (X-4-4)] {};
			\node[fill=red!30,inner xsep=1mm,inner ysep=.6mm, fit=(X-5-4) (X-5-4) (X-5-4) (X-5-4)] {};
			\node[fill=red!30,inner xsep=1mm,inner ysep=.6mm, fit=(X-3-2) (X-3-2) (X-3-2) (X-3-2)] {};
			\node[fill=red!30,inner xsep=1mm,inner ysep=.6mm, fit=(X-4-3) (X-4-3) (X-4-3) (X-4-3)] {};
			\node[fill=red!30,inner xsep=1mm,inner ysep=.6mm, fit=(X-6-4) (X-6-4) (X-6-4) (X-6-4)] {};
			\node[fill=red!30,inner xsep=1mm,inner ysep=.6mm, fit=(X-1-1) (X-1-1) (X-1-1) (X-1-1)] {};
		\end{scope}
	\end{tikzpicture}
	\caption{$p = 0.4545$.}
	\end{subfigure}~~~~\vrule
	\begin{subfigure}{.30\textwidth}
	\centering
	\begin{tikzpicture}[scale=.65,every node/.style={scale=.65}]
		\matrix (X) [matrix of math nodes,nodes={align=center}] at (0,0) {
			0.4455&	1&		1&		1\\
			0.1985&	0.6925&	1&		1\\
			0.0884&	0.4186&	0.8295&	1\\
			0.0394&	0.2355&	0.6016&	0.9055\\
			0.0175&	0.1268&	0.3986&	0.7370\\
			0.0078&	0.0662&	0.2479&	0.5494\\
			0.0035&	0.0338&	0.1471&	0.3822\\
		};
		\begin{scope}[on background layer]
			\node[fill=red!30,inner xsep=1mm,inner ysep=.6mm, fit=(X-2-2) (X-2-2) (X-2-2) (X-2-2)] {};
			\node[fill=red!30,inner xsep=1mm,inner ysep=.6mm, fit=(X-3-3) (X-3-3) (X-3-3) (X-3-3)] {};
			\node[fill=red!30,inner xsep=1mm,inner ysep=.6mm, fit=(X-4-4) (X-4-4) (X-4-4) (X-4-4)] {};
			\node[fill=red!30,inner xsep=1mm,inner ysep=.6mm, fit=(X-5-4) (X-5-4) (X-5-4) (X-5-4)] {};
			\node[fill=red!30,inner xsep=1mm,inner ysep=.6mm, fit=(X-3-2) (X-3-2) (X-3-2) (X-3-2)] {};
			\node[fill=red!30,inner xsep=1mm,inner ysep=.6mm, fit=(X-4-3) (X-4-3) (X-4-3) (X-4-3)] {};
			\node[fill=red!30,inner xsep=1mm,inner ysep=.6mm, fit=(X-6-4) (X-6-4) (X-6-4) (X-6-4)] {};
			\node[fill=red!30,inner xsep=1mm,inner ysep=.6mm, fit=(X-1-1) (X-1-1) (X-1-1) (X-1-1)] {};
		\end{scope}
	\end{tikzpicture}
	\caption{$p = 0.4455$.}
	\end{subfigure}
	\caption{The values of $\beta_p(t,r)$ for $r = 1, 2, 3, 4$ and $t = 1, 2, \ldots$ with different $p$. The coloured numbers are the largest eight values small than $1$.}
	\label{fig:cond}
\end{figure}

We can see that our algorithms only depend on the order of the values of $\beta_p(\cdot, \cdot)$, it is possible that the optimal $\{t_b\}_{b \in \mathcal{L}}$ for an incorrect $p$ is the same as the one for a correct $p$.
As shown in Fig.~\ref{fig:bp_table}, those boundaries $0$s and $1$s are not affected by $p \in (0,1)$.
That is, we only need to investigate the stability of $\beta_p(t,r)$ for $t \ge r > 0$.
We calculate some values of $\beta_p(t,r)$ corrected to $4$ digital places in Fig.~\ref{fig:cond} for $M = 4$, and $p = 0.1, 0.45$ and their $1\%$ relative changes.
We can see that the order of the values are mostly the same when we change $p$ a little bit.

\begin{figure}
	\centering
	\begin{subfigure}{.30\textwidth}
	\centering
	\begin{tikzpicture}[scale=.61,every node/.style={scale=.61}]
		\matrix (X) [matrix of math nodes,nodes={align=center}] at (0,0) {
			1.0000&	-&		-&		-\\
			2.0000&	0.9474&	-&		-\\
			3.0000&	1.9286&	0.8967&	-\\
			4.0000&	2.9189&	1.8585&	0.8479\\
			5.0000&	3.9130&	2.8388&	1.7898\\
			6.0000&	4.9091&	3.8268&	2.7596\\
			7.0000&	5.9062&	4.8187&	3.7412\\
		};
	\end{tikzpicture}
	\caption{$p = 0.1$}
	\end{subfigure}~~~~\vrule
	\begin{subfigure}{.30\textwidth}
	\centering
	\begin{tikzpicture}[scale=.61,every node/.style={scale=.61}]
		\matrix (X) [matrix of math nodes,nodes={align=center}] at (0,0) {
			1.0000&	-&		-&		-\\
			2.0000&	0.7097&	-&		-\\
			3.0000&	1.5714&	0.4899&	-\\
			4.0000&	2.4906&	1.2070&	0.3296\\
			5.0000&	3.4375&	2.0325&	0.9059\\
			6.0000&	4.4000&	2.9157&	1.6288\\
			7.0000&	5.3721&	3.8326&	2.4384\\
		};
	\end{tikzpicture}
	\caption{$p = 0.45$}
	\end{subfigure}
	\caption{The condition numbers of $\beta_p(t,r)$ for $r = 1, 2, 3, 4$ and $t = 1, 2, \ldots$.
	}
	\label{fig:cond2}
\end{figure}

We can also check with the \emph{condition number} \cite{condition_number} to verify the stability.\footnote{%
Roughly speaking, the relative change in the function output is approximately equal to the condition number times the relative change in the function input.
A small condition number of $\beta_p(t,r)$ means that the effect of the inaccurate $p$ is small.
As shown Fig.~\ref{fig:cond}, the values of $\beta_p(t,r)$ drops quickly when $t$ increases.
In the view throughput, which is proportional to the sum of these values, we can torture a larger relative change, i.e., a larger condition number, when $\beta_p(t,r)$ is small.
}
We calculate some condition numbers of $\beta_p(t,r)$ in Fig.~\ref{fig:cond2} by the formula stated in Theorem~\ref{thm:cond}. 

\begin{theorem} \label{thm:cond}
	Let $p \in (0,1)$ and $t \ge r > 0$.
	The condition number of $\beta_p(t,r)$ with respect to $p$ is $\frac{p^{t-r+1} (1-p)^{r-1} t!}{I_p(t-r+1,r) (t-r)! (r-1)!}$, or equivalently, 
	$\frac{\sum_{j = 0}^{r-1} (-1)^j \binom{r-1}{j} p^{t-r+j+1}}{\sum_{j = 0}^{r-1} (-1)^j \binom{r-1}{j} p^{t-r+j+1} / (t-r+j+1)}$. 
\end{theorem}

\begin{IEEEproof}
	See Appendix~\ref{sec:proof:thm:cond}.
\end{IEEEproof}

\subsection{Impact of Inaccurate Channel Models}
\label{sec:inaccurate}

To demonstrate the impact of inaccurate channel model, we consider three different channels to present our observations.
\begin{itemize}
	\item ch1: independent packet loss with constant loss rate $p = 0.45$.
	\item ch2: burst packet loss modelled by the GE model illustrated in Fig~\ref{fig:ge_model} with the parameters used in \cite{ge_adaptive}, which are $\pgb = \pbg = \pg = 0.1$, $\pb = 0.8$. 
	\item ch3: independent packet loss with varying loss rate $p = 0.45 + 0.3 \sin(2\pi c/1280)$ where $c$ is the number of batches transmitted.
\end{itemize}
All the three channels have the same average packet loss rate $0.45$.
The formula of ch3 is for demonstration purpose only.

We now demonstrate the impact of inaccurate $p$ on the throughput.
We consider a line network where all the links use the same channel (ch1, ch2 or ch3).

\begin{figure}
	\centering
	\includegraphics{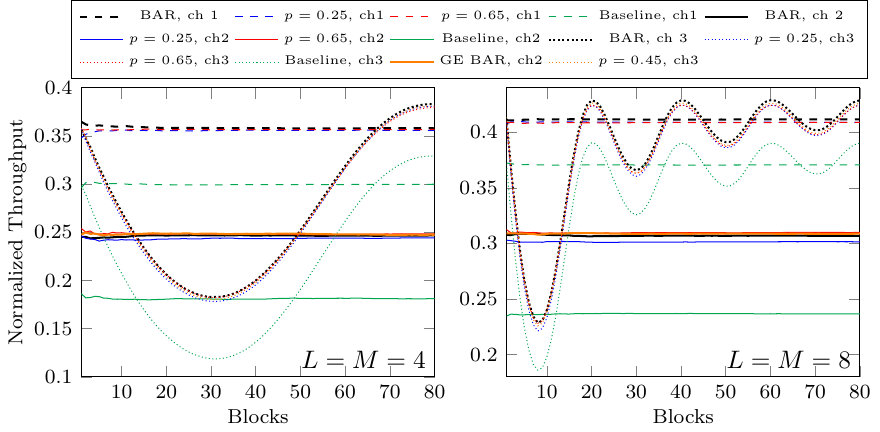}
	\caption{Throughput with inaccurate channel conditions.}
	\label{fig:plot1}
\end{figure}

In Fig.~\ref{fig:plot1}, we plot the normalized throughput of the first $80$ received blocks at the $4$-th hop where $L = M = 4$ or $8$.
We use BAR with \eqref{eq:ert} for each network although ch2 is a bursty channel.
The black curves with legend BAR are the throughput of BAR where the loss rate is known. 
For ch1 and ch2, this loss rate $p$ is a constant $0.45$.
The red and blue curves are the throughput of BAR when we guess $p = 0.65$ and $0.25$ respectively, which is $\pm 0.2$ from the average loss rate $0.45$.
As there is no feedback, we do not change our guess on $p$ for these curves.
We can see that the throughput is actually very close to the corresponding black curves.
This suggest that in the view of throughput, BAR is not sensitive to $p$.
Even with a wild guess on $p$, BAR still outperforms baseline recoding, as illustrated by the green curves.
Regarding ch2, we also plot the orange curve with legend GE BAR, which is the throughput achieved by BAR with \eqref{eq:ert_ge}. 
We can see that the gap between the throughput achieved by BAR with \eqref{eq:ert} and \eqref{eq:ert_ge} is very small.
As a summary of our demonstration: 
\begin{enumerate}
	\item We can use BAR with \eqref{eq:ert} for bursty channels 
		and the loss in throughput is not significant.
	\item BAR with an inaccurate constant $p$ 
		can achieve a throughput close to the one when we have the exact real-time loss rate.
	\item We can see a significant throughput gain from baseline recoding by using BAR even with inaccurate channel models.
\end{enumerate}

\subsection{Feedback Design}

Although an inaccurate $p$ can give an acceptable throughput, we can further enhance the throughput by adapting the varying $p$. 
To achieve this goal, we need to use feedback.

We adopt a simple feedback strategy which let the next node returns the number of received packets of the batches for the current node to estimate $p$.
Although the next node does not know the number of lost packets per batch, it knows the number of received packets per batch.
So, we do not need to introduce more overhead to the packets transmitted by the current node.

When we estimate $p$, we have to know the number of packets lost during a certain time frame.
If the time frame is too small, the estimation is too sensitive so that the estimated $p$ changes rapidly and unpredictably.
If the time frame is too long, we captured too much out-dated information about the channel so the estimated $p$ changes too slow and may not be able to adapt to the real loss rate.
Recall that the block size is not large as we want to keep the delay small.
We use a block as an atomic unit of the time frame.
The next node gives feedback on the number of received packets per block.
The current node uses the feedback of the blocks in the time frame to estimate $p$.
We perform an estimation of $p$ per received feedback.
This way, the estimated $p$ is the same for each block so that we can apply BAR with \eqref{eq:ert}.

If the feedback is sent via a reliable side channel, then we can assume that the current node can always receive the feedback.
However, if the feedback is sent via an unreliable channel, say, the reverse direction of the same channel the data packets were sent, then we need to consider feedback loss.
Let $\Lambda$ be a set of blocks in a time frame with received feedback.
We handle the case of feedback loss by considering the total number of packets transmitted for the blocks in $\Lambda$ as the total number of packets transmitted during the time frame.
This way, we can also start the estimation before a node sent enough blocks to fill up a time frame.
Suppose no feedback is received for every block in a time frame, then we reuse the previously estimated $p$ for BAR.

At the beginning of the transmission, we have no feedback yet so we have no information to estimate $p$.
To outperform baseline recoding without the knowledge of $p$, we can use the approximation of BAR given by Algorithm~\ref{alg:approx}.
Once we have received at least one feedback, we can start estimating $p$.

\begin{figure*}
	\hspace*{-1cm}
	\includegraphics{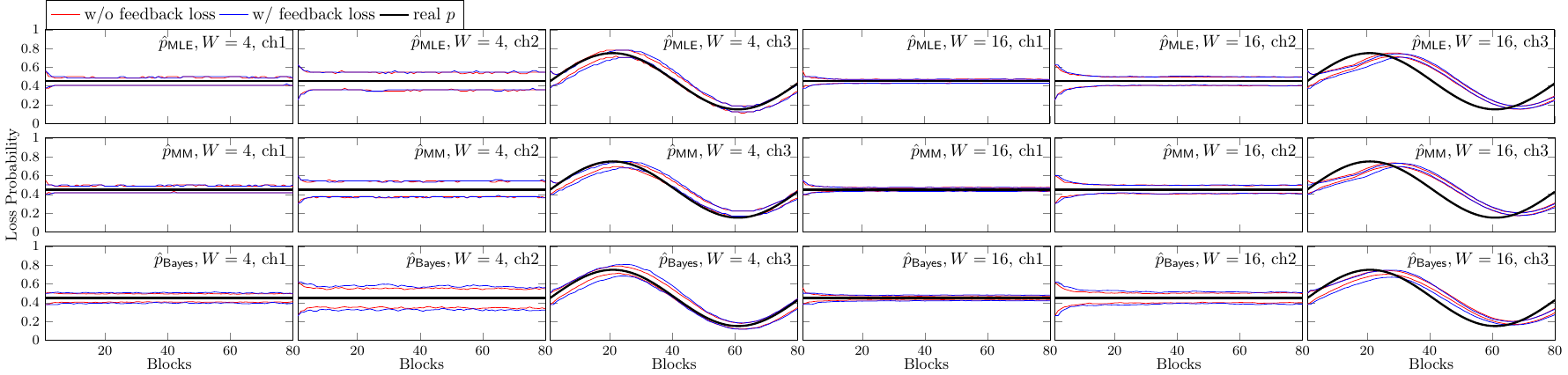}
	\caption{The $25\%$ and $75\%$ percentiles of the estimation of $p$ by different schemes where $L = M = 4$ in $1000$ runs.}
	\label{fig:plot2}
\end{figure*}

\subsection{Estimators}

Let $x$ and $n$ be the total number of packets received by the current node and the total number of packets transmitted by the previous node respectively in a time frame of observation.
That is, the number of packets lost in the time frame is $n-x$.
We introduce three types of estimators for our numerical evaluations.

\emph{1) Maximum likelihood estimator (MLE):}
The MLE, denoted by $\mle$, estimates $p$ by maximizing the likelihood function.
$\mle = (n-x)/n$ is a well-known result which can be obtained via derivative test.
This form collides with the sample average, so by the law of large number, $\mle \to p$ when $n \to \infty$ if $p$ does not change over time.

\emph{2) Minimax estimator:}
The minimax estimator achieves the smallest maximum risk among all estimators.
With the popular mean squared error (MSE) as the risk function, it is a Bayes estimator with respect to the least favourable prior distribution.
As studied in \cite{minimax1,minimax2}, such prior distribution is a beta distribution $\bet(\sqrt{n}/2, \sqrt{n}/2)$.
The minimax estimator of $p$, denoted by $\minimax$, is the posterior mean, which is $\frac{\sqrt{n}}{1+\sqrt{n}} \frac{n-x}{n} + \frac{1}{1+\sqrt{n}} \frac{1}{2}$, or equivalently, $\frac{n-x + 0.5 \sqrt{n}}{n+\sqrt{n}}$.

\emph{3) Weighted Bayesian update:}
Suppose the prior distribution is $\bet(a, b)$, where the hyperparameters can be interpreted as a pseudo-observation having $a$ successes and $b$ failures. 
Given a sample of $s$ successes and $f$ failures from a binomial distribution, the posterior distribution is $\bet(a + s, b + f)$. 
To fade out the old samples captured by the hyperparameters, we introduce a scaling factor $0 \le \gamma \le 1$ and let the posterior distribution be $\bet(\gamma a + s, \gamma b + f)$.
This factor can also prevent the hyperparameters grow indefinitely. 
The estimation of $p$, denoted by $\wbu$, is the posterior mean with $s = n-x$ and $f = x$, which is $\frac{\gamma a + n-x}{\gamma(a+b) + n}$.
To prevent a bias when there is no enough sample, we should select a non-informative prior as the initial hyperparameters.
Specifically, we use the Jeffreys prior, which is $\bet(1/2, 1/2)$.

We first show the estimation of $p$ by different schemes in Fig.~\ref{fig:plot2}.
We use BAR with \eqref{eq:ert} and $L = M = 4$. 
The size of the time frame is $W$ blocks.
For $\mle$ and $\minimax$, the observations in the whole time frame has the same weight.
For $\wbu$, the effect of each observation is deceasing exponentially fast.
We consider an observation is out of the time frame when it is scaled into $10\%$ of the original value.
That is, we define the scaling factor by $\gamma = \sqrt[W]{0.1}$.
In each subplot, the black curve is the real-time $p$.
The red and blue curves are for the estimation without feedback loss and for that with feedback loss respectively.
In each case, the two curves are the 
$25\%$ and $75\%$ percentiles from $1000$ runs respectively.

We can see that a larger $W$ has a slower respond to the change of $p$ in ch3.
Among the estimators, $\wbu$ has the fastest respond speed as its observations in a time frame is not fairly weighted.
Also, although ch1 and ch2 have the same average loss rate, the estimation has a larger variance when the channel is bursty.

\subsection{Throughput Evaluations}

\begin{figure}
	\centering
	\hspace*{-.5cm}
	\includegraphics{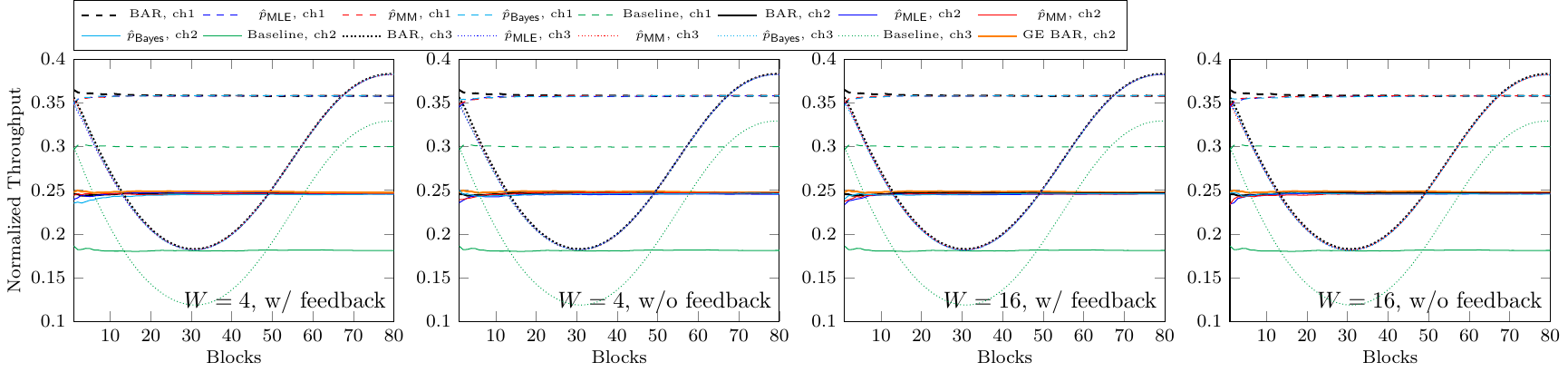}
	\caption{Throughput with estimated $p$ via feedback where $L = M = 4$.}
	\label{fig:plot3}
\end{figure}

As we have discussed in Section~\ref{sec:inaccurate}, the $p$ we guess 
does not have a significant impact on the throughput.
We now show the throughput achieved by the estimation schemes in Fig.~\ref{fig:plot3}.
We are not wild guessing $p$ anymore so it is not surprise that we can achieve nearly the same throughput as when we know the real $p$ for ch1 and ch2.
If we look closely, we can see from Fig.~\ref{fig:plot1} that for ch3, there is a small gap between the throughput of BAR when we know the real-time $p$ and the one of BAR when we use a constant $p$.
Although the estimation may not be accurate at all time, we can now adapt to the change of $p$ so we can finally achieve a throughput nearly the same as when we know the real-time $p$.
On the other hand, no matter the feedback is loss or not, the plots shown in Fig.~\ref{fig:plot3} are basically the same.

\section{Concluding Remarks}
\label{sec:conclusion}

We proposed blockwise adaptive recoding (BAR) in this paper which can adapt to variation of the incoming channel condition.
In a practical perspective, we discussed how to calculate the components of BAR and how to solve BAR efficiently.
We also investigated the impact of inaccurate channel model on the throughput achieved by BAR.
Our evaluations showed that
\begin{enumerate}
	\item BAR is not sensitive to the channel model: A wild guess on the loss rate can still outperform baseline recoding.
	\item For bursty channels, the throughput achieved by BAR with an independent loss model is nearly the same as the one with the real channel model.
		This is, we can use the independent loss model for BAR in practice and apply the techniques in this paper to reduce the computational costs of BAR.
	\item Feedback can enhance the throughput a little bit for channels with dynamic loss rate.
		On the other hand, feedback loss barely has an effect to the throughput of BAR.
		So, we can send the feedback through a lossy channel without the need of retransmission.
		Unless we need to use an accurate estimated loss rate in other applications, we can use MLE with a small time frame for BAR to reduce computational time.
\end{enumerate}
These encouraging results suggest that BAR is suitable to be deployed in real-world applications. 

\appendices

\section{Discussion on Algorithm~\ref{alg:approx}}

\subsection{Linear Time Selection}
\label{sec:approx:linear}

Here we discuss the way to add one to the number of recoded packets for those batches having the highest ranks in $\mathcal{O}(|\mathcal{L}|)$ time.
		The linear time worst case can be achieved by using introselect \cite{introselect} or quickselect \cite{quickselect} with median of medians \cite{median} pivot strategy. 
		We use the selection algorithm to find the $r$-th largest element, and we also make use of its intermediate steps.
		During an iteration, one of the following three cases will occur.
		If the algorithm decides to search the part larger than the pivot, then it means that the discarded part does not contain the largest $r$ elements. 
		If the part smaller than the pivot is selected, then the discarded part is part of the largest $r$ elements.
		If the pivot is exactly the $r$-th largest element, then the part larger than the pivot together with the pivot are part of the largest $r$ elements.

\subsection{Counting Technique}
\label{sec:approx:count}

Now we discuss how to search those batches having the highest ranks in $\mathcal{O}(|\mathcal{L}|+M)$ time.
	It can be done by using part of the counting sort algorithm \cite{seward54}.
	We first compute the histogram of the number of times each rank occurs, which takes $\mathcal{O}(M)$ time for initialization and $\mathcal{O}(|\mathcal{L}|)$ time for scanning the block.
		Then, we can scan and count the frequencies of the histogram from the highest rank, and eliminate the part where the count excesses $\ell \bmod |\mathcal{L}|$.
		This takes $\mathcal{O}(M)$ time.
	Lastly, we scan the ranks of the batches again in $\mathcal{O}(|\mathcal{L}|)$ time.
	If it is included in the modified histogram, we add $1$ to the corresponding $t_b$ and minus $1$ to the corresponding frequency in the histogram.

\subsection{Performance Guarantee and Bounded Error}
\label{sec:proof:thm:approx2}

We start the discussion with the following theorem.

\begin{theorem} \label{thm:approx2}
	Let $\text{SOL}$ and $\text{OPT}$ be the solution given by Algorithm~\ref{alg:approx} and the optimal solution of \eqref{eq:B} respectively, then
	\begin{equation*}
		\left\{
			\begin{IEEEeqnarraybox*}[][c]{rCl}
				\IEEEstrut
				\text{SOL} & \ge & (1-p) \text{OPT}, \\
				\text{OPT}-\text{SOL} & \le & (1-p) \sum_{b \in \mathfrak{L}} \sum_{j = r_b + \ell}^{r_b + |\mathfrak{L}|\ell' - 1} \beta_p(j,r_b),
				\IEEEstrut
		\end{IEEEeqnarraybox*}
		\right.
	\end{equation*}
	where $\ell' = (t_\text{max}^\mathcal{L} - \sum_{b \in \mathcal{L}} r_b)/|\mathfrak{L}|$ and $\ell = \lfloor \ell' \rfloor$.
\end{theorem}

\begin{IEEEproof}
	We first show that the algorithm has a relative performance guarantee factor $1-p$.
	As stated in Theorem~\ref{thm:approx}, when $t_\text{max}^\mathcal{L} \le \sum_{b \in \mathcal{L}} r_b$, the algorithm guarantees an optimal solution.
	So, we only consider $t_\text{max}^\mathcal{L} > \sum_{b \in \mathcal{L}} r_b$.
	Let $\{t_b\}_{b \in \mathcal{L}}$ be the approximation given by the algorithm. 

	Note that any linear combinations of $r$ independent vectors cannot obtain more than $r$ independent vectors.
	So, the expected rank of a batch at the next hop must be no larger than the rank of the batch at the current hop, and, it is also non-negative.
	That is,
	\begin{equation} \label{thm:approx:exp}
		0 \le E(r_b,t) \le r_b, \forall t \ge 0, b \in \mathcal{L}.
	\end{equation}
	This gives a bound of the optimal solution by
	\begin{equation} \label{thm:approx:opt}
		0 \le \text{OPT} \le \sum_{b \in \mathcal{L}} r_b.
	\end{equation}

	We consider the exact formula of the approximation:
	\begin{IEEEeqnarray}{rCl}
		\text{SOL} & = & (1-p) \sum_{b \in \mathcal{L}} r_b + (1-p) \sum_{b \in \mathcal{L}} \sum_{j = r_b}^{t_b-1} \beta_p(j, r_b) \label{thm:approx:rho1} \\
				& \ge & (1-p) \sum_{b \in \mathcal{L}} r_b \label{thm:approx:rho2} \\
				& \ge & (1-p) \text{OPT}, \label{thm:approx:rho3}
	\end{IEEEeqnarray}
	where
	\begin{itemize}
		\item \eqref{thm:approx:rho1} is stated in Lemma~\ref{lem:exp_rank}\eqref{lem:exp_rank_exact};
		\item \eqref{thm:approx:rho2} holds as $\beta_p(j, r_b) \ge 0$ for all $j, r_b$, which is by \eqref{eq:beta_range};
		\item \eqref{thm:approx:rho3} follows the inequality \eqref{thm:approx:opt}.
	\end{itemize}

	Lastly, we show the bounded error.
	Let $\{t_b^\ast\}$ be a solution to \eqref{eq:B}.
	By Corollary~\ref{cor:first_packet}, we can write $t_b^\ast = r_b + \ell_b$ where $\ell_b \ge 0$ for all $b \in \mathcal{L}$.
	Note that the constraint of \eqref{eq:B}, i.e., $\sum_{b \in \mathcal{L}} t_b^\ast = t_\text{max}^\mathcal{L}$, suggests that
	\begin{equation} \label{thm:approx:c0}
		\ell_b \le t_\text{max}^\mathcal{L} - \sum_{b \in \mathcal{L}} r_b = |\mathfrak{L}|\ell'.
	\end{equation}

	On the other hand, it is easy to see that the approximation must give either $t_b = r_b + \ell$ or $t_b = r_b + \ell + 1$.
	That is, we have $t_b \ge r_b + \ell$.
	By Lemma~\ref{lem:exp_rank}\eqref{lem:exp_rank_exact}, we have
	\begin{equation} \label{thm:approx:c1}
		\text{SOL} \ge (1-p) \sum_{b \in \mathfrak{L}} \left[ r_b + \sum_{j = r_b}^{r_b+\ell-1} \beta_p(j, r_b) \right].
	\end{equation}

	We consider the difference between $\text{OPT}$ and $\text{SOL}$:
	\begin{IEEEeqnarray}{Cl}
		& \text{OPT} - \text{SOL} \nonumber \\
		\le & (1-p) \sum_{b \in \mathfrak{L}} \left( \sum_{j = r_b}^{r_b+\ell_b-1} \beta_p(j,r_b) - \sum_{j = r_b}^{r_b+\ell-1} \beta_p(j,r_b) \right) \label{thm:approx:c2} \\
		= & (1-p) \sum_{b \in \mathfrak{L}} \left( \sum_{\substack{j = r_b+\ell,\\\ell_b > \ell}}^{r_b + \ell_b - 1} \beta_p(j,r_b) - \sum_{\substack{j = r_b + \ell_b,\\\ell_b < \ell}}^{r_b+\ell-1} \beta_p(j, r_b) \right) \nonumber \\
		\le & (1-p) \sum_{b \in \mathfrak{L}} \sum_{\substack{j = r_b+\ell,\\\ell_b > \ell}}^{r_b + \ell_b - 1} \beta_p(j,r_b) \label{thm:approx:c3} \\
		\le & (1-p) \sum_{b \in \mathfrak{L}} \sum_{j = r_b + \ell}^{r_b + |\mathfrak{L}|\ell' - 1} \beta_p(j, r_b), \label{thm:approx:c4}
	\end{IEEEeqnarray}
	where
	\begin{itemize}
		\item \eqref{thm:approx:c2} is the difference between the exact form of $\text{OPT}$ by Lemma~\ref{lem:exp_rank}\eqref{lem:exp_rank_exact} after substituting the lower bound of $\text{SOL}$ shown in \eqref{thm:approx:c1};
		\item the condition $\ell_b > \ell$ in the summation of \eqref{thm:approx:c3} can be removed, as we have $r_b + \ell_b - 1 < r_b + \ell$ if $\ell_b \le \ell$;
		\item \eqref{thm:approx:c4} follows \eqref{thm:approx:c0} and the fact shown in \eqref{eq:beta_range} that the extra $\beta_p(j, r_b)$ terms are non-negative. 
	\end{itemize}
	The proof is done.
\end{IEEEproof}

If the relative performance guarantee factor $1-p$ is tight, we need both equalities in \eqref{thm:approx:rho2} and \eqref{thm:approx:rho3} hold.
First, by \eqref{eq:beta_range}, we know that $\beta_p(j,r_b)$ is always non-negative.
The equality in \eqref{thm:approx:rho2} holds if and only if $\sum_{j = r_b}^{t_b-1} \beta_p(j,r_b) = 0$ for all $b \in \mathcal{L}$.
The sum equals $0$ only when
\begin{itemize}
	\item $r_b = 0$ and $t_b \ge 0$ according to \eqref{eq:beta0}; or
	\item $t_b - 1 < r_b$ which forms an empty sum.
\end{itemize}

The equality in \eqref{thm:approx:rho3} holds if and only if $\text{OPT} = \sum_{b \in \mathcal{L}} E(r_b,t_b^\ast) = \sum_{b \in \mathcal{L}} r_b$.
Note that \eqref{thm:approx:exp} shows that $E(r_b, t_b^\ast)$ is upper bounded by $r_b$.
This implies that we need $E(r_b, t_b^\ast) = r_b$ for all $b \in \mathcal{L}$.
When $t_b^\ast \le r_b$, we can apply Lemma~\ref{lem:exp_rank}\eqref{lem:exp_rank_rec} to obtain $E(r_b, t_b^\ast) = (1-p)t_b^\ast$, which equals $r_b$ if and only if $r_b = 0$, as we assumed $0 < p < 1$ in this paper.
By Lemma~\ref{lem:exp_rank2}, $E(r_b,t)$ is a monotonic increasing function in terms of $t$ for all $r_b \ge 0$.
So when $r_b \neq 0$, we need $t_b^\ast > r_b$, which implies that $t_\text{max}^\mathcal{L} > \sum_{b \in \mathcal{L}} r_b$.
Then, the approximation will also give $t_b > r_b$ for some $b \in \mathcal{L}$ in this case, and the equality in \eqref{thm:approx:rho2} does not hold.

That is, we have $\text{SOL} = (1-p) \text{OPT}$ only when $r_b = 0$ for all $b \in \mathcal{L}$.
In this case, we have $\text{SOL} = \text{OPT} = 0$.
In practice, the probability of having $r_b = 0$ for all $b \in \mathcal{L}$ is very small.
So, we can consider that the bound is not tight in most cases but it guarantees that the approximation is good when the packet loss probability is small.

\section{Lazy Evaluations in Algorithm~\ref{alg:opt2}} \label{sec:corrupted_heap}

In Algorithm~\ref{alg:opt2}, we need to query the minimum of $\beta_p(t_a-1,r_a)$ and the maximum of $\beta_p(t_b,r_a)$, where $a, b \in \mathcal{L}$.
During an iteration, suppose we choose to increase $t_b$ by $1$ and decrease $t_a$ by $1$.

It is clear that we need to decrease the key $\beta_p(t_b,r_b)$ into $\beta_p(t_b+1,r_b)$ in the max-heap, and increase the key $\beta_p(t_a-1,r_a)$ into $\beta_p(t_a-2, r_a)$ in the min-heap.
However, we can omit the updates for the batch $a$ in the max-heap and the batch $b$ in the min-heap, which will be discussed below.

\begin{lemma} \label{lem:heap}
	If the batch $a$ is selected by the max-heap or the batch $b$ is selected by the min-heap in any future iteration, then the optimal solution is reached.
\end{lemma}

\begin{IEEEproof}
	Suppose the batch $a$ with key $\mathcal{A}$ is selected by the max-heap in a future iteration.
	Note that $\mathcal{A}$ was once the smallest element in $\mho_k$ for some $k$.
	So at the current state where $k' > k$, every element in $\mho_{k'}$ must be no smaller than $\mathcal{A}$.
	Equivalently, we have $(1-p)\beta_p(t_\kappa,r_\kappa) \le (1-p)\beta_p(t_\rho-1,r_\rho)$ for all $\kappa, \rho \in \mathcal{L}$.
	By Theorem~\ref{thm:not_opt}, the optimal solution is reached.
	The min-heap counterpart can be proved in a similar fashion.
\end{IEEEproof}

Suppose we omit the update for the batch $\rho$ in the heap.
We call the key of the batch $\rho$ a \emph{corrupted key}, or the key of the batch $\rho$ is \emph{corrupted}.
A key which is not corrupted is called an \emph{uncorrupted key}.
A heap with corrupted keys is called a \emph{corrupted heap}.\footnote{
	We do not have a guaranteed maximum portion of corrupted keys as an input.
	Also, we do not adopt the carpooling technique.
	This suggests that the heap here is not a soft heap \cite{softheap}.
}

In our scenario which is mentioned at the beginning of this section, the key of a batch is corrupted in a corrupted max-heap if and only if the same batch was once the minimum of the counterpart original min-heap, and vice versa.

\begin{lemma} \label{lem:root}
	If the root of a corrupted heap is a corrupted key, then the optimal solution is reached.
\end{lemma}

\begin{IEEEproof}
	We only consider a corrupted max-heap in the proof.
	We can use similar arguments to show that a corrupted min-heap also works.

	In a future iteration, suppose the batch $a$ is selected by the corrupted max-heap.
	We consider the real maximum in the original max-heap.
	There are three cases.

	Case I: the batch $a$ is also the root of the original max-heap.
	As the key of $a$ is corrupted, it means that the batch was once selected by the counterpart min-heap.
	By Lemma~\ref{lem:heap}, the optimal solution is reached.

	Case II: the root of the original max-heap is a batch $a'$ where the key of $a'$ is also corrupted.
	Similar to Case I, the batch $a'$ was once selected by the counterpart min-heap, and we can apply Lemma~\ref{lem:heap} to finish this case.

	Case III: the root of the original max-heap is a batch $a''$ where the key of $a''$ is not corrupted.
	In this case, the uncorrupted key of $a''$ is also in the corrupted max-heap.
	Note that the corrupted key of $a$ is no larger than the actual key of $a$ in the original max-heap.
	This means that the key of $a$, $a''$ and the corrupted key of $a$ are having the same value.
	It is equivalent to let the original max-heap select the batch $a$, as every element in $\mho_{k'}$ must be no smaller than the key of $a''$, where $k'$ represents the state of the current iteration.
	Then, the problem is reduced to Case I.

	Combining the three cases, the proof is done.
\end{IEEEproof}

\begin{theorem}
	The updates for the batch $a$ in the max-heap and the batch $b$ in the min-heap can be omitted.
\end{theorem}

\begin{IEEEproof}
	When we omit the updates, the heap itself becomes a corrupted heap.
	We have to make sure that when a batch having corrupted key is selected, the termination condition of the algorithm is also met.

	We can express the key of the batch $\pi$ in a corrupted max-heap and min-heap by $\beta_p(t_\pi+s_\pi, r_\pi)$ and $\beta_p(t_\pi-1-u_\pi, r_\pi)$ respectively, where $s_\pi, u_\pi$ are non-negative integers.
	When $s_\pi$ or $u_\pi$ is $0$, the key is uncorrupted in the corresponding corrupted heap.
	By Lemma~\ref{lem:bp}\eqref{lem:bp_dec}, we have
	\begin{IEEEeqnarray*}{rCl}
		\beta_p(t_\pi+s_\pi, r_\pi) & \le & \beta_p(t_\pi, r_\pi),\\
		\beta_p(t_\pi-1, r_\pi) & \le & \beta_p(t_\pi-1-u_\pi, r_\pi).
	\end{IEEEeqnarray*}
	That is, the root of the corrupted max-heap is no larger than the root of the original max-heap.
	Similar for the min-heap.
	Mathematically, we have
	\begin{IEEEeqnarray}{rCl}
		\max_{\pi \in \mathcal{L}} \beta_p(t_\pi+s_\pi, r_\pi) & \le & \max_{\pi \in \mathcal{L}} \beta_p(t_\pi, r_\pi), \label{eq:thm:corrupted1} \\
		\min_{\pi \in \mathcal{L}} \beta_p(t_\pi-1, r_\pi) & \le & \min_{\pi \in \mathcal{L}} \beta_p(t_\pi-1-u_\pi, r_\pi). \label{eq:thm:corrupted2}
	\end{IEEEeqnarray}

	Suppose a corrupted key is selected.
	By Lemma~\ref{lem:root}, we know that the optimal solution is reached.
	So, we can apply the contrapositive of Theorem~\ref{thm:not_opt} and know that
	\begin{equation} \label{eq:thm:corrupted3}
		(1-p) \beta_p(t_\kappa, r_\kappa) \le (1-p) \beta_p(t_\rho-1,r_\rho)
	\end{equation}
	for all $\kappa, \rho \in \mathcal{L}$.
	We can omit the condition $t_\rho \ge 1$ because by \eqref{eq:beta_range} and \eqref{eq:beta1}, we have $\beta_p(-1,\cdot) = 1 \ge \beta_p(\cdot,\cdot)$.
	The inequality \eqref{eq:thm:corrupted3} is equivalent to
	\begin{equation*}
		\max_{\pi \in \mathcal{L}} \beta_p(t_\pi, r_\pi) \le \min_{\pi \in \mathcal{L}} \beta_p(t_\pi-1, r_\pi).
	\end{equation*}
	We can mix this inequality with \eqref{eq:thm:corrupted1} and \eqref{eq:thm:corrupted2} to show that when a corrupted key is selected, we have
	\begin{equation*}
		\max_{\pi \in \mathcal{L}} \beta_p(t_\pi+s_\pi, r_\pi) \le \min_{\pi \in \mathcal{L}} \beta_p(t_\pi-1-u_\pi, r_\pi),
	\end{equation*}
	which is the termination condition shown in Algorithm~\ref{alg:opt2} after we replaced the heaps into corrupted heaps.

	We just showed that once a corrupted key selected, the termination condition is reached.
	In the other words, before a corrupted key is selected, every previous selection must be an uncorrupted key.
	That is, the details inside the iterations are not affected.
	If an uncorrupted key is selected where it also satisfies the termination condition, then it means that no corrupted key is touched, and the corrupted heap still acts as a normal heap at this point.

	The correctness of the algorithm when we use a corrupted heap is proved.
	Moreover, we do not need to mark down which key is corrupted.
	This is, we can omitted the mentioned heap updates for a normal heap.
\end{IEEEproof}

We do not need to mark down which key is corrupted while the algorithm still works, so we can simply omit the mentioned updates as lazy evaluations.
As there are two heaps in algorithm, we can reduce from four to two heap updates.

\section{Linear Programming Formulation of BAR}
\label{sec:linear}

In \cite{wang2021smallsample}, a distributionally robust optimization \cite{gao2016distributionally} for adaptive recoding is formulated as a linear programming problem.
It is based on an observation that when the expected rank function $E(r,t)$ is concave with respect to $t$, we can reformulate it by
\begin{equation*}
	E(r,t) = \min_{i \in \{0, 1, \ldots, \bar\imath\}} (\Delta_{r,i} t + \xi_{r,i})
\end{equation*}
if we fix an artificial upper bound $t \le \bar\imath$, where $\Delta_{r,t} := E(r,i+1)-E(r,i)$ and $\xi_{r,i} := E(r,i) - i\Delta_{r,i}$.
In \eqref{eq:B}, we implicitly have $t \le t_\text{max}^\mathcal{L}$, so we can make use of this expression to write \eqref{eq:B} as
\begin{equation*}
\begin{IEEEeqnarraybox}[][c]{rCl}
	\max_{t_b, e_b \ge 0, \forall b \in \mathcal{L}} & \quad & \sum_{b \in \mathcal{L}} e_b\\
	\mathrm{s.t.} && \sum_{b \in \mathcal{L}} t_b = t^{\mathcal{L}}_\text{max}\\
	&& \negthickspace\negthickspace\negthickspace\negthickspace e_b \le E(r_b, i) + (E(r_b, i+1) - E(r_b, i)) (t_b - i),\\
	&& \qquad \qquad \qquad \forall b \in \mathcal{L}, \forall i \in \{0, 1, \ldots, t_\text{max}^\mathcal{L}\},
\end{IEEEeqnarraybox}
\end{equation*}
where $t_b$ is allowed to be a non-integer.
A non-integer $t_b$ means that we first generate $\lfloor t_b \rfloor$ recoded packets, then we generate one more recoded packet with probability $t_b - \lfloor t_b \rfloor$.
Note that there are $|\mathcal{L}|t_\text{max}^\mathcal{L}$ constraints for $e_b$.

To turn such a non-deterministic solution into a deterministic one, we perform the following steps:
\begin{enumerate}
\item Collect the batches having non-integer number of recoded packets into a set $S$.
\item Calculate $R = \sum_{b \in S} (t_b - \lfloor t_b \rfloor)$. Note that $R$ must be an integer.
\item For every $b \in S$, remove the fractional part of $t_b$.
\item Randomly select $R$ batches from $S$ and add one recoded packet to each of these batches.
\end{enumerate}

We have an integer $R$ because $\sum_{b \in \mathcal{L}} t_b = t_\text{max}^\mathcal{L}$.
Also, we have $R < |S|$.
Referring to the idea of Algorithm~\ref{alg:opt1}, we have the same value of $\Delta_{r_b,\lfloor t_b \rfloor}$ for all $b \in S$.
After removing the fractional part of $t_b$ for all $b \in S$, it becomes the subproblem \eqref{eq:Bk} with $k = t_\text{max}^\mathcal{L}-R$.
The last step follows Algorithm~\ref{alg:opt1} so that the output is a solution to \eqref{eq:B} where $t_b$ for all $b \in \mathcal{L}$ are all integers.

\section{BAR with Known and Unchanged Incoming Link Condition}
\label{sec:longrun}

Every batch arrives the current node with a rank.
The ranks of all the batches during the transmission can form a rank distribution.
As this distribution is for the input batches, we call it the \emph{input rank distribution}.

If we know the input rank distribution, we can consider a block large enough to contain all the incoming batches.
We can have the following benefits:
\begin{enumerate}[i)]
	\item by Theorem~\ref{thm:2blocks}, we can achieve the highest expected rank at the next node;
	\item as there is only one block, we only need to solve \eqref{eq:B} once; and
	\item we can solve \eqref{eq:B} before we receive a batch, i.e., we can decide the number of recoded packets once we finished receiving a batch, which the delay is the same as a block of size $1$.
\end{enumerate}

We can calculate the input rank distribution before we receive all the batches if we know
\begin{enumerate}[i)]
	\item the packet loss probability of the incoming channel;
	\item the input rank distribution at the previous node; and
	\item the decision of the recoding scheme at the previous node.\footnote{The previous node can use any recoding scheme, which is out of control by the current node.}
\end{enumerate}

However, the previous node may not know its input rank distribution, then it needs the above three points from its previous node to calculate it.
The problem recursively expends to the source node.
That is, it becomes a centralized problem, which is not practical.
On the other hand, it is not practical to wait for all incoming batches, as the delay would be tremendous.

One solution is to use small blocks at the beginning and record the statistics of the ranks of the incoming batches.
When the number of incoming batches is large enough, the empirical distribution formed by the collected statistics is close to the exact input rank distribution.
Then, we can use the empirical input rank distribution to approximate the case of having received all the incoming batches.

According to Theorem~\ref{thm:diff1}, we can consider a solution such that $|t_b - t_{b'}| \le 1$ for all $b, b' \in \mathcal{B}_r$, $r = 0, 1, \ldots, M$.
Let $S = \min_{b \in \mathcal{B}_R} t_b$.
Suppose $t_b = S$ or $S+1$ for all $b \in \mathcal{B}_R$.
Let $\mathcal{B}'_R = \{b \in \mathcal{B}_R \colon t_b = S+1\}$. 
Then, the sum of expected rank functions for the batches having rank $R$ is
\begin{equation*}
	|\mathcal{B}_R \setminus \mathcal{B}'_R| E(R,S) + |\mathcal{B}'_R| E(R,S+1)
	= |\mathcal{B}_R| \left( \frac{|\mathcal{B}_R| - |\mathcal{B}'_R|}{|\mathcal{B}_R|} E(R,S) + \frac{|\mathcal{B}'_R|}{|\mathcal{B}_R|} E(R,S+1) \right).
\end{equation*}

To simplify the expression, define
\begin{equation*}
	E\left(R, S+\frac{|\mathcal{B}'_R|}{|\mathcal{B}_R|}\right)\\
	:= \frac{|\mathcal{B}_R| - |\mathcal{B}'_R|}{|\mathcal{B}_R|} E(R,S) + \frac{|\mathcal{B}'_R|}{|\mathcal{B}_R|} E(R,S+1).
\end{equation*}

In this case, we say that the batches having rank $R$ transmits $S+|\mathcal{B}'_R|/|\mathcal{B}_R|$ recoded packets, although the number is a non-integral rational number.
With this definition, we can denote by $t_r$ the number of recoded packets to be transmitted for the batches having rank $r$.

Now, we can apply Theorem~\ref{thm:diff1} to reformulate \eqref{eq:B}.
We can rewrite the objective of \eqref{eq:B} by
\begin{equation*}
	\sum_{b \in \mathcal{L}} E(r_b, t_b) = \sum_{r = 0}^M |\mathcal{B}_r| E(r, t_r), 
\end{equation*}
where $t_r$ is a non-negative rational number.
Theorem~\ref{thm:diff1} also states that there is at most one $r$ can satisfy $|t_b - t_{b'}| = 1$ for all $b, b' \in \mathcal{B}_r$.
This means that there is at most one non-integral $t_r$.

Similarly, we can rewrite the constraint by
\begin{equation*}
	\sum_{b \in \mathcal{L}} t_b = \sum_{r = 0}^M |\mathcal{B}_r| t_r 
	= t_\text{max}^\mathcal{L}.
\end{equation*}
That is, we have the following optimization problem:
\begin{equation}
\tag{IP} \label{eq:IP}
\begin{IEEEeqnarraybox}[][c]{rCl}
	\max_{\substack{t_r \in \mathbb{Q}^+ \cup \{0\},\\\forall r = 0, 1, \ldots, M}} & \quad & \sum_{r = 0}^M |\mathcal{B}_r| E(r, t_r)\\
	\mathrm{s.t.} && \sum_{r = 0}^M |\mathcal{B}_r| t_r = t_\text{max}^\mathcal{L},\\
				&& \text{there is at most one non-integral } t_r.
\end{IEEEeqnarraybox}
\end{equation}

When we have collected enough statistics, we can consider a block $\mathcal{L}$ which contains all the already received batches.
The ratio $|\mathcal{B}_r| : |\mathcal{L}|$ can be used to approximate the portion of batches having rank $r$ among all the incoming batches.
Note that if we scale the size of $|\mathcal{B}_r|$ for all $r = 0, 1, \ldots, M$ in \eqref{eq:IP}, we also have to scale the value of $t_\text{max}^\mathcal{L}$.
As the scaling factor is a constant which can be moved away from the maximization, we can use \eqref{eq:IP} for the block $\mathcal{L}$ as an approximation to the same problem for the a block containing all the incoming batches.

Now, suppose we have a set of $\{t_r\}_{r = 0}^M$ solving \eqref{eq:IP}.
When a new batch having rank $R$ arrives, we can immediately know that we should transmit $t_R$ recoded packets if $t_R$ is an integer.
If $t_R$ is not an integer, then we can first transmit $\lfloor t_R \rfloor$ recoded packets.
After that, we have a $t_R - \lfloor t_R \rfloor$ chance to transmit one more recoded packet.

However, we still have the following issues:
\begin{enumerate}[i)]
	\item the algorithms in Section~\ref{sec:algo} take a longer time to solve the problem when $|\mathcal{L}|$ is large;
	\item a batch having rank $R$ where $|\mathcal{B}_R| = 0$ may arrive in the future, but we do not have a reasonable $t_R$ for it.
\end{enumerate}

Issue (i) is obvious as the time complexity of any algorithm in Section~\ref{sec:algo} includes the term $\mathcal{O}(|\mathcal{L}|)$.
Issue (ii) may occur as we are only having an empirical distribution.
However for \eqref{eq:IP}, $|\mathcal{B}_R| = 0$ means that any value of $t_R$ will not affect the optimal solution and the constraint.
So, we can actually provide a reasonable $t_R$, i.e., consider the batch having rank $R$ really appears but takes no contribution to the optimization problem.

When $t_\text{max}^\mathcal{L} \le \sum_{r = 0}^M r|\mathcal{B}_r|$, Lemma~\ref{lem:first_packet} tells us every feasible solution satisfying $t_b \le r_b$ for all $b \in \mathcal{L}$ can solve \eqref{eq:IP}.
Although a similar approach used in Algorithm~\ref{alg:opt1} can satisfy the requirement, it may leave some bachtes having non-zero rank transmitting nothing.
This clearly cannot resolve issue (ii).

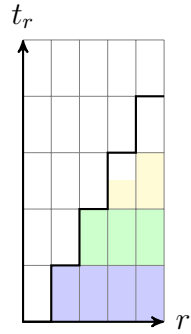
\begin{figure}
	\centering
	\begin{tikzpicture}[scale=.75]
		\fill[blue!20] (.5,0) rectangle ++(2,1);
		\fill[green!20] (1,1) rectangle ++ (1.5,1);
		\fill[yellow!20] (2,2) rectangle ++ (.5,1);
		\fill[yellow!20] (1.5,2) rectangle ++ (.5,.5);
		\draw[xscale=.5,step=1,gray,very thin] (0,0) grid (5,5);
		\draw[thick] (0,0) -| ++(.5,1) -| ++(.5,1) -| ++(.5,1) -| ++(.5,1) -- ++(.5,0);
		\draw[<->,thick] (0,5) node (yaxis) [above] {$t_r$} |- (2.5,0) node (xaxis) [right] {$r$};
	\end{tikzpicture}
	\caption{
		This figure illustrates an example of Algorithm~\ref{alg:opt3a}.
		We can treat the bold staircase as an uneven celling of a water container which has a flat bottom.
		Every iteration fills the height of a stair if there are enough timeslots, i.e., the blue and green cells are filled in the first and second iterations respectively.
		To fill a cell at the $r$-th column, we need $|\mathcal{B}_r|$ timeslots.
		When the timeslots are not enough to fill the whole stair, we fill the cells one by one from the highest rank, which are the yellow cells.
		At last if the remaining timeslots cannot fill a cell completely, we use up all the remaining timeslots to fill it partially.
	}
	\label{fig:alg3a}
\end{figure}

To handle this problem, we propose Algorithm~\ref{alg:opt3a} which acts similar to a water filling algorithm.\footnote{
	Comparing to the traditional water filling algorithms used in communication systems, our water container is upside down.
	We have a flat bottom and an uneven ceiling.
}
The algorithm increases all $t_r$ where $t_r < r$ by $1$ if there are enough unassigned timeslots.
When the number of timeslots are not enough, it allocates the remaining timeslots from the highest rank, which is $M$, in a descending order.
Fig.~\ref{fig:alg3a} illustrates the idea of this algorithm.

This assignment ensures that all ranks are considered, and we have a feasible solution satisfying $t_r \le r$ for all $r = 0, 1, \ldots, M$.
It is equivalent to $t_b \le r_b$ for all $b \in \mathcal{L}$, thus the correctness is implied by Lemma~\ref{lem:first_packet}.
It is easy to see that the worst case time complexity of Algorithm~\ref{alg:opt3a} is $\mathcal{O}(M)$.

\begin{figure}
\removelatexerror
\begin{algorithm}[H]
	\footnotesize
	\caption{Reasonable Solution for \eqref{eq:IP} when $t_\text{max}^\mathcal{L} \le \sum_{r = 0}^M r|\mathcal{B}_r|$}
	\label{alg:opt3a}
	\KwData{$t_\text{max}^\mathcal{L}; \{|\mathcal{B}_r|\}_{r = 0}^M$ such that $t_\text{max}^\mathcal{L} \le \sum_{r = 0}^M r|\mathcal{B}_r|$}
	\KwResult{An assignment $\{t_r^\ast\}_{r = 0}^M$ solving \eqref{eq:IP} when $t_\text{max}^\mathcal{L} \le \sum_{r = 0}^M r|\mathcal{B}_r|$}
	$t \leftarrow t_\text{max}^\mathcal{L}$ \;
	$s \leftarrow \sum_{r = 1}^M |\mathcal{B}_r|$ \;
	$a \leftarrow 0$ \;
	\For{$i = 1, 2, \ldots, M$}{
		\If{$t < s$}{
			\Break \;
		}
		$t \leftarrow t - s$ \;
		$s \leftarrow s - |\mathcal{B}_i|$ \;
		$a \leftarrow a + 1$ \;
	}
	$t_r \leftarrow \min\{a, r\}, \forall r = 0, 1, \ldots, M$ \;
	\For{$j = M, M-1, \ldots, a+1$}{
		\If{$t < |\mathcal{B}_j|$}{
			$t_j \leftarrow t_j + t/|\mathcal{B}_j|$ \;
			\Break \;
		}\Else{
			$t_j \leftarrow t_j + 1$ \;
			$t \leftarrow t - |\mathcal{B}_j|$ \;
		}
	}
	\Return The assignment $\{t_r\}_{r = 0}^M$ \;
\end{algorithm}
\end{figure}

When $t_\text{max}^\mathcal{L} > \sum_{r = 0}^M r|\mathcal{B}_r|$, we modify Algorithm~\ref{alg:opt1} a bit to resolve the both issues.
First, we do not assign the timeslot one by one and batch by batch.
We assign one timeslot to all batches in $\mathcal{B}_r$ for some $r$ in a single iteration when it is possible, i.e., we consider the change of $t_r$.
Second, no matter $\mathcal{B}_R$ is empty or not, we calculate the value of $t_R$, $R = 0, 1, \ldots, M$.

Algorithm~\ref{alg:opt3} is the improved algorithm for empirical input rank distributions, which the output follows the guideline shown in Theorem~\ref{thm:diff1}.

\begin{figure}
\removelatexerror
\begin{algorithm}[H]
	\footnotesize
	\caption{Solver of BAR for Empirical Input Rank Distributions}
	\label{alg:opt3}
	\KwData{$t_\text{max}^\mathcal{L}; \{|\mathcal{B}_r|\}_{r = 0}^M$}
	\KwResult{An assignment $\{t_r^\ast\}_{r = 0}^M$ solving \eqref{eq:IP}}
	\If{$t_\text{max}^\mathcal{L} \le \sum_{r = 0}^M r|\mathcal{B}_r|$}{
		\Return The output of Algorithm~\ref{alg:opt3a} \;
	}
	$t \leftarrow t_\text{max}^\mathcal{L} - \sum_{r = 0}^M r|\mathcal{B}_r|$ \;
	$t_r \leftarrow r, \forall r = 0, 1, \ldots, M$ \;
	\While{$t > 0$}{
		$r \leftarrow$ an element in $\argmax_{r = 0, 1, \ldots, M} \beta_p(t_r,r)$ \;
		\If{$|\mathcal{B}_r| > t$}{
			$t_r \leftarrow t_r + t/|\mathcal{B}_r|$ \;
			$t \leftarrow 0$ \;
		}\Else{
			$t_r \leftarrow t_r + 1$ \;
			$t \leftarrow t - |\mathcal{B}_r|$ \;
		}
	}
	\Return The assignment $\{t_r\}_{r = 0}^M$ \;
\end{algorithm}
\end{figure}

\begin{theorem}
	Algorithm~\ref{alg:opt3} outputs $\{t_r^\ast\}_{r = 0}^M$ which solves \eqref{eq:IP} in $\mathcal{O}(M+\sum_{r = 0}^M \lceil t_r^\ast \rceil \log M)$ time.
\end{theorem}

\begin{IEEEproof}
	We have discussed the optimality when $t_\text{max}^{\mathcal{L}} \le \sum_{b \in \mathcal{L}} r_b = \sum_{r = 0}^M r|\mathcal{B}_r|$ already.
	Now, consider $t_\text{max}^{\mathcal{L}} > \sum_{b \in \mathcal{L}} r_b = \sum_{r = 0}^M r|\mathcal{B}_r|$.
	If $\mathcal{B}_R$ is empty for some $R = 0, 1, \ldots, M$, then any value of $t_R$ will not affect the objective value nor the constraint.
	On the other hand, by Theorem~\ref{thm:diff1}, we can consider the batches having the same rank in some consecutive iterations in Algorithm~\ref{alg:opt1}, i.e., we can change $t_r$ in a single iteration.
	As \eqref{eq:B} and \eqref{eq:IP} are equivalent under this setting, the correctness of Algorithm~\ref{alg:opt1} shown in Theorem~\ref{thm:alg:opt1} implies the correctness of Algorithm~\ref{alg:opt3}.

	Next, we calculate the time complexity.
	If $t_\text{max}^{\mathcal{L}} = \sum_{r = 0}^M r|\mathcal{B}_r|$, then we switch to run Algorithm~\ref{alg:opt3a} instead, so the time complexity is $\mathcal{O}(M)$.
	Otherwise, we first run an initialization including the heapify of the binary heaps, which takes $\mathcal{O}(M)$ time.
	Then, we totally have $\sum_{r = 0}^M \lceil t_r^\ast \rceil$ iterations.
	Each iteration query an element in $\argmax_{r = 0, 1, \ldots, M} \beta_p(t_r,r)$, which takes $\mathcal{O}(\log M)$ time by using a binary heap.
	The remaining operations in an iteration take constant time to run.
	So, the overall time complexity is $\mathcal{O}(M+\sum_{r = 0}^M \lceil t_r^\ast \rceil \log M)$.
\end{IEEEproof}

If $|\mathcal{B}_r| \neq 0$ for all $r = 0, 1, \ldots, M$, which is the usual case, then we can write down an upper bound of $\sum_{r = 0}^M \lceil t_r^\ast \rceil$.
Let $B^\dagger = \min_{r = 0, 1, \ldots, M} |\mathcal{B}_r| \ge 1$.
Note that the solution given by Algorithm~\ref{alg:opt3} has at most one $r$ such that $t_r^\ast$ is not an interger, so we have
\begin{equation*}
	\sum_{r = 0}^M \lceil t_r^\ast \rceil \le \sum_{r = 0}^M t_r^\ast + 1 \le \frac{1}{B^\dagger} \sum_{r = 0}^M |\mathcal{B}_r| t_r^\ast + 1 = \frac{t^\mathcal{L}_\text{max}}{B^\dagger} + 1,
\end{equation*}
where the last equality follows the constraint of \eqref{eq:IP}.

Suppose we have a $t_r$ which is not an integer for some $r$, then we need to decide if we should transmit $\lfloor t_r \rfloor$ or $\lfloor t_r \rfloor + 1$ packets, which we need a random number generator.
If it is expensive to obtain a random number, we can sacrifice the optimality by using a deterministic number of packets.
As the number of timeslots is limited, we cannot send more packets than what we are allowed to send.
So, one choice is to drop the fractional part of $t_r$.

\section{Proof of Theorem~\ref{thm:2blocks}}
\label{sec:proof:thm:2blocks}

	Denote the problem \eqref{eq:B} for the block $\mathcal{L} \cup \mathcal{L}'$ by (B').
	Note that a batch cannot exist in both blocks, i.e., $\mathcal{L} \cap \mathcal{L}' = \emptyset$, and so $|\mathcal{L} \cup \mathcal{L}'| = |\mathcal{L}| + |\mathcal{L}'|$.

	Suppose we solve \eqref{eq:B} for the blocks $\mathcal{L}$ and $\mathcal{L}'$ separately.
	Let $\{t_b\}$ be the solution, where $b \in \mathcal{L} \cup \mathcal{L}'$.

	As $\{t_b\}$ is feasible to both problems, we have
	\begin{equation*}
		\sum_{b \in \mathcal{L}} t_b = t^{\mathcal{L}}_\text{max} \qquad \text{and} \qquad \sum_{b \in \mathcal{L}'} t_b = t^{\mathcal{L}'}_\text{max}.
	\end{equation*}
	This implies that
	\begin{equation*}
		\sum_{b \in \mathcal{L} \cup \mathcal{L}'} t_b = t^{\mathcal{L}}_\text{max} + t^{\mathcal{L}'}_\text{max} = t^{\mathcal{L} \cup \mathcal{L}'}_\text{max},
	\end{equation*}
	which means that $\{t_b\}_{b \in \mathcal{L} \cup \mathcal{L}'}$ is in the feasible region of (B').

	The sum of expected rank by maximizing separately is
	\begin{equation*}
		\sum_{b \in \mathcal{L}} E(r_b,t_b) + \sum_{b \in \mathcal{L}'} E(r_b,t_b) = \sum_{b \in \mathcal{L} \cup \mathcal{L}'} E(r_b,t_b),
	\end{equation*}
	which is equal to the objective value of (B') with the feasible solution $\{t_b\}_{b \in \mathcal{L} \cup \mathcal{L}'}$.
	Therefore, the optimal value of (B') is larger than or equal to the sum of expected rank of maximizing two blocks separately.

\section{Proof of Lemma~\ref{lem:bp}}
\label{sec:proof:lem:bp}

By \cite[Eq.~8.17.20, 8.17.21]{NIST:DLMF}, we have the following properties when $a, b$ are positive integers and $0 \le x \le 1$:
\begin{IEEEeqnarray}{rCl}
	I_x(a,b) - I_x(a+1,b) & = & \binom{a+b-1}{a} x^a (1-x)^b; \label{eq:betainc1} \\
	I_x(a,b+1) - I_x(a,b) & = & \binom{a+b-1}{b} x^a (1-x)^b. \label{eq:betainc2}
\end{IEEEeqnarray}

\begin{IEEEproof}[Proof of \eqref{lem:bp_rec}]
	It is trivial for $i = 0$.
	For $i > 0$, recall the recursive formula of binomial coefficients \cite[Eq.~1.2.7]{NIST:DLMF}:
	\begin{equation*}
		\binom{t+1}{i} = \binom{t}{i-1} + \binom{t}{i}, i = 1, 2, \ldots, t.
	\end{equation*}
	Applying the formula, we have
	\begin{IEEEeqnarray*}{Cl+x*}
		& B_p(t+1,i)\\
		= & \binom{t+1}{i} (1-p)^i p^{t+1-i}\\
		= & (1-p) \binom{t}{i-1} (1-p)^{i-1} p^{t-(i-1)} + p \binom{t}{i} (1-p)^i p^{t-i}\\
		= & (1-p) B_p(t,i-1) + p B_p(t,i). & \IEEEQEDhere
	\end{IEEEeqnarray*}
\end{IEEEproof}

\begin{IEEEproof}[Proof of \eqref{lem:bp_dec}]
	Case I: $t < r$.
	By \eqref{eq:beta_range} and \eqref{eq:beta1}, $\beta_p(t+1,r) \le 1 = \beta_p(t,r)$, and the equality holds if and only if $t+1 \le r-1 < r$.

	Case II: $t \ge r > 0$.
	By \eqref{eq:betainc0} and \eqref{eq:betainc1},
	\begin{IEEEeqnarray*}{Cl}
		& \beta_p(t,r) - \beta_p(t+1,r)\\
		= & I_p(t-r+1,r) - I_p(t-r+2,r)\\
		= & \binom{t}{t-r+1} p^{t-r+1} (1-p)^r\\
		> & 0.
	\end{IEEEeqnarray*}

	Case III: $t \ge r = 0$.
	By \eqref{eq:beta0}, the equality always hold.
\end{IEEEproof}

\begin{IEEEproof}[Proof of \eqref{lem:bp_diag}]
	Case I: $t < r$.
	By \eqref{eq:beta1}, the equality always hold.

	Case II: $t \ge r > 0$.
	By \eqref{eq:betainc0} and \eqref{eq:betainc2},
	\begin{IEEEeqnarray*}{Cl}
		& \beta_p(t+1,r+1) - \beta_p(t,r)\\
		= & I_p(t-r+1,r+1) - I_p(t-r+1,r)\\
		= & \binom{t}{r} p^{t-r+1} (1-p)^r\\
		> & 0.
	\end{IEEEeqnarray*}

	Case III: $t \ge r = 0$.
	By \eqref{eq:beta_range} and \eqref{eq:beta0}, $\beta_p(t,r) = 0 < \beta_p(t+1,r+1)$.
\end{IEEEproof}

\begin{IEEEproof}[Proof of \eqref{lem:bp_inc}]
	Case I: $t = -1$.
	By definition, $\beta_p(t,r+1) = \beta_p(t,r) = 1$.

	Case II: $t \ge 0$.
	Recall that $\beta_p(t,r)$ is the partial sum of the probability mass of the binomial distribution $\bin(t,1-p)$.
	By summing one more term, i.e., $\beta_p(t,r+1)$, the partial sum must be larger than or equal to $\beta_p(t,r)$.
	Note that $B_p(t,i) \neq 0$ when $0 \le i \le t$, so the equality holds if and only if $\beta_p(t,r) = 1$, if and only if $t < r$ by \eqref{eq:beta1}.
\end{IEEEproof}

\begin{IEEEproof}[Proof of \eqref{lem:bp_max} and \eqref{lem:bp_min}]
	Inductively by \eqref{lem:bp_dec}, we have
	\begin{equation} \label{eq:lem:bp1}
		\beta_p(t_a+u,r_a) \le \beta_p(t_a,r_a) \le \beta_p(t_a-v,r_a)
	\end{equation}
	for all $a \in \Lambda$ where $u, v$ are non-negative integers such that $t_a-v \ge -1$.
	By \eqref{eq:beta_range},
	\begin{equation} \label{eq:lem:bp2}
		0 \le \min_{b \in \Lambda} \beta_p(t_b,r_b) \le \beta_p(t_a,r_a) \le \max_{b \in \Lambda} \beta_p(t_b,r_b) \le 1
	\end{equation}
	for all $a \in \Lambda$. Combining \eqref{eq:lem:bp1} and \eqref{eq:lem:bp2}, the proof is done.
\end{IEEEproof}

\section{Proof of Lemma~\ref{lem:exp_rank}}
\label{sec:proof:lem:exp_rank}

	By Lemma~\ref{lem:exp_rank2}, we have $E(r,t+1) = E(r,t) + (1-p)\beta_p(t,r)$.
	If $t < r$, we have
	\begin{equation*}
		\beta_p(t,r) = \sum_{i = 0}^{r-1} B_p(t,i) = 1,
	\end{equation*}
	which proves \eqref{lem:exp_rank_rec}.

	For \eqref{lem:exp_rank_exact}, note that we have the initial condition
	\begin{equation*}
		E(r,0) = B_p(0,0) \min\{r,0\} = 0 = (1-p) \sum_{j = 0}^{(0)-1} \beta_p(j,r).
	\end{equation*}
	We can evaluate Lemma~\ref{lem:exp_rank2} recursively and obtain the first equality in \eqref{lem:exp_rank_exact}.

	By \eqref{lem:exp_rank_rec}, we can show that when $t < r$, we have
	\begin{equation} \label{eq:lem:exp_rank_r}
		E(r,t) = t(1-p).
	\end{equation}
	This implies that when $t \ge r$, we have
	\begin{equation} \label{eq:lem:exp_rank_t}
		E(r,t) = (1-p) \left( r + \sum_{j = r}^{t-1} \beta_p(j,r) \right).
	\end{equation}
	When $t < r$, the summation term $\sum_{j = r}^{t-1} \beta_p(j,r)$ in \eqref{eq:lem:exp_rank_t} equals $0$.
	So, we can combine \eqref{eq:lem:exp_rank_r} and \eqref{eq:lem:exp_rank_t} and give
	\begin{equation*}
		E(r,t) = (1-p) \left( \min\{r,t\} + \sum_{j = r}^{t-1} \beta_p(j,r) \right). \IEEEQEDhereeqn
	\end{equation*}

\section{Proof of Theorem~\ref{thm:adp_recode}}
\label{sec:proof:thm:adp_recode}

	Suppose $t_m > t_n$ for some $r_m < r_n$, i.e.,
	\begin{equation} \label{eq:thm:adp1}
		t_m > t_n \ge r_n > r_m.
	\end{equation}
	Define
	\begin{equation*}
		t_b' = \begin{cases}
			t_m & \text{if } b = n,\\
			t_n & \text{if } b = m,\\
			t_b & \text{otherwise}
		\end{cases}
	\end{equation*}
	for all $b \in \mathcal{L}$.
	Consider the difference of
	\begin{IEEEeqnarray*}{Cl}
		& \sum_{b \in \mathcal{L}} E(r_b,t_b') - \sum_{b \in \mathcal{L}} E(r_b,t_b)\\
		= & [E(r_m,t_n)+E(r_n,t_m)] - [E(r_m,t_m)+E(r_n,t_n)]\\
		= & [E(r_n,t_m)-E(r_n,t_n)] + [E(r_m,t_n)-E(r_m,t_m)]\\
		= & (1-p) \left[ \left( \sum_{j = 0}^{t_m-1} \beta_p(j,r_n) - \sum_{j = 0}^{t_n-1} \beta_p(j,r_n) \right) \right.\\
		  & \left. +\: \left( \sum_{j = 0}^{t_n-1} \beta_p(j,r_m) - \sum_{j = 0}^{t_m-1} \beta_p(j,r_m) \right) \right] \yesnumber \label{eq:thm:adp2} \\
		= & (1-p) \left[ \sum_{j = t_n}^{t_m-1} \beta_p(j,r_n) - \sum_{j = t_n}^{t_m-1} \beta_p(j,r_m) \right]\\
		= & (1-p) \sum_{j = t_n}^{t_m-1} \left( \beta_p(j,r_n) - \beta_p(j,r_m) \right)\\
		> & 0, \yesnumber \label{eq:thm:adp3}
	\end{IEEEeqnarray*}
	where 
	\begin{itemize}
		\item \eqref{eq:thm:adp2} follows Lemma~\ref{lem:exp_rank}\eqref{lem:exp_rank_exact};
		\item \eqref{eq:thm:adp3} follows Lemma~\ref{lem:bp}\eqref{lem:bp_inc} together with \eqref{eq:thm:adp1}.
	\end{itemize}

	The above result contradicts $\{t_b\}_{b \in \mathcal{L}}$ solves \eqref{eq:B}, which gives that $t_m \le t_n$ for all $r_m < r_n$.

	Next, suppose $t_m = t_n$ for some $r_m < r_n$, i.e.,
	\begin{equation} \label{eq:thm:adp4}
		t_m = t_n \ge r_n > r_m.
	\end{equation}
	Define 
	\begin{equation*}
		t_b'' = \begin{cases}
			t_n+1 & \text{if } b = n,\\
			t_m-1 & \text{if } b = m,\\
			t_b & \text{otherwise}
		\end{cases}
	\end{equation*}
	for all $b \in \mathcal{L}$.
	Again, we compare the difference of
	\begin{IEEEeqnarray*}{Cl}
		& \sum_{b \in \mathcal{L}} E(r_i,t_i'') - \sum_{b \in \mathcal{L}} E(r_i,t_i)\\
		= & [E(r_n,t_n+1) + E(r_m,t_m-1)]\\
		  & -\: [E(r_n,t_n) + E(r_m,t_m)]\\
		= & [E(r_n,t_n+1) - E(r_n,t_n)]\\
		  & -\: [E(r_m,t_m) - E(r_m,t_m-1)]\\
		= & (1-p) [\beta_p(t_n,r_n) - \beta_p(t_m-1,r_m)] \yesnumber \label{eq:thm:adp5} \\
		\ge & (1-p) [\beta_p(t_m,r_m+1) - \beta_p(t_m-1,r_m)] \yesnumber \label{eq:thm:adp6} \\
		> & 0, \yesnumber \label{eq:thm:adp7}
	\end{IEEEeqnarray*}
	where
	\begin{itemize}
		\item \eqref{eq:thm:adp5} follows Lemma~\ref{lem:exp_rank}\eqref{lem:exp_rank_rec};
		\item \eqref{eq:thm:adp6} follows \eqref{eq:thm:adp4} and Lemma~\ref{lem:bp}\eqref{lem:bp_inc};
		\item \eqref{eq:thm:adp7} follows Lemma~\ref{lem:bp}\eqref{lem:bp_diag} together with \eqref{eq:thm:adp4}.
	\end{itemize}

	It contradicts $\{t_b\}_{b \in \mathcal{L}}$ solves \eqref{eq:B}.
	So, we have $t_m \neq t_n$ for all $r_m < r_n$.

	Combining the two cases, the proof is done.

\section{Proof of Theorem~\ref{thm:not_opt}}
\label{sec:proof:thm:not_opt}

	We first prove the sufficient condition.
	If $\{t_b\}_{b \in \mathcal{L}}$ does not solve \eqref{eq:B}, then it means that there exists another configuration $\{t_b'\}_{b \in \mathcal{L}}$ which can give a higher objective value.
	As $\sum_{b \in \mathcal{L}} t_b = \sum_{b \in \mathcal{L}} t_b' = t_\text{max}^{\mathcal{L}}$, there exists distinct $\kappa, \rho \in \mathcal{L}$ such that $t_\kappa' > t_\kappa$ and $t_\rho' < t_\rho$.
	Note that $t_\rho' \ge 0$ so we must have $t_\rho \ge 1$.
	Define
	\begin{equation*}
		\Theta = \{\kappa \colon t_\kappa' > t_\kappa\} \quad \text{and} \quad \Phi = \{\rho \colon t_\rho' < t_\rho\},
	\end{equation*}
	where
	\begin{equation} \label{eq:lem:exp_largest_sum0}
		\sum_{\theta \in \Theta} (t_\theta' - t_\theta) = \sum_{\phi \in \Phi} (t_\phi - t_\phi') > 0.
	\end{equation}

	Using the fact that $\{t_b'\}_{b \in \mathcal{L}}$ gives a larger objective value and by Lemma~\ref{lem:exp_rank}\eqref{lem:exp_rank_exact}, we have
	\begin{equation} \label{eq:lem:exp_largest_sum1}
		\sum_{\theta \in \Theta} \sum_{t = t_\theta}^{t_\theta'-1} (1-p) \beta_p(t,r_\theta) > \sum_{\phi \in \Phi} \sum_{t = t_\phi'}^{t_\phi-1} (1-p) \beta_p(t,r_\phi).
	\end{equation}

	Now, we fix $\kappa, \rho$ such that
	\begin{equation*}
		\kappa \in \argmax_{\theta \in \Theta} \beta_p(t_\theta,r_\theta) \quad \text{and} \quad \rho \in \argmin_{\phi \in \Phi} \beta_p(t_\phi-1,r_\phi).
	\end{equation*}

	We have
	\begin{IEEEeqnarray*}{Cl}
		& \sum_{\theta \in \Theta} (t_\theta'-t_\theta) (1-p) \beta_p(t_\kappa,r_\kappa)\\
		\ge & \sum_{\theta \in \Theta} (t_\theta'-t_\theta) (1-p) \beta_p(t_\theta,r_\theta)\\
		\ge & \sum_{\theta \in \Theta} \sum_{t = t_\theta}^{t_\theta'-1} (1-p) \beta_p(t,r_\theta) \yesnumber \label{eq:lem:exp_largest_sum2} \\
		> & \sum_{\phi \in \Phi} \sum_{t = t_\phi'}^{t_\phi-1} (1-p) \beta_p(t,r_\phi) \yesnumber \label{eq:lem:exp_largest_sum3} \\
		\ge & \sum_{\phi \in \Phi} (t_\phi - t_\phi') (1-p) \beta_p(t_\phi-1,r_\phi) \yesnumber \label{eq:lem:exp_largest_sum4} \\
		\ge & \sum_{\phi \in \Phi} (t_\phi - t_\phi') (1-p) \beta_p(t_\rho-1,r_\rho),
	\end{IEEEeqnarray*}
	where
	\begin{itemize}
		\item \eqref{eq:lem:exp_largest_sum2} and \eqref{eq:lem:exp_largest_sum4} follows Lemma~\ref{lem:bp}\eqref{lem:bp_dec};
		\item \eqref{eq:lem:exp_largest_sum3} is the inequality shown in \eqref{eq:lem:exp_largest_sum1}.
	\end{itemize}
	Applying \eqref{eq:lem:exp_largest_sum0}, we have
	\begin{equation*}
		(1-p) \beta_p(t_\kappa,r_\kappa) > (1-p) \beta_p(t_\rho-1,r_\rho),
	\end{equation*}
	which proves the sufficient condition.

	Now we consider the necessary condition, where we have
	\begin{equation} \label{eq:lem:largest_sum0}
		(1-p) \beta_p(t_\kappa,r_\kappa) > (1-p) \beta_p(t_\rho-1,r_\rho)
	\end{equation}
	for some distinct $\kappa, \rho \in \mathcal{L}$.
	Let
	\begin{equation*}
		t_b' = \begin{cases}
			t_\kappa+1 & \text{if } b = \kappa,\\
			t_\rho-1 & \text{if } b = \rho,\\
			t_b & \text{otherwise}
		\end{cases}
	\end{equation*}
	for all $b \in \mathcal{L}$.
	Then, we consider the following:
	\begin{IEEEeqnarray*}{Cl}
		& \sum_{b \in \mathcal{L}} E(r_b,t_b')\\
		= & \sum_{b \in \mathcal{L}\setminus\{\kappa,\rho\}} E(r_b,t_b) + E(r_\kappa,t_\kappa+1) + E(r_\rho,t_\rho-1)\\
		= & \sum_{b \in \mathcal{L}\setminus\{\kappa,\rho\}} E(r_b,t_b) + E(r_\rho,t_\rho-1)\\
		  & +\: [E(r_\kappa,t_\kappa) + (1-p)\beta_p(t_\kappa,r_\kappa)] \yesnumber \label{eq:lem:largest_sum1} \\
		> & \sum_{b \in \mathcal{L}\setminus\{\kappa,\rho\}} E(r_b,t_b) + E(r_\kappa,t_\kappa)\\
		  & +\: [E(r_\rho,t_\rho-1) + (1-p) \beta_p(t_\rho-1,r_\rho)] \yesnumber \label{eq:lem:largest_sum2} \\
		= & \sum_{b \in \mathcal{L}\setminus\{\rho\}} E(r_b,t_b) + E(r_\rho,t_\rho) \yesnumber \label{eq:lem:largest_sum3} \\
		= & \sum_{b \in \mathcal{L}} E(r_b,t_b),
	\end{IEEEeqnarray*}
	where
	\begin{itemize}
		\item \eqref{eq:lem:largest_sum1} and \eqref{eq:lem:largest_sum3} follow Lemma~\ref{lem:exp_rank}\eqref{lem:exp_rank_rec};
		\item \eqref{eq:lem:largest_sum2} follows \eqref{eq:lem:largest_sum0}.
	\end{itemize}
	This means that $\{t_b\}_{b \in \mathcal{L}}$ is not an optimal solution of \eqref{eq:B}.

\section{Proof of Theorem~\ref{thm:diff1}}
\label{sec:proof:thm:diff1}

	We only consider Algorithm~\ref{alg:opt1} in this proof.

	We first consider the case that $t_\text{max}^\mathcal{L} > \sum_{b \in \mathcal{L}} r_b$.
	Note that \eqref{eq:B} is the same as $(\text{B}^{(t_\text{max}^\mathcal{L})})$.
	We are going to prove the following proposition by induction: there exists a set of $\{t_b\}_{b \in \mathcal{L}}$ solving \eqref{eq:B} such that
	\begin{itemize}
		\item $|t_b - t_{b'}| \le 1$ for all $b, b' \in \mathcal{B}_r$, $r = 0, 1, \ldots, M$; and
		\item there is at most one $R$ which can achieve $|t_b - t_{b'}| = 1$ for some $b, b' \in \mathcal{B}_R$, where $\mathcal{B}_R \cap \argmax_{b \in \mathcal{L}} \beta_p(t_b, r_b) \neq \emptyset$.
	\end{itemize}

	The base case is the subproblem $(\text{B}^{(k_0)})$ where $k_0 = \sum_{b \in \mathcal{L}} r_b$.
	Algorithm~\ref{alg:opt1} initializes that $t_b = r_b$ for all $b \in \mathcal{L}$ and we fall into the \textbf{while} loop.
	At this point, we have $|t_b - t_{b'}| = 0$ for all $b, b' \in \mathcal{B}_r$, $r = 0, 1, \ldots, M$, which is the solution of $(\text{B}^{(k_0)})$.

	Next, we consider the inductive step.
	Assume the proposition is true for the subproblem \eqref{eq:Bk} where $k = k_0, k_0+1, \ldots, t_\text{max}^\mathcal{L}-1$.
	To find the solution solving $(\text{B}^{(k+1)})$, the algorithm selects a batch $b \in \argmax_{b \in \mathcal{L}} \beta_p(t_b,r_b)$, and increase the corresponding $t_b$ by $1$.
	The freedom to select the batch in $\argmax_{b \in \mathcal{L}} \beta_p(t_b,r_b)$ allows the following arguments.

	Case A: there is a $R$ such that $|t_b - t_{b'}| = 1$ for some $b, b' \in \mathcal{B}_R$ where $\mathcal{B}_R \cap \argmax_{b \in \mathcal{L}} \beta_p(t_b, r_b) \neq \emptyset$.
	Let $b \in \mathcal{B}_R \cap \argmax_{b \in \mathcal{L}} \beta_p(t_b, r_b)$, and $b' \in \mathcal{B}_R$ be another batch such that $|t_b - t_{b'}| = 1$.
	We have two cases.

	Case A(I): $b' \in \argmax_{b \in \mathcal{L}} \beta_p(t_b, r_b)$.
	We can select the batch $\argmin_{\alpha \in \{b, b'\}} t_\alpha$ for the algorithm.
	Then, we have $|t_b - t_{b'}| = 0$ after the iteration.

	Case A(II): $b' \not \in \argmax_{b \in \mathcal{L}} \beta_p(t_b, r_b)$.
	By Lemma~\ref{lem:bp}\eqref{lem:bp_dec}, we know that $t_{b'} = t_b + 1$.
	So, we can choose $b$ for the algorithm and we have $|t_b - t_{b'}| = 0$ after the iteration.

	The above two cases show that we can maintain $|t_b - t_{b'}| \le 1$ for all $b, b' \in \mathcal{B}_R$ after the iteration.

	Consider the moment when $\mathcal{B}_R \cap \argmax_{b \in \mathcal{L}} \beta_p(t_b, r_b) \neq \emptyset$.
	Let $S = \min_{b \in \mathcal{B}_R \cap \argmax_{b \in \mathcal{L}} \beta_p(t_b, r_b)} t_b$.
	An iteration will only remove at most one batch in $\mathcal{B}_R \cap \argmax_{b \in \mathcal{L}} \beta_p(t_b, r_b)$.
	If the set $\mathcal{B}_R \cap \argmax_{b \in \mathcal{L}} \beta_p(t_b, r_b)$ is empty after the iteration, we must have
	\begin{enumerate}[i)]
		\item $|\mathcal{B}_R \cap \argmax_{b \in \mathcal{L}} \beta_p(t_b, r_b)| = 1$; and
		\item $\beta_p(S, R) \neq \beta_p(S+1, R)$.
	\end{enumerate}
	By Lemma~\ref{lem:bp}\eqref{lem:bp_dec}, condition (ii) above is equivalent to $\beta_p(S, R) > \beta_p(S+1, R)$.
	So, we can apply Case A(II) to show that if $\mathcal{B}_R \cap \argmax_{b \in \mathcal{L}} \beta_p(t_b, r_b) = \emptyset$ after the iteration, then we have $|t_b - t_{b'}| = 0$ for all $b, b' \in \mathcal{B}_R$.
	That is, the uniqueness of $R$ preserves.

	Case B: $|t_b - t_{b'}| = 0$ for all $b, b' \in \mathcal{B}_r$, $r = 0, 1, \ldots, M$.
	Let $b \in \argmax_{b \in \mathcal{L}} \beta_p(t_b,r_b)$.
	Consider another batch $b' \in \argmax_{b \in \mathcal{L}} \beta_p(t_b,r_b)$ where $r_b = r_{b'}$ and $b \neq b'$.
	If $b'$ exists, then no matter $b$ or $b'$ is selected, we must have $|t_b - t_{b'}| = 1$ after the iteration, and the next iteration will fall into Case A.
	If $b'$ does not exist, then it means that $|\mathcal{B}_{r_b}| = 1$, and the result is trivial.
	The next iteration will still stay in Case B.

	Recall that after an iteration, the set of $\{t_b\}_{b \in \mathcal{L}}$ solves $(\text{B}^{(k+1)})$.
	Case A and B imply that the proposition is true for the subproblem $(\text{B}^{(k+1)})$.
	By induction, the proof for the case $t_\text{max}^\mathcal{L} > \sum_{b \in \mathcal{L}} r_b$ is done.

	Now, we consider the case that $t_\text{max}^\mathcal{L} \le \sum_{b \in \mathcal{L}} r_b$.
	Suppose we have sorted the batches by their ranks in ascending order, and the \textbf{foreach} loop in Algorithm~\ref{alg:opt1} follows the rank in ascending order.
	Then, the output of the algorithm is in the form described below.
	There exists a rank $R$ such that
	\begin{equation*}
		t_b = \begin{cases}
			r_b & \text{if } r_b < R,\\
			0 & \text{if } r_b > R,\\
			0 \text{ or } t \text{ or } r_b & \text{if } r_b = R,
		\end{cases}
	\end{equation*}
	where $0 < t < r_b$ and there is at most one $b$ which have $t_b = t$.

	It is obvious that $|t_b - t_{b'}| = 0$ for all $b, b' \in \mathcal{B}_r$ where $r \in \{0, 1, \ldots, M\} \setminus \{R\}$.
	Now, we consider the batches in $\mathcal{B}_R$.
	We are going to redistribute the assigned timeslots.

	Let $T = \sum_{b \in \mathcal{B}_R} t_b$, which is the total number of timeslots assigned to the batches having rank $R$.
	We can reassign the timeslots by the following steps:
	\begin{enumerate}
		\item assign $\lfloor T/|\mathcal{B}_R| \rfloor$ timeslots to each batch in $\mathcal{B}_R$;
		\item select $T \bmod |\mathcal{B}_R|$ distinct batches from $\mathcal{B}_R$, and assign one more timeslots to each of them.
	\end{enumerate}

	It is easy to see that the above assignment
	\begin{itemize}
		\item uses up all the $T$ timeslots; and
		\item $|t_b - t_{b'}| \le 1$ for all $b, b' \in \mathcal{B}_R$.
	\end{itemize}

	That is, the solution is feasible, which is also optimal by Lemma~\ref{lem:first_packet}.
	Also, we have the condition that $|t_b - t_{b'}| \le 1$ for all $b, b' \in \mathcal{B}_r$, $r = 0, 1, \ldots, M$, and there is at most one $r$ which can achieve the equality.
	The proof is done.

\section{Proof of Theorem~\ref{thm:cond}}
\label{sec:proof:thm:cond}

	Let $\mathbb{B}(a, b; y) := \int_0^y x^{a-1} (1-x)^{b-1} \,dx$ be the incomplete beta function.
	We have the beta function $\mathbb{B}(a, b) := \mathbb{B}(a, b; 1)$.

	From \eqref{eq:betainc0}, we have $\beta_p(t, r) = I_p(t-r+1, r) = \frac{\mathbb{B}(t-r+1, r; p)}{\mathbb{B}(t-r+1, r; 1)}$.
	By direct calculation, the condition number is
	\begin{IEEEeqnarray*}{rCl}
		\left| \frac{p \frac{d\beta_p(t,r)}{dp}}{\beta_p(t,r)} \right| &
		= & \left| \frac{p \frac{d}{dp} \int_0^p x^{t-r} (1-x)^{r-1} \,dx}{\mathbb{B}(t-r+1, r; 1) I_p(t-r+1,r)} \right|\\
		& = & \frac{p^{t-r+1} (1-p)^{r-1}}{\mathbb{B}(t-r+1, r; 1) I_p(t-r+1,r)} \yesnumber \label{eq:cond_ans} \\
		& = & \frac{p^{t-r+1} (1-p)^{r-1}}{\int_0^p x^{t-r} (1-x)^{r-1} \,dx}\\
		& = & \frac{p^{t-r+1} \sum_{j = 0}^{r-1} (-1)^j \binom{r-1}{j} p^j}{\int_0^p x^{t-r} \sum_{j = 0}^{r-1} (-1)^j \binom{r-1}{j} x^j \,dx}\\
		& = & \frac{\sum_{j = 0}^{r-1} (-1)^j \binom{r-1}{j} p^{t-r+j+1}}{\sum_{j = 0}^{r-1} (-1)^j \binom{r-1}{j} \int_0^p x^{t-r+j} \,dx}\\
		& = & \frac{\sum_{j = 0}^{r-1} (-1)^j \binom{r-1}{j} p^{t-r+j+1}}{\sum_{j = 0}^{r-1} (-1)^j \binom{r-1}{j} p^{t-r+j+1} / (t-r+j+1)},
	\end{IEEEeqnarray*}
	where the absolute value disappeared as both numerator and denominator are non-negative.
	The first form of the condition number can be obtained by substituting $\mathbb{B}(t-r+1, r; 1) = \frac{(t-r)!(r-1)!}{t!}$ into \eqref{eq:cond_ans}.

\bibliographystyle{IEEEtran}
\bibliography{adaptive-bib}

\end{document}